\documentclass[12pt,letterpaper]{article}

\usepackage{graphicx,array}
\usepackage{url}
\usepackage{color}
\usepackage{latexsym}
\usepackage{amsthm}
\usepackage{amsmath}
\usepackage{amssymb}
\usepackage{amsfonts}
\usepackage[numbers,sort&compress]{natbib}
\usepackage{bm}
\usepackage{slashed}
\usepackage{mathrsfs}
\usepackage{enumerate}
\usepackage{tikz}
\usepackage{siunitx}
\usepackage{mdframed}
\usepackage{setspace}  
\usepackage{esvect}
\usepackage{esint}
\usepackage{subcaption}
\usepackage[normalem]{ulem}
\usepackage{setspace}
\usepackage{multirow,makecell}

\usepackage[utf8x]{inputenc}%
\usepackage{tcolorbox}%

\usepackage{hyperref} 

\hypersetup{
    colorlinks=true,       
    linkcolor=red,          
    citecolor=blue,        
    filecolor=magenta,      
    urlcolor=blue           
}
\usepackage[all]{hypcap} 

\usepackage{natbib}
\newcommand{\nc}{\newcommand}

\nc{\beq}{\begin{equation}}  
\nc{\eeq}{\end{equation}}  
\nc{\beqa}{\begin{eqnarray}}  
\nc{\eeqa}{\end{eqnarray}}  
\nc{\bit}{\begin{itemize}}  
\nc{\eit}{\end{itemize}}

\usepackage{floatrow}
\newfloatcommand{capbtabbox}{table}[][\FBwidth]

\usepackage{blindtext}

\title{ 
{\bf Axion Dark Matter
\\
 from Cosmic String Network}
\author{\large Heejoo Kim$^{\,\dagger}$, Junghyeon Park$^{\,\dagger}$, and Minho Son$\,^\dagger$}
\date{\small \it 
$^\dagger$Department of Physics, Korea Advanced Institute of Science and Technology, \\
291 Daehak-ro, Yuseong-gu, Daejeon 34141, Republic of Korea\\
}
}

\begin{document}

\maketitle

\setlength{\parskip}{0.2ex}

\begin{abstract}	
We perform lattice simulations to estimate the axion dark matter abundance radiated from the global cosmic strings in the post-inflationary scenario. The independent numerical confirmation on the recently observed logarithmic growth in both the number of strings per Hubble patch and the spectral index of the power law scaling for the axion spectrum is reported. These logarithmic scalings are checked against two different prescriptions for generating initial random field configurations, namely fat-string type and thermal phase transition. We discuss a possible strong correlation between the axion spectrum and the string evolutions with different initial conditions to support the insensitivity of scaling behaviors against different initial data and we provide a qualitative understanding of it. 
The impact of various combinations of the power law of the axion spectrum, nonlinearities around the QCD scale, and average inter-string distances on the axion abundance are discussed.
Additionally, we introduce a new novel string identification method, based on the tetrahedralization of the space, which guarantees the connectedness of the strings and provides a convenient way of assigning the core location. 
Finally we derive a lower bound on the axion mass.
\end{abstract}

\thispagestyle{empty}  
\newpage  
  
\setcounter{page}{1}

\begingroup
\hypersetup{linkcolor=black,linktocpage}
\tableofcontents
\endgroup
\newpage


\section{Introduction}
\label{sec:intro}

Axion as dark matter has been increasingly attractive, although it was originally introduced in some extensions of the Standard Model to resolve a mysterious fine-tuning in quantum chromodynamics (QCD)~\cite{Peccei:1977hh,Weinberg:1977ma,Wilczek:1977pj,Dine:1982ah,Preskill:1982cy,Abbott:1982af}. 
While the QCD axion dark matter scenario is a certainly interesting possibility (\cite{Marsh:2015xka} for a review), its dominant production mechanism is still inconclusive. The topological production such as axions radiated from the global cosmic string network~\cite{Kibble:1984hp} formed during the phase transition of the Peccei-Quinn (PQ) symmetry breaking~\cite{Peccei:1977hh} after inflation in the early universe can be predicted in terms of the symmetry breaking scale. However, it entirely relies on numerical simulations as the effective field theory (EFT) description does not work due to an evolution occurring between two vastly different scales for the PQ symmetry breaking and QCD crossover and a nonlinearity of the equation of motion. While a precise estimation of the topological production with respect to the misalignment mechanism has been subject to a large uncertainty, any progress for the improvement should be crucial in understanding its origin and role of the QCD axion as the dark matter. Especially, the outcome will have a large impact on the direct search for the QCD axion as the ballpark for the natural axion mass will change.

Using the advanced computing resources, some state-of-the-art simulations have been performed by several groups on the largest lattice grids of the order $\mathcal{O}(2^{12\sim13})$ per dimension.
An intriguing discovery of those recent simulations was the strong evidence for a scaling regime, or an attractor solution, exhibiting a logarithmic growth in the number of long strings per Hubble patch, denoted by $\xi$~\cite{Gorghetto:2018myk,Kawasaki:2018bzv,Vaquero:2018tib,Klaer:2019fxc,Buschmann:2019icd,Gorghetto:2020qws,Buschmann:2021sdq,OHare:2021zrq}. In this work, we independently confirm the logarithmic growth within the simulation time coverage.
Since the cosmic strings will eventually decay into axions in late times, the evolution of $\xi$ is an important property that has to be estimated as precisely as possible to determine the axion abundance. If the logarithmic growth in $\xi$ holds until the late times for the QCD crossover, its prediction will be very different from the constant scaling scenario of the order one value~\cite{Hindmarsh:2019csc}. 
Similar constant scaling behaviors observed in early simulations on smaller lattice grids~\cite{Yamaguchi:1998gx,Yamaguchi:1999yp,Yamaguchi:1999dy,Yamaguchi:2002sh,Hiramatsu:2010yu,Kawasaki:2014sqa} may not be helpful to support either scenario due to a limited dynamic time range (see related discussions in~\cite{Bennett:1985qt,Bennett:1986zn,Battye:1993jv,Battye:1994au,Martins:1995tg}). 

The differential axion energy spectrum is another crucial property for an accurate estimation of the axion abundance. In the absence of any non-trivial new scale between the scale of the PQ symmetry breaking and the QCD phase transition, it is expected to follow the power law fall-off behavior. The characteristic scaling feature of the axions radiated from the topological cosmic strings can be captured by the instantaneous emission. However, there has been a disagreement in the determination of the power law scaling $\sim k^{-q}$ in the instantaneous axion emission spectrum. While the static lattice simulation performed in~\cite{Gorghetto:2020qws} finds a hint on the logarithmic growth of the spectral index $q$ in time, an independent simulation in~\cite{Buschmann:2021sdq} with an adaptive mesh refinement (AMR) \footnote{The AMR method can parametrically extend the simulation time by locally increasing the lattice resolution with recursively created finer grids around strings on-the-fly. The AMReX~\cite{zhang2020amrex,AMReX_JOSS}, a software for block-structured AMR, was adopted in~\cite{Buschmann:2021sdq,Benabou:2023ghl}. GRChombo~\cite{Clough:2015sqa} is another AMR based open-source code originally developed for numerical relativity simulation and it was recently applied to cosmic strings~\cite{Drew:2019mzc}.}~\cite{BERGER1984484} finds a consistency with the no-log hypothesis. The former predicts the dominant topological production over the misalignment mechanism (which favors a heavier axion mass than the natural ballpark for the QCD axions from misalignments) whereas the latter predicts a comparable size. 
Two simulation results in~\cite{Gorghetto:2020qws,Buschmann:2021sdq} were obtained with different setups, differing by how the spacetime were discretized, how the initial conditions were generated, and how they were subsequently evolved over time. This makes the transparent comparison quite challenging. 
In this work, we perform independent static lattice simulations on the grids of $N^3 = 4096^3$ by taking two commonly used approaches in literature: one using the fat-string type pre-evolution to prepare for initial conditions for the physical string evolution~\cite{Gorghetto:2018myk,Gorghetto:2020qws} and the other scheme where the field evolution occurs at finite temperature, which will be referred to as the thermal pre-evolution (for instance, our setup for this approach is similar to those in~\cite{Kawasaki:2018bzv,Buschmann:2021sdq}). 
While the fat-string pre-evolution may be criticized for that it is not the same theory as the one for the physical string evolution, the direct comparison between simulations adopting different schemes should greatly help establishing the validity of the final observations or the insensitivity to the aforementioned relaxation schemes. 
The result in~\cite{Kawasaki:2018bzv} has not been reported in similar presentations to~\cite{Gorghetto:2018myk,Gorghetto:2020qws}, making it difficult to be used for a meaningful comparison.
The AMR technique, adopted in~\cite{Buschmann:2021sdq}, allows to cover much longer dynamic time range than what one has ever imagined to be possible. However, whether it eventually leads us to the better determination in the region beyond the time coverage of the static lattice simulation does not seem to be straightforward. Extended comprehensive comparison in the context of AMR will be presented in our companion paper.

The parametric behavior of the axion abundance originated from the scaling regime can be significantly affected by the nonlinearities around the QCD scale~\cite{Gorghetto:2020qws}. In particular, it varies depending on the value of $q$ ($q>1$ vs $q=1$) and axion field values (large vs small relative to the symmetry breaking scale) around the QCD scale (equivalent to the relevance of the nonlinearities) and the Infrared (IR) momentum cutoff (suggested by average inter-string distances). As different combinations of each scenario match to different observations in literature, it should be highly useful to work out all relevant cases. Following the similar strategy to~\cite{Gorghetto:2020qws}, we discuss an impact of the modified IR momentum cutoff and axion field values on the axion abundance. We extend our discussion to the case with $q=1$ to estimate the relevance of the nonlinearities around the QCD scale in that situation. This will make the comparisons between different scenarios transparent.

While there are already various string identification algorithms available in literature, in this work, we propose a new novel string identification algorithm, based on a tetrahedralization of the space, which guarantees the connectedness of strings. As the name stands, the algorithm works on a  tetrahedron instead of a cube. It has a few practical advantages that will be discussed below. Compared to existing techniques~\cite{Fleury:2015aca,Yamaguchi:2002zv,Yamaguchi:2002sh,Hiramatsu:2010yu}, its performance time is comparable with them.

The most severe limiting factors in increasing the simulation size are CPU and memory resources. We have greatly optimized our own independent codes (made from the scratch) running on OpenMPI to speed up each simulation. Nevertheless, obtaining a large ensemble, performing at the same time all kinds of measurements such as $\xi$, energy budget, power spectrum, instantaneous emission, string tension, string velocity and checking out dependencies on various prescriptions for each measurement, from the lattice simulation on the grids of $N^3=4096^3$  is challenging. Therefore, we extensively used lattice simulations on smaller grids such as $N^3=1024^3$ and $N^3=2048^3$ as well for numerous sanity checks and to get guideline results for an educated selection of the benchmark setups for the simulation on the bigger sized lattice of $N^3 = 4096^3$.
Our final result on the axion spectrum, which supports the soft axions of the order of the Hubble scale around the QCD crossover, strongly suggests making simulations on a factor of two larger lattice.\\

Our paper is organized as follows. In Section~\ref{sec:cos:strings:model}, we introduce a model for formation of the global axion strings. In Section~\ref{sec:sim:setup}, brief descriptions on the discretization of the equation of motion and random field configurations are given. In Section~\ref{sec:string:id}, a new novel string identification algorithm based on a tetrahedralization is introduced. In Section~\ref{sec:SR}, we present our independent evidence for the logarithmically growing scaling solution in the number of strings per Hubble patch. 
Also the scaling of the average inter-string distance is discussed. In Section~\ref{sec:cosmo:strings}, detailed discussion on the energy budget and string properties such as the string tension and velocity are given. 
In Section~\ref{sec:axion:spectrum}, we provide our fit results on the spectral index of the power law in the axion spectrum. 
In Section~\ref{sec:axion:abundance}, we discuss the prospects for the axion abundance for the spectral index from our simulation.
In Section~\ref{sec:correlation}, a possible correlation between axion spectrum and string evolutions with different initial field configurations is discussed. In Appendices, all simulation details are given. Extra detailed discussions and materials are given for various subjects. Especially, the parametric behavior of the axion abundance for several distinctive scenarios are worked out.\\

Note added: while this work was being finalized, another work addressing similar aspects of axions from cosmic strings has appeared~\cite{Saikawa:2024bta}.

\section{Formation of cosmic strings}
\label{sec:cos:strings:model}
Topological cosmic strings are generated upon the phase transition of the $U(1)_{PQ}$ symmetry by a complex scalar field $\phi$ in the early Universe~\cite{Kibble:1976sj,Kibble:1980mv,Vilenkin:1981kz}. The PQ symmetry breaking is conveniently parametrized by the Lagrangian,
\begin{equation}\label{eq:Lag:orig}
  \mathcal{L} = \partial_\mu \phi ^* \partial^\mu \phi - \frac{m_r^2}{2 f_a^2} \left ( |\phi|^2 - \frac{f_a^2}{2} \right )^2~,
\end{equation}
where the quartic coupling corresponds to $m_r^2/(2 f_a^2)$, often denoted by $\lambda$ in literature, and $f_a$ is the symmetry breaking scale. The dynamical evolution of the cosmic strings in an expanding universe will be explored, assuming a radiation domination with metric given by 
\begin{equation}
  ds^2 = dt^2 - R^2(t) d \vec{x}^2~,
\end{equation}
where $\vec{x}$ denotes comoving coordinates.
The equation of motion from the Lagrangian in Eq.~(\ref{eq:Lag:orig}) in the expanding universe is given by
\begin{equation}\label{eq:phi:eom}
  \ddot{\phi} + 3 H \dot{\phi} - \frac{1}{R^2} \nabla ^2 \phi + \frac{m_r^2}{f_a^2} \phi \left ( |\phi|^2 - \frac{f_a^2}{2} \right ) = 0~,
\end{equation}
where dot is the differentiation with respect to the cosmic time $t$ and the gradient $\nabla$ is evaluated in the comoving coordinate. 
The equation of motion in Eq.~(\ref{eq:phi:eom}) admits a solitonic solution whose configuration corresponds to topological cosmic strings. 
Cosmic strings formed at the PQ scale will evolve for a long dynamic time range all the way up to the QCD crossover where the abundance of the axion dark matter radiated from the cosmic strings are estimated. The dynamics of the cosmic strings are traced through evolutions of the field. As they are highly nonlinear in between two largely separated scales, an EFT approach is not available and it has to proceed through a numerical simulation on a lattice. The numerical simulation of the equation of motion in Eq.~(\ref{eq:phi:eom}) on a lattice is limited since it involves three different scaling behaviors. While the lattice spacing scales as $R \propto \sqrt{t}$ in the radiation-dominated universe, the string core width $m_r^{-1}$ stays constant and the Hubble length $H^{-1}$ scales as $\sim t$ during the evolution. 
At some point, the simulation runs out of the resolution within the string core or the correlation length $\sim H^{-1}$ becomes comparable with the size of the simulation box and the boundary conditions become relevant.

The field evolution on the lattice in terms of $\phi$ is subject to the convergence condition, or Courant–Friedrichs–Lewy (CFL) condition, $\Delta t < R\Delta x$. This condition requires arbitrarily small time step size in the early stage of the evolution where the scale factor is small. It can be relaxed by re-expressing the equation of motion in terms of the rescaled field $\psi (\tau,\, \vec{x}) = R \phi (t,\, \vec{x})$ in the comoving spacetime coordinate. The CFL condition in this situation is relaxed to $\Delta \tau < \Delta x$. All the simulation details are given in Appendices~\ref{app:sec:eom} and~\ref{app:sec:detail:lattice}.

\section{Simulation Setup}
\label{sec:sim:setup}

\subsection{Discretization}
The equation of motion in Eq.~(\ref{eq:phi:eom}) is discretized in the conformal frame according to the leap-frog method whose truncation error is of the order of $\mathcal{O}((\Delta x)^3)$.
The lattice spacing $\Delta x$ is determined by $m_r$ for a given lattice size $N$. It is set to reach the maximum dynamic time range allowed by the lattice size, namely $\Delta x = \frac{m_r^{-1}}{n_c} \left ( \frac{2 n_c n_H}{N} \right )^{1/2}$ where $n_c$ is the number of lattice grids inside the core and $n_H$ is the number of Hubble patches in the simulation box per dimension at the final time. $n_c = 1$ and $n_H = 4^{1/3}$ will be our default choice in the simulation unless otherwise specified. $n_H = 4^{1/3}$ ensures at least four Hubble volumes in the simulation box at the final time. See Appendix~\ref{sec:app:setup} for the detail.
The time interval $\Delta \tau$ is chosen to be $\Delta \tau/\Delta x = 1/3$ so that it satisfies the CFL condition. Our simulation data sets are summarized in Table~\ref{tab:xi:fit:scaling}.

\subsection{Initial conditions and Relaxation}
\label{sec:initialconditions}

The string configurations produced out of initial conditions in the very beginning of the simulation are highly noisy. Given the limited dynamic time range of the physical string simulation, a pre-evolution for the purpose of relaxing the string network to a relatively clean level seems to be a necessary step. The relaxation step may be justified if the string network eventually approaches toward a scaling regime~\cite{Albrecht:1984xv,Bennett:1987vf,Allen:1990tv} and becomes insensitive to the initial configurations.

The fat-string evolution~\cite{Press:1989yh,Moore:2001px} was adopted in recent works~\cite{Gorghetto:2018myk,Gorghetto:2020qws} for the purpose of the relaxation (see Appendices~\ref{sec:app:option1} and~\ref{sec:app:option2} for details). The outcome of the fat-string pre-evolution feeds into the physical string evolution as an initial condition. It is called fat-string since the string core width is forced to scale as $R$ during the pre-evolution so that $\frac{m_r^{-1}}{R}$ stays constant. It is implemented in the simulation by the quartic coupling scaling as $R^{-2}$ for the fixed $f_a$. 
$R\propto t$ adopted in~\cite{Gorghetto:2020qws} is a convenient choice for the scale factor during the fat-string pre-evolution as it ensures the constant ratio of the Hubble length to the string core width which we can freely choose. The random field configurations were generated such that the Fourier amplitude $\tilde\phi(\vec{k})$ below the momentum cutoff $k < k_\text{max} = m_r$ follows normal distributions. 
The initial configuration is rapidly adjusted via the equation of motion in the broken phase. Similarly to~\cite{Gorghetto:2020qws}, the pre-evolution stops when a requested number of strings per Hubble patch, denoted by $\xi_0$, is achieved. 

Another approach is to evolve the equation of motion from the epoch right before the phase transition where the symmetry was restored (see Appendix~\ref{sec:app:option3} for the detail).
It is realized by adding the temperature dependent term to the potential~\cite{Kawasaki:2018bzv,Buschmann:2021sdq} (and accordingly modifying the equation of motion), $\Delta \mathcal{L} = -\frac{m_r^2}{6 f_a^2} T^2 |\phi|^2$ where 
$T^4 = f_a^4\cdot \frac{H^2}{m_r^2}\cdot 4 \zeta^2$ with $ \zeta^2 = \frac{45 }{2\pi^2 g_*}\frac{M_p^2 m_r^2}{f_a^4}$ and the Hubble parameter $H^2 = \frac{\pi^2}{90} g_* \frac{T^4}{M_p^2}$.
The correlation length over the core size at the critical temperature $T_c$ ($= \sqrt{3} f_a$) is given by $\left . \frac{m_r}{H} \right |_{T_c} = \frac{2}{3} \zeta$. The choice $\zeta = \frac{3}{\sqrt{2}} = 2.12123$ corresponds to the case with $H|_{T_c} = f_a$.
While $\zeta \sim \mathcal{O}(1)$ ensures the order one size of the correlation length in the unit of the core width at the phase transition where cosmic strings are formed, the natural size of it is roughly $\zeta \sim 10^7$ for $g_* \sim \mathcal{O}(10^2)$, the order one size of the quartic coupling, and $f_a \sim \mathcal{O}(10^{10}\ \text{GeV})$ (see Eq.~(\ref{app:eq:zeta}) for the explicit expression).
The order one size of $\zeta$ may correspond to a large symmetry breaking scale of $f_a \sim 10^{17}$ GeV, a large number of relativistic degrees of freedom of $g_* \sim 10^{14}$, or a combination of them for the fixed order one quartic coupling. This might be considered as a drawback of this approach. The random initial conditions for field configurations were generated, assuming a Gaussian random field configuration following the thermal distributions as in~\cite{Kawasaki:2018bzv} in the symmetric phase. We impose the cut on the momentum $k < k_\text{max} =10 m_r$ at the initial time $\tau_i  = 0.1 \tau_c$ ($\tau_c$ as the time at the phase transition). Since $k_\text{max} \propto \tau^{-1}$, $k|_{\tau_i} \leq 10\, m_r$ implies $k|_{\tau_c} \leq m_r$ with the choice of $\tau_i = 0.1 \tau_c$, and this looks consistent with the cut imposed in the aforementioned fat-string type relaxation. Unlike the approach exploiting the fat-string technique, the generated initial configurations smoothly transit to those of the broken phase via the equation of motion with the thermal potential naturally realizing the phase transition.
More details regarding initial conditions are given in Appendix~\ref{app:sec:IC}.

The higher momentum modes above $m_r$ scale can be considered as a short distance or ultraviolet (UV) physics smaller than the string core size. Having the momentum cutoff imposed, the simulation results become insensitive to $\Delta x$ or equivalently the lattice size $N$ (for instance, see Fig.~\ref{fig:app:xi:Nindependence:option2:vs:option3}).
Otherwise, the higher momentum modes set by a smaller $\Delta x$ for a larger $N$ kick in, the spectrum becomes more noisy, and it takes longer time to be relaxed.
The impact of UV modes above $m_r$ on initial conditions in the properties of the string network will be discussed in the Appendix~\ref{app:sec:dependency}.

\section{String identification on tetrahedron}
\label{sec:string:id}

A string is basically identified by examining the pattern of phases of the complex scalar fields. While there are various options for string identification algorithm, we yet introduce a new novel string identification algorithm which we find numerically very efficient and guarantees the connectedness of strings. It starts with a tetrahedralization, sub-dividing the lattice space into a set of tetrahedrons. 
%
%
\begin{figure}[tp]
\begin{center}
\includegraphics[width=0.27\textwidth]{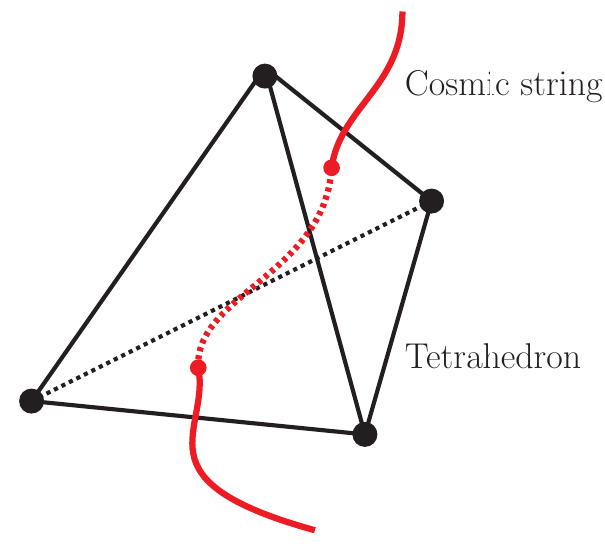} \hspace{2cm}
\includegraphics[width=0.25\textwidth]{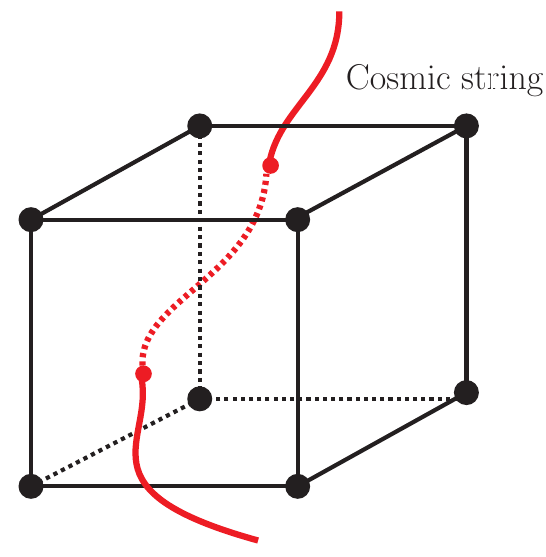}
\caption{\small Illustration of a cosmic string configuration in the tetrahedron (left) and in a cube consisting of six plaquettes (right).}
\label{fig:cartoon:tetrahedron}
\end{center}
\end{figure}
%
%
Each tetrahedron, illustrated in the left panel of Fig.~\ref{fig:cartoon:tetrahedron}, has four triangular faces, and each triangular face has three vertices which we can label as $v_{i=1,2,3}$. Let us define $\theta_{123} = \theta_{12} + \theta_{23} + \theta_{31}$ where $\theta_{ij} = \arg(\phi_j \phi^*_i)$ which measures the phase difference of fields at two adjacent vertices. A string core on the triangular face is declared for the configuration with $\theta_{123} = 2\pi$ ($-2\pi$). This last statement for the string declaration can be replaced with more numerically efficient and equivalent statement, as was explained in Appendix~\ref{app:sec:stringID}. With the assignment of $\omega_{ij} = \Re\phi_i \Im\phi_j - \Im\phi_i\Re\phi_j$ and $\omega_{123} \equiv \omega_{12} + \omega_{23} + \omega_{31}$, $\theta_{123} = 2\pi$ ($- 2\pi$) can be shown to be equivalent to the case where all $\omega_{ij}$'s are non-negative (non-positive) while $\omega_{123}$ is positive (negative):
\begin{equation}
\begin{split}
\theta_{123} = 2\pi & \leftrightarrow \omega_{ij} \geq 0 \quad \text{and}\quad \omega_{123} > 0~,
\\[3pt]
\theta_{123} = - 2\pi & \leftrightarrow \omega_{ij} \leq 0 \quad \text{and}\quad \omega_{123} < 0~.
\end{split}
\end{equation}
The location of the string core inside the triangle $r_\text{core} $ is assigned via the linear interpolation, or $r_\text{core} =(\omega_{23} v_1 + \omega_{31} v_2 + \omega_{12} v_3)/\omega_{123}$ such that the field value vanishes at the core. This may be considered as a leading approximation of a more complete expansion in higher orders of vertices. In this prescription, the string core is strictly inside the triangle. This approach is well-defined as each tetrahedron includes either none or a pair of faces on which string cores are declared. This property also proves that strings are always connected without any discontinuity. See~\cite{Yamaguchi:2002zv,Yamaguchi:2002sh,Hiramatsu:2010yu} for other types of string identifications that ensure the connectedness of strings.
In a typical string identification, each cubic cell is associated with one volume and three faces (plaquette)~\footnote{For $n$ cells in the simulation box, there are total $6^n$ faces and each face is shared by two adjacent cells.}. Upon tetrahedralization of the volume out of existing lattice points as vertices, each cubic cell contains 6 tetrahedrons and it is associated only with 12 triangular faces (only 4 times more). Although there are multiple choices of assigning the 6 tetrahedrons, it will be interesting to study if this method can improve the spatial resolution out of the existing vertices. While our approach has more faces to examine, our criteria based on the signs of $\omega_{ij}$ for the string core declaration and the prescription for the string core location save computational times a lot. Overall, our approach for the string identification takes a similar amount of the computation time compared to others in literature.

\section{Scaling regime}
\label{sec:SR}

\begin{figure}[tp]
\begin{center}
\includegraphics[width=0.48\textwidth]{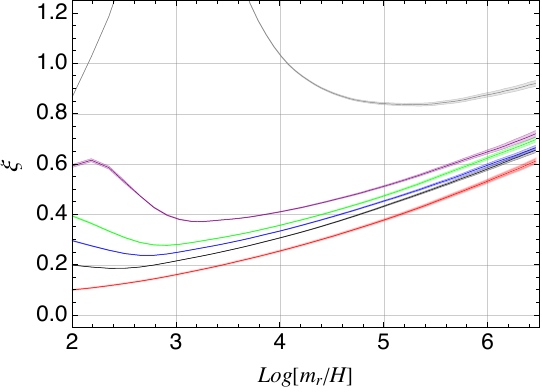}\quad
\includegraphics[width=0.48\textwidth]{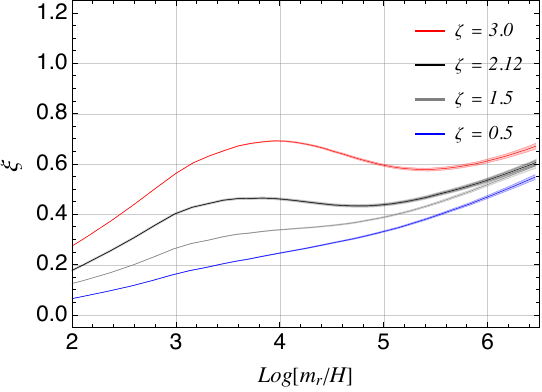}
\caption{\small The evolutions of the averaged $\xi$ in the logarithmic scales from simulations on the lattice of $N^3=1024^3$ with various initial conditions (distinguished by different curves within the same panel) made using fat-string pre-evolutions (left) and thermal pre-evolutions (right).}
\label{fig:L1024:xi20:option2:vs:zeta1:option3}
\end{center}
\end{figure}

A pair of string cores found on a tetrahedron forms a line segment, and connecting them forms a cosmic string.
In our simulation, the total string length in the simulation box is obtained by summing length of all the line segments. The average number of strings per Hubble patch is measured by
\begin{equation}\label{eq:xi:L:t:scaling}
 \xi = \frac{\ell_\text{tot}(L) t^2}{L^3}~,
\end{equation}
where $L$ is the size of the simulation box and $\ell_\text{tot}(L)$ is the total string length inside the box. The maximum dynamic time range is limited by $\log \frac{m_r}{H} \leq \log \frac{N}{n_c n_H}$.
Early simulations have shown strong hints that the cosmic string network approaches a scaling regime where the reduction of $\xi$ due to the radiation is balanced with the enhancement in $\xi$ by more strings entering the Hubble volume over time. 
Once the string network is believed to flow into the scaling regime at some point, its property will have to be insensitive to the initial configurations and the relaxation schemes. 
However, apparently the dynamic time range limited by the lattice size is not long enough to reach the asymptote that we expect it to match the scaling solution, and choosing benchmark parameter sets for further analyses is rather subjective. 
We performed a large number of simulations on the lattice of $N^3 = 1024^3$ for various initial conditions and two different relaxation schemes explained in Section~\ref{sec:initialconditions}, and outcomes are illustrated in Fig.~\ref{fig:L1024:xi20:option2:vs:zeta1:option3}.
%
%
While the curves of $\xi$ in the logarithmic time $\log\frac{m_r}{H}$ in Fig.~\ref{fig:L1024:xi20:option2:vs:zeta1:option3} clearly show a strong evidence for the scaling regime (also looks consistent with the result in~\cite{Gorghetto:2018myk,Gorghetto:2020qws}), they demonstrate that the residual dependency on initial conditions and relaxation types is not settled down within the time coverage. It takes different amounts of time to be relaxed to a relatively clean level, especially, the $\xi$ evolutions started at finite temperature, in the right panel of Fig.~\ref{fig:L1024:xi20:option2:vs:zeta1:option3}, have relatively short time range for a reliable fit~\footnote{We empirically observe that the curves with this type of the relaxation requires a larger ensemble size for the smooth curves than the case with the fat-string pre-evolution.}.
%
%
%
\begin{figure}[tp]
\begin{center}
\includegraphics[width=0.48\textwidth]{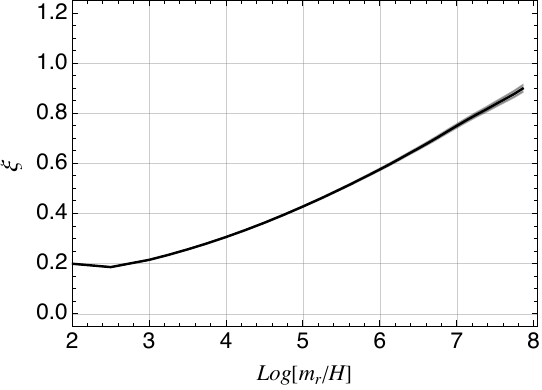}\quad
\includegraphics[width=0.48\textwidth]{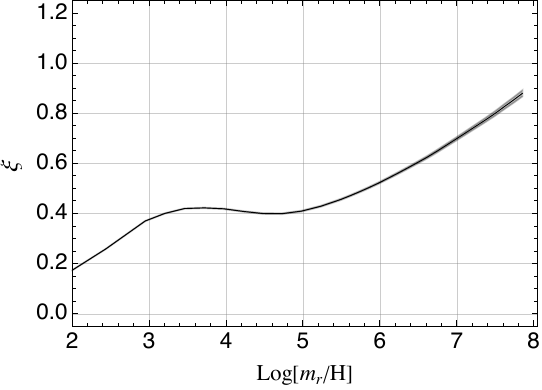}
\caption{\small The evolutions of the averaged $\xi$ in the logarithmic scales from simulations on the lattice of $N^3=4096^3$ for two relaxations whose initial conditions were prepared using fat-string pre-evolutions (left) and thermal pre-evolutions (right). The width of each curve represents the statistical errors.}
\label{fig:L4096:xi20:option2:vs:zeta1:option3}
\end{center}
\end{figure}
%

For the physical simulation whose relaxed initial conditions were prepared by the fat-string pre-evolution, following the suggestion in~\cite{Gorghetto:2020qws}, we similarly chose $\xi_0 = 0.2$ at the initial time of the physical string simulation as our benchmark curve for a bigger-sized lattice simulation. This curve with $\xi_0 = 0.2$ can be considered in between over-populated and under-populated string networks, and thus it is hoped to be close to the scaling solution. For simulations starting right before the epoch of the phase transition at finite temperature, the curve corresponding to $\zeta = 2.12$ is chosen as our benchmark. While looking at curves in the right panel of Fig.~\ref{fig:L1024:xi20:option2:vs:zeta1:option3}, the one with $\zeta = 1.5$ may be considered to be more consistent with the selection criteria for a benchmark curve in terms of over- and under-populated string networks, $\zeta = 2.12$ can provide more consistent comparison with results in literature. The $\xi$ curves for two benchmark points from the simulation on the lattice of $N^3 = 4096^3$ are illustrated in Fig.~\ref{fig:L4096:xi20:option2:vs:zeta1:option3}. While it clearly shows the strong dependency on the relaxation methods for $\log\frac{m_r}{H} < 5$ (also supported by the results in Fig.~\ref{fig:L1024:xi20:option2:vs:zeta1:option3}), $\xi$ from two benchmarks exhibit similar scaling behaviors for $\log\frac{m_r}{H} > 5$. 

If the long-term scaling behavior is assumed to be characterized by a single linear logarithmic term to which all trajectories with different initial conditions are attracted over time, deviations from the linear term in early times may be parametrized in terms of various mixtures of $\log^{-n}$ ($n \geq 1$) terms. We fit the $\xi$ curves with an ansatz including up to $\log^{-2}$~\cite{Gorghetto:2020qws},
\begin{equation}\label{eq:xi:fit:m2top1}
   \xi = \frac{d_2}{\log^2\frac{m_r}{H}} + \frac{d_1}{\log\frac{m_r}{H}} + c_0 + c_1 \log\frac{m_r}{H}~.
\end{equation}
We refer to this as global log-scaling hypothesis. Along this line of reasoning, all curves are fitted together by taking $c_0$ and $c_1$ as global parameters while taking $d_1$ and $d_2$ as local ones which are specific to each curve. Variations in $c_0$ and $c_1$ when adding $\log^{-3}$ as well to Eq.~(\ref{eq:xi:fit:m2top1}) or when changing the fitting range (two choices, $\log \geq 4.5$ and $\log \geq 5$ are considered) may be taken as the part of the systematic error.
Details of the fitting procedure are presented in Appendix~\ref{app:sec:relaxation}.
While we find that fitted curves are sensitive to both the fitting ansatz and the fitting range, our results from the fit over the range $\log \geq 5$ are presented in Table~\ref{tab:xi:fit:4096}~\footnote{Enlarging the range to $\log \geq 4.5$ gives $c_1 = 0.26$, and including $\log^{-3}$ term, while keeping the same range of $\log \geq 5$, leads to $c_1 = 0.23$ for the simulation with the fat-string pre-evolution. As the scaling regime occurs at a bit later times in the simulation using the thermal pre-evolution, a similar exercise may not be meaningful.}. 
\begin{table}[tbh]
\centering
  \renewcommand{\arraystretch}{1.05}
      \addtolength{\tabcolsep}{0.35pt} 
\scalebox{0.95}{
\begin{tabular}{lc|c|c}  
\multicolumn{2}{c|}{Pre-evolution type} & Fit with global log-scaling hypothesis & Interval for fit
\\[5pt] 
\hline
    Fat-string &  &  $\xi \sim -0.81 + 0.21 \log\frac{m_r}{H}$ & $\log\frac{m_r}{H} = [5.0,\,  -] $
\\[8pt]
    Thermal  &  & $\xi \sim -1.15 + 0.26 \log\frac{m_r}{H}$ & $\log\frac{m_r}{H} = [5.0,\,  -] $
\end{tabular}
}
\caption{\small The fitted curve taking the global log-hypothesis from two benchmark simulations differing by the relaxation schemes. The combined data from simulations on the grids of $N^3 = 1024^3$ and $4096^3$ (results in Figs.~\ref{fig:L1024:xi20:option2:vs:zeta1:option3} and~\ref{fig:L4096:xi20:option2:vs:zeta1:option3}) were used in the fit.}
\label{tab:xi:fit:4096}
\end{table}

\subsection{Scaling of inter-string distances}
\label{sec:inter:string:dist}
\begin{figure}[tp]
\begin{center}
\includegraphics[width=0.48\textwidth]{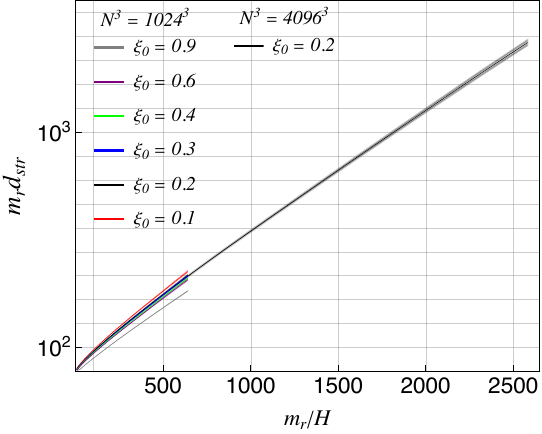}\quad
\includegraphics[width=0.48\textwidth]{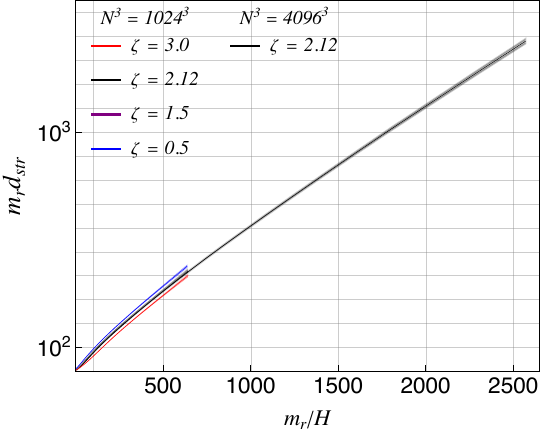}
\caption{\small The evolutions of the averaged $m_r d_\text{str.}$ from simulations on the lattice of $N^3=1024^3$ and $4096^3$. Initial conditions were prepared using fat-string pre-evolutions (left) and thermal pre-evolutions (right). The width of each curve represents the statistical errors.}
\label{fig:L1024:interstrings}
\end{center}
\end{figure}

\begin{figure}[tp]
\begin{center}
\includegraphics[width=0.47\textwidth]{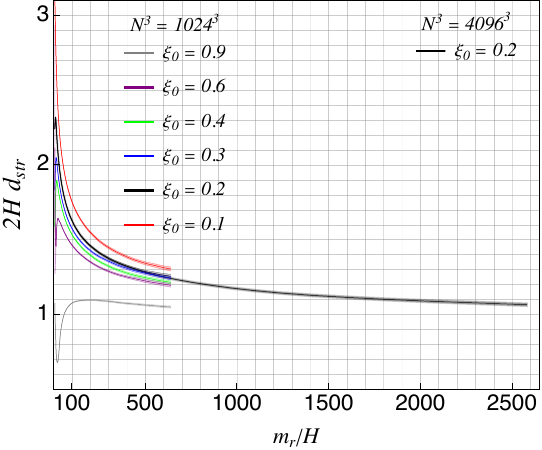}\quad
\includegraphics[width=0.47\textwidth]{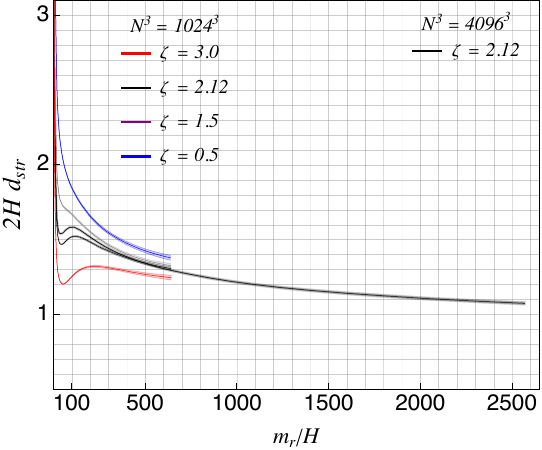}
\caption{\small The evolutions of the averaged $2H d_\text{str.}$ from simulations on the lattice of $N^3=1024^3$ and $4096^3$. Initial conditions were prepared using fat-string pre-evolutions (left) and thermal pre-evolutions (right). The width of each curve represents the statistical errors.}
\label{fig:L1024:2Hd}
\end{center}
\end{figure}

The average inter-string distance can be measured by $d_\text{str.} = (L^3/\ell_\text{tot}(L))^{1/2}$. Using the definition for $\xi$ in Eq.~(\ref{eq:xi:L:t:scaling}), the inter-string distance is expected to scale like $d_\text{str.} = 1/(2H\sqrt{\xi})$. 
Taking the log-scaling hypothesis, the simulation data is equivalently fitted with $d_\text{str.} = t/\sqrt{\xi}$ where $\xi$ is given by Eq.~(\ref{eq:xi:fit:m2top1}), and $d_\text{str.} \propto t/\sqrt{\log t}$ is expected at late times.
On contrary, the scenario favoring the constant scaling of $\xi$ at late times is realized as a linear scaling in time for the inter-string distances, or $d_\text{str.} \propto x_* t$ with a constant $x_*$.
If the long-term behavior of $d_\text{str.}$ is assumed to be attracted to a global linear term in time $t$, possible deviations from the linear term in early times may be accounted by extra time-dependent terms. 
We refer to this as global constant-scaling hypothesis as it implies that $\xi =  1/(4H^2 d^2_\text{str.})$ asymptotes to a constant. 
The evolution of $d_\text{str.}$ in the $m_r^{-1}$ unit is illustrated in Fig.~\ref{fig:L1024:interstrings}. As is evident in Fig.~\ref{fig:L1024:interstrings}, the curves of $d_\text{str.}$ from different initial conditions do not look possible for them to be attracted toward the single linear term in $t$ at late times. While fitting by taking all parameters as local ones for each curve can be carried out, it looks challenging to distinguish between the log-scaling and constant-scaling hypotheses since the evolution of $d_\text{str.}$ at late times is dominated by the $t$ term. Instead, the quantity $2H d_\text{str.}$, measured by evaluating $2H(L^3/\ell_\text{tot}(L))^{1/2}$, may be better choice to isolate the scaling behavior of the slope of $d_\text{str.}$ itself, or $1/\sqrt{\xi}$. Its evolution is illustrated in Fig.~\ref{fig:L1024:2Hd}. Although $m_r d_\text{str.}$ at late times looks like following almost straight lines in time $t$, the asymptotic behavior of $2H d_\text{str.}$ in Fig.~\ref{fig:L1024:2Hd} does not support the constant scaling behavior (as it should be obvious from our observed logarithmic growth behavior of $\xi$).

%

\section{Cosmological evolution of string network}
\label{sec:cosmo:strings}

\subsection{Energy density}
\label{sec:energybudget}
The cosmological evolutions of the energy densities of individual components, namely axions, radial modes, and strings, demonstrate the interplay among those over time, and provide a dynamic picture of how the string network approaches the scaling regime.
The total Hamiltonian density of the PQ field $\phi$ is given by 
$T_{00} = {\dot\phi}^*{\dot\phi} + \frac{1}{R^2}\nabla\phi^* \cdot \nabla \phi + V$ where $\nabla$ is the gradient with respect to the comoving coordinate. 
It can be expressed explicitly in terms of the radial mode $r$, axion mode $a$, and their interactions,
\begin{equation}\label{eq:T:tensor:comp}
 T_{00} = \frac{1}{2}\left ( 1 + \frac{r}{f_a} \right )^2 \left ( \dot{a}^2 + \frac{1}{R^2}  \left | \nabla a \right |^2 \right ) 
 +\frac{1}{2}  \left ( \dot{r}^2 + \frac{1}{R^2} \left | \nabla r \right |^2 \right ) + V~.
\end{equation} 
The total energy density is the spatial average of the total Hamiltonian $\rho_\text{tot} = \langle T_{00} \rangle$ over the physical coordinates.
The energy densities of axions and radial modes are measured by evaluating
\begin{equation}\label{eq:rhoa:rhor:exp}
\begin{split}
  \rho_{a} &=   2\times\langle \frac{1}{2} \dot{a}^2 \rangle~,
  \\[3pt]
  \rho_r &= \langle \frac{1}{2}  \left ( \dot{r}^2 + \frac{1}{R^2} \left | \nabla r \right |^2 \right ) + V \rangle~,
\end{split}
\end{equation}
where fields are properly masked according to the prescription described in Appendix~\ref{app:sec:masking} to screen off the effect from string cores. Assuming that the total energy density of the PQ field is conserved, the energy density of strings is extracted by subtracting $\rho_a$ and $\rho_r$ in Eq.~(\ref{eq:rhoa:rhor:exp}) from the total energy density, $\rho_s = \rho_\text{tot} - \rho_a - \rho_r$.
The gradient of axions in Eq.~(\ref{eq:T:tensor:comp}) mainly contributes to the string tension and it is taken as part of the string energy density.
%
%
%
\begin{figure}[tp]
\begin{center}
\includegraphics[width=0.47\textwidth]{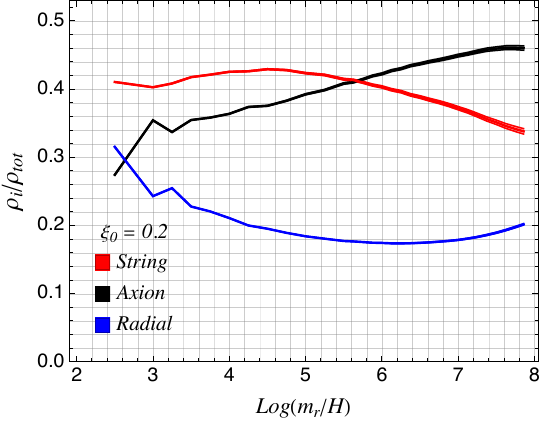}\quad
\includegraphics[width=0.47\textwidth]{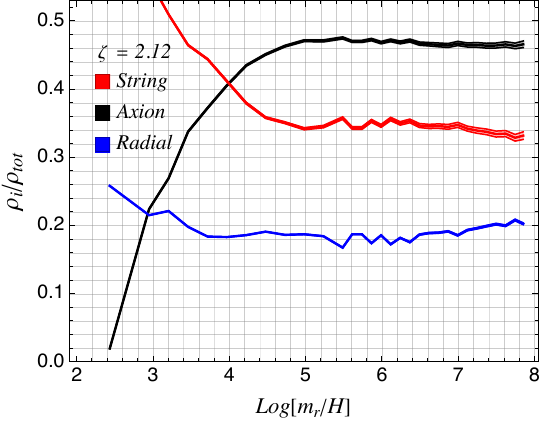}
\\[5pt]
\includegraphics[width=0.47\textwidth]{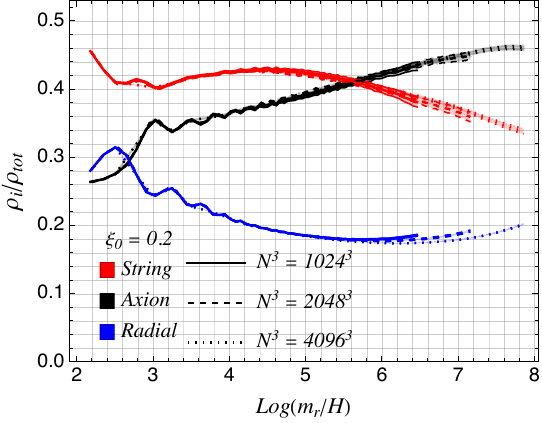}\quad
\includegraphics[width=0.47\textwidth]{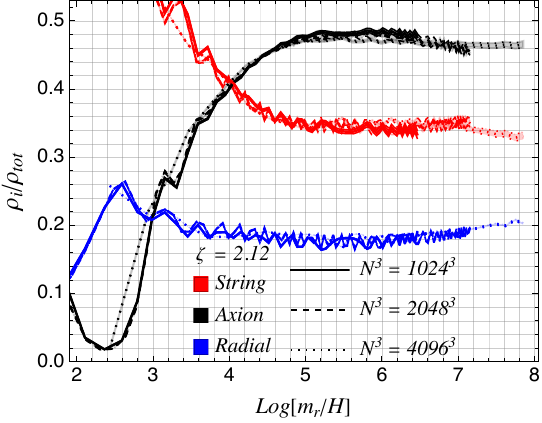}
\caption{\small Energy budgets of axions, radial modes, and strings from the simulations with $N^3 = 4096^3$ (top) and also with smaller lattice size (bottom) whose initial data was relaxed by the fat-string pre-evolution (left) and thermal pre-evolution (right).}
\label{fig:Ebudget:benchmark:4096}
\end{center}
\end{figure}
%

Energy densities of individual components for two different simulation setups are illustrated in Fig.~\ref{fig:Ebudget:benchmark:4096}. 
In the region before the string network enters the scaling regime, the separation between different components should be ambiguous, and their physical meanings become more transparent at late times.
Contrary to similar scaling behaviors in $\xi$ curves in Fig.~\ref{fig:L4096:xi20:option2:vs:zeta1:option3} for $\log\frac{m_r}{H} > 5$ for both types of pre-evolutions, the apparent evolution of the energy density for each component looks dramatically different.  Nevertheless, all energy densities for axions, strings, and radial modes near the end of simulation amazingly agree well. Since two simulations in Fig.~\ref{fig:Ebudget:benchmark:4096} differ by the way the initial field configurations were generated and subsequent relaxation schemes, the distinctive behaviors until late times may indicate a source of large systematic errors or an ambiguity in selecting the would-be attractor solution.
Unlike the energy budget in~\cite{Gorghetto:2018myk} where the different prescription for the fat-string pre-evolution was used~\footnote{The simulation in~\cite{Gorghetto:2018myk} implements the fat-string pre-evolution with the scale factor scaling as $R\propto \sqrt{t}$ whereas, in our prescription, the scale factor $R \propto t$ as in~\cite{Gorghetto:2020qws} is used during the pre-evolution.}, the contributions from strings and axions in our simulation cross around $\log \sim 5.5$, as is seen in the left panels of Fig.~\ref{fig:Ebudget:benchmark:4096}, and the axion-to-string energy density ratio seems to follow the expectation in the scaling regime, $\rho_a/\rho_s \sim \log\frac{m_r}{H}$. The contribution from radial modes is supposed to decrease at late times whereas $\rho_r$ in Fig.~\ref{fig:Ebudget:benchmark:4096} shows a rising behavior around $\log\frac{m_r}{H} \sim 6$ in both benchmark simulations. The similar observation was made in~\cite{Gorghetto:2020qws} except that the overall rate is smaller, for instance, $\rho_r \sim 20$\% in the scaling regime in Fig.~\ref{fig:Ebudget:benchmark:4096} versus $\sim 14$\% in~\cite{Gorghetto:2020qws}. The latter~\cite{Gorghetto:2020qws} compares better with our simulation with $\xi_0 \sim 0.1$ as is indicated in the left of Fig.~\ref{fig:app:Ebudget:varyingN} in Appendix~\ref{app:sec:Ebudget} where the radial modes appear taking roughly 14\% of the total energy density. However, the crossing point between strings and axions for the case with $\xi_0 = 0.1$ is delayed to more late times, around $\log\frac{m_r}{H} \sim 6.5$. The bottom-left panel of Fig.~\ref{fig:Ebudget:benchmark:4096} overlays three simulation results only differing by the lattice size which equivalently can be taken to extract the dependency on the lattice spacing in the string core length unit $m_r \Delta$~\footnote{Since the bigger lattice size $N$ is associated with the smaller lattice spacing $\Delta$, increasing $N$ is equivalent to taking the continuum limit from the smaller-sized-lattice point of view at a fixed time.}, and it strongly indicates that the rising behavior after around $\log\frac{m_r}{H} \sim 6$ could be an artifact due to the finite size of the lattice spacing. A similar conclusion from the simulation using the thermal pre-evolution is rather difficult due to undamped noises as is evident in the bottom-right panel of Fig.~\ref{fig:Ebudget:benchmark:4096}.

In the right panels of Fig.~\ref{fig:Ebudget:benchmark:4096}, the energy budgets from simulations using thermal pre-evolution are presented in a similar manner. While energy densities from the simulation with $N^3 = 4096^3$ were coarsely recorded starting only from $\log\frac{m_r}{H} \sim 2.3$ to save the computational time, more dense time samplings starting from earlier times were selected to record the data in the simulations with smaller lattice sizes.
While the bottom-right panel of Fig.~\ref{fig:Ebudget:benchmark:4096} shows apparently different evolution patterns of axions and strings compared to those using fat-string pre-evolution, the energy budgets of two simulations using the fat-string and thermal pre-evolutions amazingly agree well at the final time, $\log\frac{m_r}{H} \sim 7.8$, as was mentioned above. This agreement may justify the choice of two benchmark simulations on the lattice with $N^3 = 4096^3$.
The analogous dependency on the lattice spacing $m_r \Delta$ does not have a clear pattern, for instance, the slightly decreasing $\rho_a$ (and slightly increasing $\rho_s$) near their final times in simulations with $N^3 = 1024^3$, $2048^3$ is flipped when the dynamic time range is extended with the bigger lattice size. 
The overall relaxation in the simulation using the thermal pre-evolution seems to take longer dynamic time and is subject to the larger uncertainty than the case using the fat-string pre-evolution.

\subsection{String tension and boost factor}
\label{sec:stringtension}
%
%
\begin{figure}[tp]
\begin{center}
\includegraphics[width=0.48\textwidth]{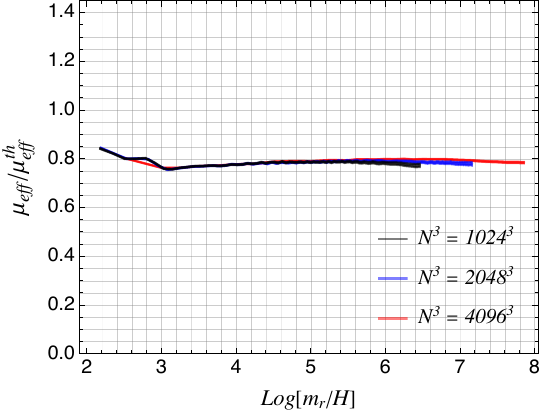}\quad
\includegraphics[width=0.48\textwidth]{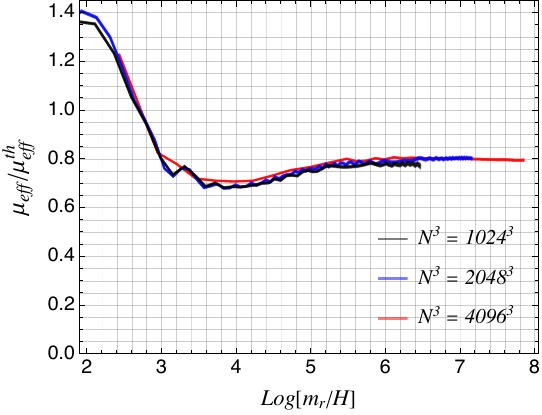}
\caption{\small The cosmological evolution of the string tension with respect to the theory expectation from two benchmark simulations whose initial field configurations were prepared by fat-string pre-evolution (left) and thermal pre-evolution (right).}
\label{fig:tension:ratio:4096}
\end{center}
\end{figure}
The total energy stored in strings $E_s$ is given by $E_s = \mu_\text{eff}\, \ell_\text{tot} (L)$ where $\mu_\text{eff}$ is an effective tension and  $\ell_\text{tot} (L)$ is the total string length in the simulation box of size $L$. The effective tension can be extracted from measurements of $\xi$ and the string energy density $\rho_s = E_s/L^3$. That is, using the relation in Eq.~(\ref{eq:xi:L:t:scaling}), $\mu_\text{eff}$ can be expressed as
\begin{equation}\label{eq:mueff}
\mu_\text{eff} = \frac{E_s}{\ell_\text{tot} (L)}  = \rho_s \frac{t^2}{\xi}~,
\end{equation}
or the energy density of strings is given by 
\begin{equation}
  \rho_s = \mu_\text{eff} \frac{\xi}{t^2}~.
\end{equation}
The string tension of the global strings in Eq.~(\ref{eq:mueff}) is logarithmically divergent. Its divergence is cut off by the UV cutoff, $r_\text{UV}$, roughly string core length $m_r^{-1}$, and the inter-string distance, which is roughly $(L^3/\ell_\text{tot}(L) )^{1/2}$, and it leads to the form that suits well for the straight string~\cite{Fleury:2015aca},
\begin{equation}\label{eq:mueff:th}
 \mu^\text{th}_\text{eff} = \langle \gamma \rangle \mu_0 \log\frac{\left (L^3/\ell_\text{tot}(L) \right )^{1/2}}{r_\text{UV}}
  = \langle \gamma \rangle \mu_0 \log \frac{1}{2 r_\text{UV} H \xi^{1/2}}~,
\end{equation}
where $\langle \gamma \rangle$ is the averaged (over the length) boost factor of string velocity, and $\mu_0$ denotes the proportionality factor. 
On the contrary, the logarithmic divergence is absent in local strings~\cite{Bennett:1985qt,Bennett:1987vf,Allen:1990tv,Vincent:1997cx,Moore:2001px} due to a compensating contribution from gauge fields. 
The logarithmic dependence on distances in global strings indicates the long-range force mediated by massless axions between strings. However, as the coupling of large wavelength axions with strings scales as $1/\log(m_r/H)$~\cite{Dabholkar:1989ju}, the string network at late times is expected to become similar to the situation of the local strings.
By comparing the measured effective string tension in Eq.~(\ref{eq:mueff}) against the theoretical expectation in Eq.~(\ref{eq:mueff:th}), the non-trivial scaling property of the string energy density can be checked.
The effective tension $\mu_\text{eff}$ with respect to $\mu_\text{eff}^\text{th}$ is illustrated in Fig.~\ref{fig:tension:ratio:4096} where $\mu_\text{eff}^\text{th}$ is given by Eq.~(\ref{eq:mueff:th}) with $\langle \gamma \rangle \mu_0 = \pi f_a^2$ and $r_\text{UV} = m_r^{-1} (4\pi)^{1/2}/2$~\cite{Gorghetto:2020qws}.
As is evident in Fig.~\ref{fig:tension:ratio:4096}, the effective string tension-to-theoretical expectation ratio is relaxed to a roughly constant value at later times, and it implies that the expression in Eq.~(\ref{eq:mueff:th}) is intriguingly applicable to more generic configurations.
An approach to the scaling regime looks more dramatic in the simulation with the thermal pre-evolution as is seen in the right panel of Fig.~\ref{fig:tension:ratio:4096}.

\begin{figure}[tp]
\begin{center}
\includegraphics[width=0.48\textwidth]{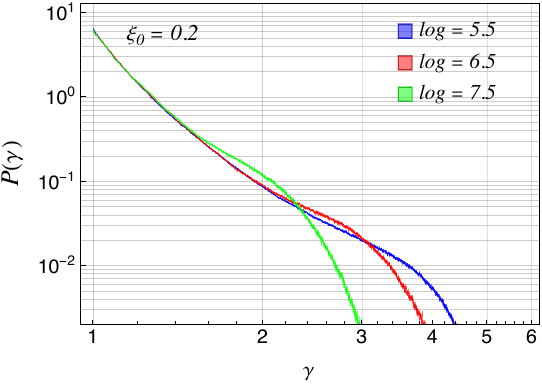}\quad
\includegraphics[width=0.48\textwidth]{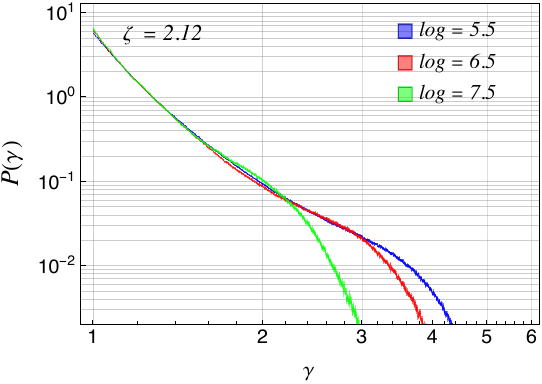}
\caption{\small The probabilistic distribution $P(\gamma)$ of the boost factor $\gamma$ from two benchmark simulations on the grid with $N^3 = 4096^3$.}
\label{fig:Pgamma:4096}
\end{center}
\end{figure}
The time variation of $\mu^\text{th}_\text{eff}$ in Eq.~(\ref{eq:mueff:th}) is given by
\begin{equation}
  \dot{\mu^\text{th}_\text{eff}} = \frac{\dot{\langle \gamma \rangle}}{\langle \gamma \rangle} \mu^\text{th}_\text{eff}
  + \langle \gamma \rangle \mu_0 \left ( \frac{1}{t} - \frac{\dot{\xi}}{2\xi} \right )~.
\end{equation}
If the averaged boost factor $\langle \gamma \rangle$ approaches the scaling regime, the prefactor $\langle \gamma \rangle \mu_0$ in Eq.~(\ref{eq:mueff:th}) is expected to behave like
\begin{equation}
 \langle \gamma \rangle \mu_0 =  \dot{\mu^\text{th}_\text{eff}}  \left ( \frac{1}{t} - \frac{\dot{\xi}}{2\xi} \right )^{-1}~,
\end{equation}
which can be explicitly checked from lattice simulations.
The average boost factor is obtained by evaluating

\begin{figure}[tp]
\begin{center}
\includegraphics[width=0.48\textwidth]{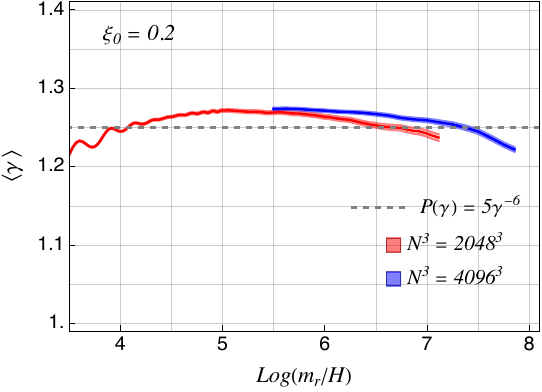}\quad
\includegraphics[width=0.48\textwidth]{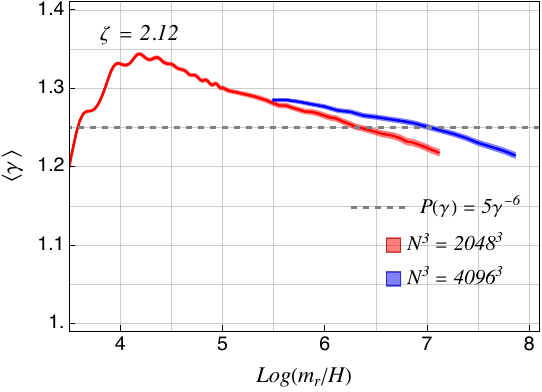}
\caption{\small The averaged boost factor $\langle \gamma \rangle$ from two benchmark simulations on grids with $N^3 = 2048^3$ and $4096^3$.}
\label{fig:Avegamma}
\end{center}
\end{figure}
\begin{equation}
  \langle \gamma \rangle = \int_{1}^\infty \gamma P (\gamma) d\gamma
  =  \int_{1}^\infty \gamma\ \frac{1}{\xi}\frac{d\xi_\gamma}{d\gamma} d\gamma~,
\end{equation}
where $\xi_\gamma$ is the fraction of $\xi$ with the boost factor below $\gamma$. The Lorentz boost factor $\gamma$ of the string core is computed from field configurations through the relation $\gamma^2 - 1 = \beta^{-2} | \dot\phi |^2$ evaluated at every string core. Recall that, in our tetrahedron-based string identification method, a pair of string core locations are assigned via the linear interpolation from field values at three vertices on each triangular face. Similarly for $\dot\phi$ at the string core. The corresponding tetrahedron includes string line segment (if exists), and the string boost factor $\gamma$ is evaluated at the mid-point of the string line segment. The probability distribution $P(\gamma)$ of the boost factor is illustrated in Fig.~\ref{fig:Pgamma:4096}. $P(\gamma)$ roughly scales like $\sim \gamma^{-6}$ as was also observed in~\cite{Gorghetto:2020qws}.
Finally, the time variation of the averaged boost factor is illustrated in Fig.~\ref{fig:Avegamma}. To save the computational time, the boost factors were computed only at the coarsely selected sampling points. In the simulation with the lattice size of $N^3 = 4096^3$, only time slices after $\log \sim 5.5$ were picked. The earlier times were covered by the simulation on the smaller lattice of $N^3 = 2048^3$ with the small overlap with those from the former. The overall qualitative behavior and averaged boost factor $\langle \gamma \rangle = 1.2 \div 1.3$ agree well with the result in~\cite{Gorghetto:2020qws} as is seen in the left panel of Fig.~\ref{fig:Avegamma}. 
In contrast, the early time behavior appears pronouncedly different for the cosmic strings in the simulation using the thermal pre-evolution although both become similar at late times. The gray dashed lines in both panels in Fig.~\ref{fig:Avegamma} represent the expectation from the ansatz $P(\gamma) = 5 \gamma^{-6}$. Two curves from simulations differing by the lattice sizes, at an instant time, can be effectively thought of as two cases differing by $m_r \Delta$. For instance, $m_r \Delta$ of the red curve at a fixed time within $\log = [5.5,\, 7]$ will be larger than that of the blue curve which is closer to the continuum limit.

\section{Axion spectrum}
\label{sec:axion:spectrum}

\begin{figure}[tp]
\begin{center}
\includegraphics[width=0.53\textwidth]{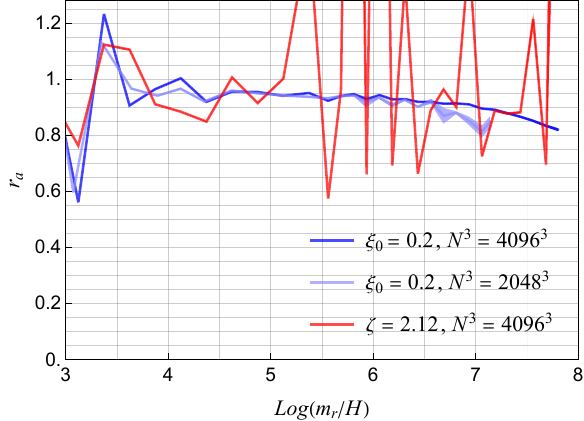}
\caption{\small Fraction of the transfer rate from strings to axions, $r_a = \Gamma_a/(\Gamma_a + \Gamma_r)$, from two benchmark simulations using the fat-string ($\xi_0 = 0.2$) and thermal pre-evolutions ($\zeta = 2.12$).}
\label{fig:raStatCorrectedL4096}
\end{center}
\end{figure}
The strings lose their energy into axions and radial modes, denoted by $\Gamma$. Individual transfer rates into axions and radial modes are represented by $\Gamma_a$ and $\Gamma_r$, respectively, and $\Gamma = \Gamma_a + \Gamma_r$. Similarly to~\cite{Gorghetto:2018myk}, our simulation confirms that strings mostly decay into axions with a small fraction going to radial modes, namely $\Gamma \sim \Gamma_a$, as is illustrated in Fig.~\ref{fig:raStatCorrectedL4096}. 
The fluctuating evolution of $r_a$ from the simulation using the thermal pre-evolution (red solid line in Fig.~\ref{fig:raStatCorrectedL4096}) is due to the oscillatory behavior in the energy budget as was illustrated in the right panels in Fig.~\ref{fig:Ebudget:benchmark:4096}.
Two evolutions of $r_a$ with $\xi_0 = 0.2$ in Fig.~\ref{fig:raStatCorrectedL4096}, differing by the lattice sizes, indicate that the late time behavior of $r_a$ will asymptote to a plateau in the continuum limit.
The evolution of the relativistic axion energy density follows the equation 
\begin{equation}
\frac{1}{R^{4}(t)} \frac{\partial}{\partial t} \left ( R^4(t) \rho_a(t) \right ) = \Gamma_a + \cdots
\end{equation} 
where $\cdots$ corresponds to the effect from the interaction between axions and radial modes. Similarly for the radial energy density except that it scales like matter. Assuming that $\Gamma_a$ eventually dominates at late times, $\Gamma \sim \Gamma_a$, in the evolution of $\rho_a$, the axion energy density can be expressed as the integration of the transfer rate, 
\begin{equation}\label{eq:rhoa:Gamma}
\rho_a(t) = \int^t dt' \left (\frac{R(t')}{R(t)} \right )^4\, \Gamma(t')~.
\end{equation} 
The inclusive axion energy density does not tell about the characteristic hardness or softness of the axions radiated from strings, and it is more informative to look into the differential distribution of the axion energy density in momentum $\partial\rho_a/\partial k$, defined by $\rho_a = \int dk\, (\partial\rho_a/\partial k)$.
In the absence of no new scales roughly between $m_r$ and $H$, the existence of the scaling regime indicates a power law fall-off behavior, $\sim k^{-q}$. 
The characteristic feature of the axion spectrum falls into roughly three categories depending on the value of $q$~\cite{Gorghetto:2018myk}.
When $q > 1$~\cite{Davis:1985pt,Davis:1986xc,Battye:1993jv,Battye:1994au}, the spectrum is more pronouncedly peaked near the order of Hubble which implies that the spectrum is IR-dominated and axions radiated from strings are soft. In contrast, when $q < 1$, the axion spectrum is relatively suppressed near the Hubble scale and more pronounced near the string core scale. Axions in this category are supposed to be hard and the spectrum is UV-dominated. The situation with $q=1$~\cite{Harari:1987ht,Hagmann:1998me} is in between the above two cases.

\begin{figure}[tp]
\begin{center}
\includegraphics[width=0.48\textwidth]{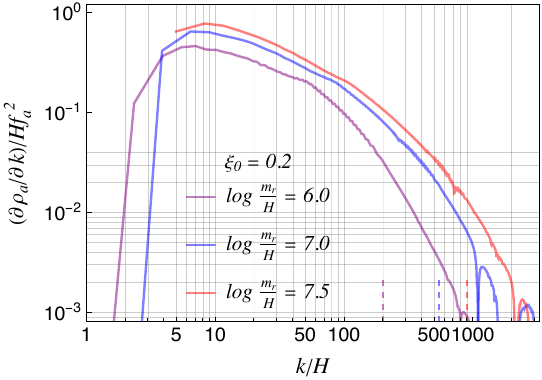}\quad
\includegraphics[width=0.48\textwidth]{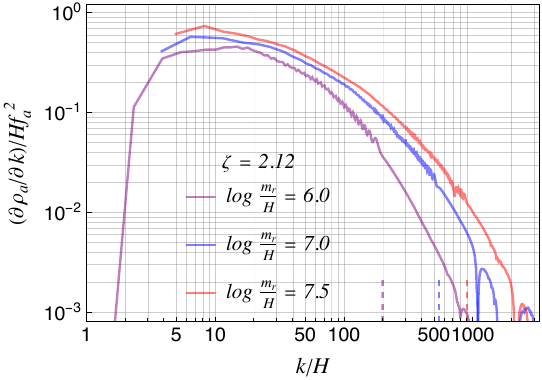}
\caption{\small The spectral axion energy density from simulations with $N^3 = 4096^3$ for two benchmark points. Dashed lines represent $m_r/(2H)$ for three choices of time snapshots.}
\label{fig:drhodk:4096}
\end{center}
\end{figure}
The distributions of the differential axion energy density, divided by $H f_a^2$ to cancel the overall time dependence, in two different simulation setups, differing by the relaxation scheme, on the lattice with $N^3 = 4096^3$ are illustrated in Fig.~\ref{fig:drhodk:4096} for three time slices, $\log = 6$ (purple), 7 (blue), 7.5 (red). The positions of $m_r/(2H)$, where the parametric resonances are expected, are also presented by dashed lines with the same color coding. The shape of $\partial \rho_a /\partial k$ at each instant time in Fig.~\ref{fig:drhodk:4096} nicely captures the power law fall-off scaling behavior between two scales $m_r$ and $H$, and the distributions rapidly drop off at the momentum below the Hubble scale $H$ and above the string core scale $m_r$ as is expected. Specifically, for the curves in Fig.~\ref{fig:drhodk:4096}, the peaks in IR region are located at $k/H \sim 5 \div 10$ and the drop-offs in UV occur around $m_r/(2H)$.

From the transfer rate from strings to axions in the differential form $\Gamma(t) =\int dk\, (\partial \Gamma(t)/\partial k)$, assuming $\Gamma \sim \Gamma_a$, the normalized instantaneous emission function $F$ can be defined as
\begin{equation}
  1 = \int \frac{dk}{H(t)}\frac{H(t)}{\Gamma(t)} \frac{\partial \Gamma}{\partial k}(k,t)
  \equiv \int\, dx F[x,y]~,
\end{equation}
where $x=k/H$ and $y=m_r/H$. The instantaneous emission $F$ is nothing but the normalized differential transfer rate measured in $k/H$, and it characterizes the distribution of the axion spectrum in momentum, contributing to the axion energy density in Eq.~(\ref{eq:rhoa:Gamma}). With the choice of $k/H$, the peak location at order of the Hubble remains constant.
In the simulation, $F$ is measured from the axion spectral energy density $\partial \rho_a/\partial k$,
\begin{equation}
 F \left [ \frac{k}{H},\, \frac{m_r}{H} \right ] = \frac{1}{\Gamma_a/H}\frac{1}{R^3} \frac{\partial}{\partial t}
 \left ( R^3 \frac{\partial \rho_a}{\partial k} \right )~,
\end{equation}
by taking data at two time slices, for instance, separated by $\log\frac{m_r}{H} = 0.25$ as our default choice. 
The time intervals used for the instantaneous emission spectrum differ in literature, and its effect on the spectrum is discussed in Appendix~\ref{app:sec:fit:F} in detail.
\begin{figure}[tp]
\begin{center}
\includegraphics[width=0.47\textwidth]{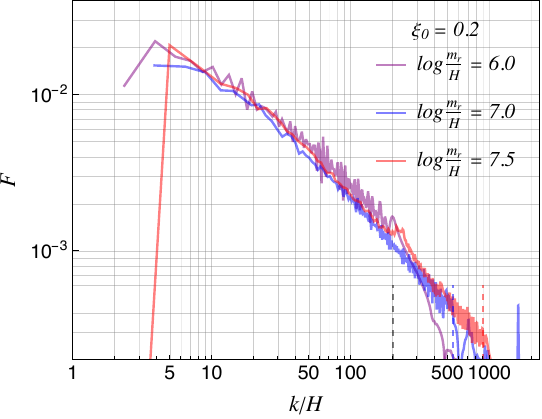}\quad
\includegraphics[width=0.47\textwidth]{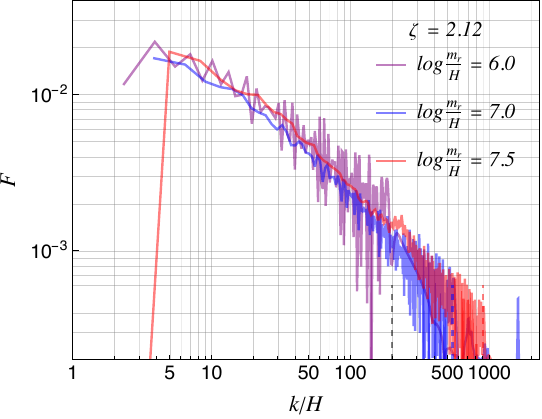}
\caption{\small The spontaneous emission $F$ from simulations on the lattice of $N^3 = 4096^3$ for $\xi_0 = 0.2$ (left) and $\zeta = 2.12$ (right) at three instant times snapshots, $\log\frac{m_r}{H} =$ 6, 7, 7.5. Dashed vertical bars with the same color coding correspond to $m_r/(2H)$ at each time instant.}
\label{fig:F:4096:benchmark}
\end{center}
\end{figure}
The instantaneous emission $F$ from our simulations on the lattice of $N^3 = 4096^3$ for two simulation setups are illustrated in Fig.~\ref{fig:F:4096:benchmark} for three logarithmic time slices, $\log = 6$, 7, 7.5. While the overall shapes in both panels of Fig.~\ref{fig:F:4096:benchmark} are similar, the emission spectra using the thermal pre-evolution are more noisy although they become cleaner at late times.

\subsection{Power law scaling}
\label{sec:spectralIndex}
The power law fall-off behavior of $F \propto k^{-q}$ between $m_r$ and $H$ is the most crucial part of the cosmic string network for the plausibility of the axion as a dark matter candidate. A precise determination of the spectral index $q$ will allow us to estimate the relic abundance of axions. To avoid contaminations from the UV and IR regions, the fit needs to be done inside an appropriate interval away from the string core and Hubble scales, or parametrized as $k = [x_\text{IR}H, \, m_r/x_\text{UV}]$. Since there is no obvious choice for the interval for fitting $F$ and it rather depends on the situation, we fit the power law fall-off region for various choices of $x_\text{UV}$, $x_\text{IR}$ to screen off UV and IR contaminations. 
It is instructive to estimate the number of sampling points within the interval for given $x_\text{IR}$, $x_\text{UV}$, and the lattice size $N$.
The physical momentum on the lattice is $\vec{k} = \frac{2\pi}{RN \Delta x} \vec{n}_k$ where $\vec{n}_k$ runs up to $\frac{N}{2}$~\footnote{It should be distinguished from the physical momentum $k$ truncated at $m_r$ when generating initial conditions for scalar fields. The maximum momentum set by the lattice spacing is $k_\text{max} = \frac{2\pi}{R(2\Delta x)}$.}, and in the Hubble unit, $x = k/H = \frac{2^{3/2} \pi}{N(\Delta x m_r)}\sqrt{y}\, n_k$. The number of sampling points used for the fit within the interval $x_\text{IR} \leq x \leq y/x_\text{UV}$ for the static lattice simulation, using $\Delta x$ in Eq.~(\ref{app:eq:Deltax:static}), is given by
\begin{equation}
 N_\text{fit}  = \frac{\text{Interval of } x}{x \text{-spacing}}
 = \frac{1}{2^{3/2} \pi} \left ( \frac{\sqrt{y}}{x_\text{UV}} - \frac{x_\text{IR}}{\sqrt{y}} \right ) \left (\frac{2 n_H N}{n_c} \right )^{1/2}~.
\end{equation}
\begin{figure}[tph]
\begin{center}
\includegraphics[width=0.48\textwidth]{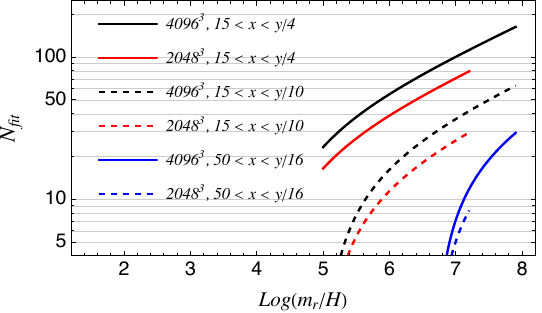}
\caption{\small The number of sampling points inside $x_\text{IR} \leq x \leq \frac{y}{x_\text{UV}}$ for fitting the power law for the lattice size of $2048^3$ and $4096^3$. $n_c = 1$, $n_H = 4^{1/3}$ were used. Curves are plotted from 5 to the maximal time coverage $\sim \log\frac{N}{n_c n_H}$.}
\label{fig:Nfit:static}
\end{center}
\end{figure}
As is illustrated in Fig.~\ref{fig:Nfit:static} for various intervals, the number of sampling points increases with the increasing time. It is because the spacing of $x$ scales as $\sim \sqrt{y}$ with the time whereas the upper end of the interval scales as $\sim y$ for the fixed lower end value. At later times, both higher and lower $k$ modes previously staying outside the interval enter into the interval. 
In terms of $k$ scaling as $\sim R^{-1}$, the lower end of the interval $\sim H$ decreases faster than $R^{-1}$ and more lower $k$  modes enter into the interval. More higher modes from above the $m_r$ scale get redshifted below $m_r$ and part of them may also enter into the interval.
As is evident in Fig.~\ref{fig:Nfit:static}, the lattice size of $N^3 = 2048^3$ seems to be enough for the loose interval, $15 \leq x \leq \frac{y}{4}$, adopted in~\cite{Gorghetto:2020qws}, and not enough for the tight interval, $40 \leq x \leq \frac{y}{16}$ adopted in~\cite{Buschmann:2021sdq}~\footnote{When using AMR, assuming the locally refined region around string cores are completely masked for simplicity, the lattice spacing which is relevant for the power spectrum analysis will be similar to the static case except for the modified lattice spacing in Eq.~(\ref{eq:dx:AMR:coarse}).
However, as the simulation time extends over beyond the coverage for the static simulation, more momentum modes enter into the interval of $x$, increasing $N_\text{fit}$. One limitation in this situation with AMR will be that the upper end of the interval should be smaller than the maximally allowed $x$ set by $\Delta x$, or $\frac{y}{x_\text{UV}} \leq x_\text{max}$ where $x_\text{max} = \frac{\pi}{HR \Delta x} = \pi \sqrt{\frac{n_c N}{2^k n_H}\, y}$, and it gives the upper limit on $\log_i \leq \log [ x_\text{UV}^2 \pi^2 (n_c/2^k) \frac{N}{n_H} ]$. For the simulations with $N^3=2048^3$, $n_c =4$, $k=4$, and $n_H = 4^{1/3}$, the limit on the time coverage is estimated to be $\log_i \leq 10.8$ for $x_\text{UV} = 4$ and $\log_i \leq 13.6$ for $x_\text{UV} = 16$.}.

The modes above $m_r$ at the phase transition correspond to the short distance physics above the PQ symmetry breaking scale. At later times, those UV modes such as radial modes are redshifted and can enter the interval. Requiring no contamination from those redshifted UV modes sets an upper bound on the $\log_i$, or $\frac{y}{x_\text{UV}} \leq \frac{m_r}{HR}$ which leads to $\log_i \leq 2 \log [\sqrt{2}\, x_\text{UV}]$. $\log_i \leq 6.2$ (3.47) for $x_\text{UV} = 16$ (4). There will be an unavoidable contamination unless the radial modes get decayed away before reaching the scaling regime. Matching this upper limit to the maximal time coverage set by the lattice size leads to $x_\text{UV} = \sqrt{N/(2 n_c n_H)}$. Assuming $n_c=1$ and $n_H = 4^{1/3}$, $x_\text{UV}$ needs to be set to $x_\text{UV} = 25.4$ for $N^3 = 2048^3$ and $x_\text{UV} = 35.9$ for $N^3 = 4096^3$. Applying this tight UV cutoff will make extracting the power law fall-off behavior challenging due to the limited sampling points.
However, upon generating the initial conditions for scalar fields, only momentum modes below $m_r$ were populated. Although higher momentum modes are generated via the equation of motion, they do not look as harm as those originally generated at the generation of the random field configurations.

\begin{figure}[tp]
\begin{center}
\includegraphics[width=0.47\textwidth]{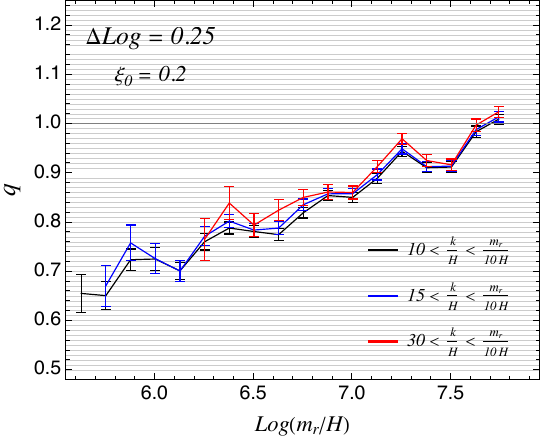}\quad
\includegraphics[width=0.47\textwidth]{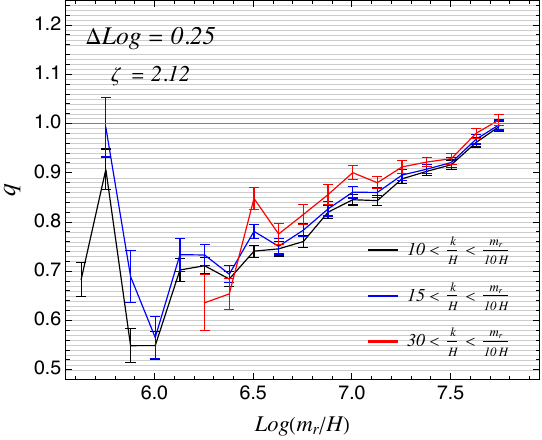}
\caption{\small The spectral index $q$ of the instantaneous emission $F$ fitted in the interval with the varying $x_\text{IR}$  while $x_\text{UV} = 10$. The error bars are statistical.}
\label{fig:q:4096:dlog0.25:IRvaried}
\end{center}
\end{figure}

The results from our simulations following the fitting prescription explained in Appendix~\ref{app:sec:fit:F} are illustrated in Figs.~\ref{fig:q:4096:dlog0.25:UVvaried} and~\ref{fig:q:4096:dlog0.25:IRvaried} for our default choice of $\Delta\log = 0.25$ (see Figs.~\ref{fig:q:4096:dlogvaried:UVvaried} and~\ref{fig:q:4096:dlogvaried:IRvaried} for different choices of $\Delta\log$). We observe a similar logarithmic growth to those in~\cite{Gorghetto:2020qws} except that our result shows a bit steeper slope than that in~\cite{Gorghetto:2020qws}, for instance, the value of $q$ reaches unity within the dynamic time range of the simulation for a similar choice of $\Delta\log$.
It will be interesting to see how $q$ further evolves beyond the time coverage of our current simulation, and this justifies improving the lattice resolution by the factor of two, namely $N^3 = 8192^3$.
In Fig.~\ref{fig:q:4096:dlog0.25:UVvaried}, the UV cutoff is varied while IR cutoff $x_\text{IR}$ is fixed to 15. In Fig.~\ref{fig:q:4096:dlog0.25:IRvaried}, the IR cutoff is varied while UV cutoff $x_\text{UV}$ is fixed to 10. 
The error bars in all curves represent the statistical ones over 100 independent simulation runs. 
Interestingly, the patterns of central values at late times look similar irrespective of types of pre-evolutions.
We suspect that the fluctuations of the central values are due to effects from the residual red-shifted high-frequency modes. 
Based on distributions in Figs.~\ref{fig:q:4096:dlog0.25:UVvaried} and~\ref{fig:q:4096:dlog0.25:IRvaried}, the interval $k = [15H, \, m_r/10]$ looks reasonable choice to get the fitted spectral index with a relatively small fluctuations.
%
\begin{figure}[tp]
\begin{center}
\includegraphics[width=0.47\textwidth]{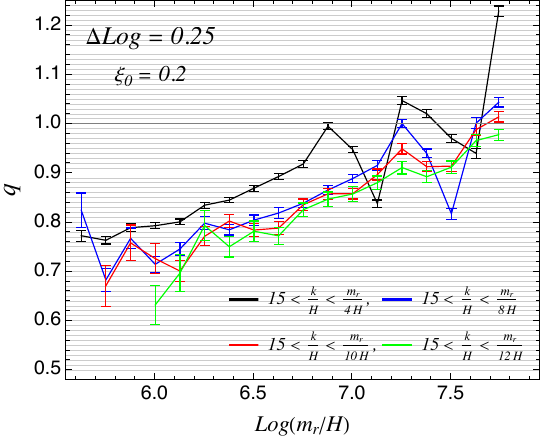}\quad
\includegraphics[width=0.47\textwidth]{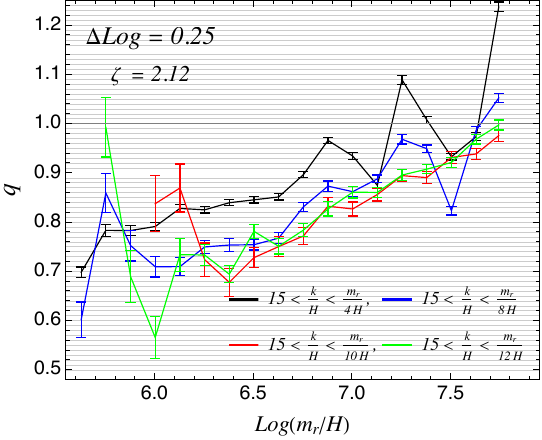}
\caption{\small The spectral index $q$ of the instantaneous emission $F$ fitted in the interval with the varying $x_\text{UV}$ while $x_\text{IR} = 15$. The error bars are statistical.} 
\label{fig:q:4096:dlog0.25:UVvaried}
\end{center}
\end{figure}
%
The fit results of the spectral index $q$, assuming the log-hypothesis, in the interval $\log = [6.5,\,  7.8]$ for two benchmark simulations on lattice with $N^3 = 4096^3$ differing by relaxation types are presented in Table~\ref{tab:q:benchmark:fit:4096}. 
\begin{table}[tbh]
\centering
  \renewcommand{\arraystretch}{1.05}
      \addtolength{\tabcolsep}{0.35pt} 
\scalebox{0.95}{
\begin{tabular}{lc|c|c}  
\multicolumn{2}{c|}{Pre-evolution type} & Fit with $(x_\text{IR},\, x_\text{UV}) = (15,\, 10)$ & Interval for fit
\\[5pt] 
\hline
    Fat-string &  &  $q \sim (-0.349 \pm 0.116) + (0.174 \pm 0.016) \log\frac{m_r}{H}$ & $\log\frac{m_r}{H} = [6.5,\,  7.8] $
\\[8pt]
    Thermal  &  & $q \sim (-0.460 \pm 0.079) + (0.186 \pm 0.011) \log\frac{m_r}{H}$ & $\log\frac{m_r}{H} = [6.5,\,  7.8] $
\end{tabular}
}
\caption{\small The fit result taking the log-hypothesis for two benchmark simulations on $N^3 = 4096^3$ differing by the relaxation schemes. The presented errors are merely statistical ones within the particular choices of cuts on $x_\text{IR}$, $x_\text{UV}$ and the fitting intervals.}
\label{tab:q:benchmark:fit:4096}
\end{table}

The outcome of fits with different assumptions on the form of the Likelihood function, UV/IR cutoffs on the momentum, and time interval $\Delta \log$ are summarized in Figs.~\ref{fig:q:summary:B} and~\ref{fig:q:summary:noB}. Discrepancies, compared to the result in Table~\ref{tab:q:benchmark:fit:4096}, may be considered as part of systematic errors. Most importantly, despite all variant situations, our 
conclusion on the spectral index predicting $q\sim O(10)$ at $\log{\frac{m_r}{H}}\sim 70$ does not change.

\section{Axion abundance}
\label{sec:axion:abundance}
The axion string network evolves until the time of the QCD crossover, denoted by $t_\star$ (roughly $\log_\star \equiv \log\frac{m_r}{H(t_\star)} \sim 70$), around which the axion mass $m_a$ becomes comparable with the Hubble, or $m_a (t_\star) \sim H_\star$ and the axion potential becomes relevant.
The impact of the logarithmically growing spectral index $q$ of the instantaneous emission $F$, obtained from fitting the data in Section~\ref{sec:spectralIndex}, on the axion abundance is dramatic. Despite the UV-dominated spectrum within the simulation time range, extrapolating the spectral index $q$ until the time of the QCD crossover indicates the IR-dominated spectrum of axions radiated from strings at those late times. The system of strings and domain walls are formed around this time~\cite{PhysRevD.45.3394,Sikivie:2006ni}. Especially, when the number of domain walls is one, $N_W = 1$, the strings-and-walls system decay into axions~\cite{Hiramatsu:2012sc} (see~\cite{Hiramatsu:2012gg,Buschmann:2019icd,OHare:2021zrq} for simulations). 
Similarly to~\cite{Gorghetto:2020qws}, our estimate will not include axions from the collapse of strings-and-walls system. 
We will focus only on the axions radiated from strings in the scaling regime and its evolution afterwards (later times than $t_\star$) through the nonlinear regime in presence of the axion potential. 
Therefore, it sets the conservative lower bound on the axion abundance (thus the upper bound on the axion decay constant).

The axion energy density from strings in the scaling regime is still bigger than the axion potential at $t = t_\star$ due to the dominant energy density stored in gradient terms. It continues evolving until the time, denoted by $t_\ell$, and it becomes comparable to the axion potential where axions transit to nonrelativistic ones.
If the average axion field values are smaller than the axion decay constant, $\langle a^2 \rangle ^{1/2} \ll f_a$, the nonlinearities due to the axion potential in the axion equations of motion can be ignored and the axion abundance will be obtained simply by evaluating $\int dk k^{-1} (\partial \rho_a/\partial k)$. Whereas, in the opposite situation, $\langle a^2 \rangle^{1/2} \gtrsim f_a$, the axion abundance at $t_\ell$ is affected by the nonlinearities and the estimation of the axion abundance is more tricky.
The parametric behavior of $n^{\text{str}}_a(t_\ell)$, axion abundance from cosmic strings, in terms of string properties takes different forms depending on the spectral index $q$, the value of ${\langle a^2 \rangle}^{1/2}/f_a$, and the IR cutoff of the momentum. The detailed derivation of all distinctive cases are postponed to Appendix~\ref{app:sec:axion:na:detail}. Here, we will discuss only relevant case to our situation.

The axion spectrum at the time $t =t_\star$ can be obtained by evaluating
\begin{equation}
 \frac{\partial \rho_a}{\partial k} 
 = \int^t dt' \frac{\Gamma(t')}{H(t')} \left ( \frac{R(t')}{R(t)} \right )^3 F \left [ \frac{k'}{H(t')},\, \frac{m_r}{H(t')} \right ]~,
\end{equation}
where the redshifted momentum $k' = k R(t)/R(t')$,  the form of $F$ for $q > 1$ is given in Appendix~\ref{app:sec:axion:na:detail}, and the transfer rate at late times is approximated by $\Gamma \sim \xi \mu_\text{eff}/t^3 \sim 8 \pi H^3 f_a^2 \xi \log \frac{m_r}{H}$. When $q >1$, the contribution to the axion abundance at $t = t_\star$ from higher momenta is power suppressed as  $\partial\rho_a/\partial k$ decreases faster than $\sim 1/k^{q}$, and only low momenta roughly $k < \sqrt{m_r H_\star}$  which leads to the scaling $(\partial\rho_a/\partial k)|_{t_\star} \propto k^{-1}$ dominantly contributes.
It has been analytically approximated in~\cite{Gorghetto:2020qws} (and supported by numerical simulation) to estimate the axion abundance around the time $t = t_\ell$, taking into account the nonlinearities. It was shown that the topological production of axions with respect to the contribution from the misalignment mechanism with the angle of unity is expected to scale as $(n_a^{\text{str}, q>1}/n_a^{\text{mis},\theta_0=1})|_{t_\ell} \propto (\xi_\star \log_\star)^{1/2 + \cdots}$ where $\xi_\star$ is the number of strings per Hubble patch at the time of the QCD crossover, $\log_\star \sim 70$ and $\cdots$ denotes dependency on axion mass evolution. The exact value of $q$ is not relevant as long as it is well above unity at the time of the QCD phase transition. Assuming that $q$ continuously increases logarithmically, the ballpark of the axion dark matter mass will be similar to roughly what was estimated in~\cite{Gorghetto:2020qws}.

As was discussed in Section~\ref{sec:inter:string:dist}, the average inter-string distance is roughly $(H\sqrt{\xi})^{-1}$ instead of $H^{-1}$. 
One may worry that axions whose momentum softer than $\sim H\sqrt{\xi}$ are suppressed and it may cause the parametrically reduced axion abundance. In Appendix~\ref{app:sec:axion:na:detail}, we have shown that it can cause a non-negligible change of the overall size of the axion abundance at late times due to the slightly modified parametric dependence on string properties. It is also demonstrated in Fig.~\ref{fig:abundance:ratio} for a set of selected parameters that will be further discussed in Section~\ref{sec:axion:mass}.
The modified IR cutoff can significantly change parametric behavior if the small average axion field values $\langle a^2 \rangle ^{1/2} \ll f_a$ are assumed. It is also true for the case with $q = 1$ similarly to~\cite{Buschmann:2021sdq} whose spectral index was claimed to be consistent with no-log hypothesis. The parametric behaviors of all distinctive cases are explicitly worked out in Appendix~\ref{app:sec:axion:na:detail}.

\subsection{Bound on axion mass}
\label{sec:axion:mass}

\begin{figure}[tp]
\begin{center}
\includegraphics[width=0.47\textwidth]{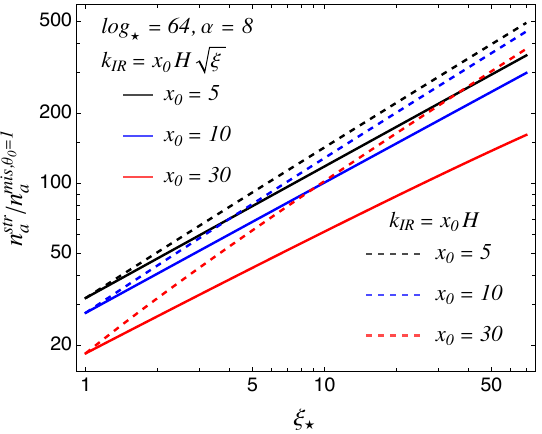}\quad
\includegraphics[width=0.47\textwidth]{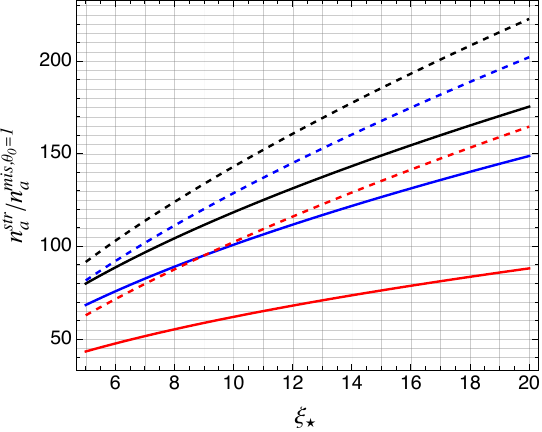}
\caption{\small The analytic prediction of the axion abundance from strings with respect to the misalignment with $\theta_0 = 1$ as a function of $\xi_\star$ for varying $x_0$ while $\log_\star = 64$ and $\alpha = 8$ are fixed. The solid lines correspond to those obtained with the IR momentum cutoff of $k_\text{IR} = x_0 H\sqrt{\xi_\star}$ whereas the dashed ones are those with $x_0 H$. $c_n=1.35$, $c_V=0.13$, $c_m=2.08$ and ${c_n}'=2.81$ were taken from~\cite{Gorghetto:2020qws}. The right panel is a zoomed-in version of the left one, and the color coding is the same as those in the left panel.}
\label{fig:abundance:ratio}
\end{center}
\end{figure}

The number of strings per Hubble patch around time $\log_\star=64$ similarly to~\cite{Gorghetto:2020qws} is estimated from our simulation result, and they are $\xi_\star=13.4$ (fat-string pre-evolution) and $\xi_\star=16.6$ (thermal pre-evolution). 
Adopting fit values $c_n=1.35$, $c_V=0.13$, $c_m=2.08$ and ${c_n}'=2.81$ from~\cite{Gorghetto:2020qws} and setting $\alpha=8$, the axion abundance with respect to the misalignment mechanism with the unit angle, given in Eq.~(\ref{app:eq:natonmis:master}) along with Eq.~(\ref{app:eq:z:sol:exact}) in Appendix~\ref{app:sec:axion:na:detail}, can be estimated to give 
$(n_a^{\text{str},q>1}/n_a^{\text{mis},\theta_0 =1})|_{t_\ell} \sim119$ (fat-string pre-evolution) and $(n_a^{\text{str},q>1}/n_a^{\text{mis},\theta_0=1})|_{t_\ell} \sim134$ (thermal pre-evolution) with the choice of $x_0=10$ as an illustration. 
While analytic predictions as a function of $\xi_\star$ for different IR cutoffs $k_\text{IR}$ (with varying $x_0$) is illustrated in Fig.~\ref{fig:abundance:ratio}, more general expressions are given in Appendix~\ref{app:sec:axion:na:detail}.
The lower bound on the axion mass can be derived by requiring that the axion dark matter abundance does not exceed the currently observed value. Taking the reference value $\Omega_a^{\text{mis}}h^2\approx0.12\left(28\, \mu\text{eV}/m_a\right)^{7/6}$ in the post-inflationary scenario as in~\cite{Gorghetto:2020qws} where we assume $ \langle \theta_0^2 \rangle \sim 5$ for translating it to $\Omega_a^{\text{mis},\theta_0=1}h^2$~\footnote{The precise estimate of the contribution from the misalignment~\cite{Fox:2004kb,Marsh:2015xka,GrillidiCortona:2015jxo} amounts to shifting the bound.},
\begin{equation} 
\begin{split}
\Omega_a^{\text{str}}h^2\approx \frac{n_a^{\text{str},q>1}}{n_a^{\text{mis},\theta_0=1}}\Omega_a^{\text{mis},\theta_0=1}h^2  \leq 0.12
\rightarrow
 &
\left\{ 
\begin{array} {lll}
  m_a \gtrsim 420~(\mu\text{eV}) &: &  \text{fat-string pre-evolution}
\\[15pt]
  m_a \gtrsim 470~(\mu\text{eV}) &: & \text{thermal pre-evolution}~.
\end{array}
\right.
\end{split}
\end{equation}
Using $m_{a}\approx 5.7\, \mu\text{eV}  ( 10^{12}\, \text{GeV}/(f_a/N_{W}))$~\cite{GrillidiCortona:2015jxo},
$f_a/N_{W}\leq 1.3\times10^{10}\text{GeV}$ (fat-string pre-evolution) and $f_a/N_W \leq 1.1 \times10^{10}\text{GeV}$ (thermal pre-evolution) for QCD axions are derived.\\

\section{Correlation between strings and axion spectrum}
\label{sec:correlation}

\begin{figure}[tp]
\begin{center}
\includegraphics[width=0.45\textwidth]{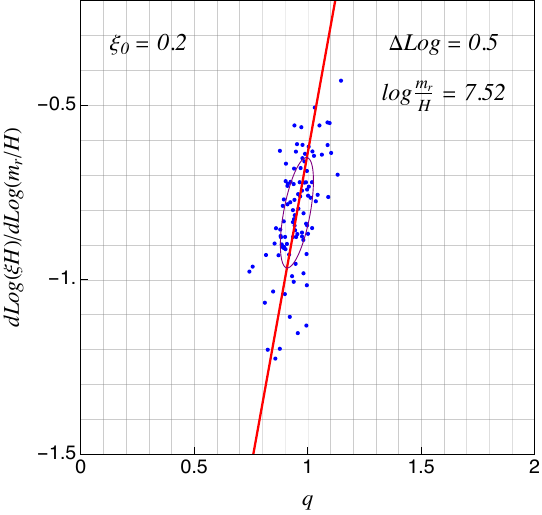}\quad
\includegraphics[width=0.45\textwidth]{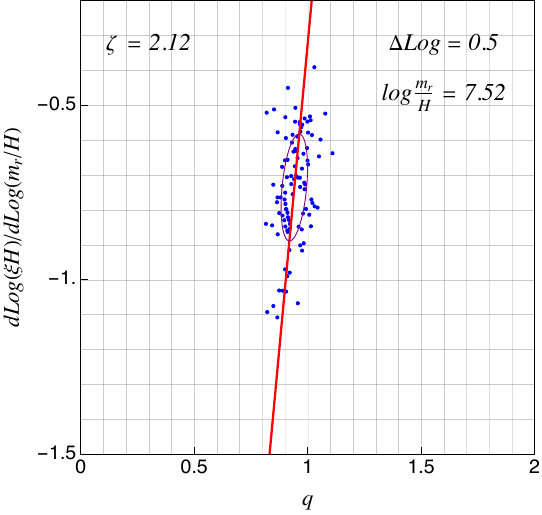}
\caption{\small The correlation between $q$ and $d\log(\xi H)/(d\log\frac{m_r}{H})$ over 100 ensemble elements from two benchmark simulations using the fat-string type pre-evolution (left) and thermal pre-evolution (right) on the lattice with $N^3 = 4096^3$.}
\label{fig:qVSxi:4096:dlog0.75}
\end{center}
\end{figure}

While our computing resources were limited to only two benchmarks, it would be interesting to explore possible observables or relations that can hint for behaviors of benchmark points with different initial conditions. In this section, we investigate one such observable, the correlation between $q$ and time variation of $\xi H$. While $\xi$ in the scaling regime is the consequence of two competing factors, namely decay of strings and more strings entering a Hubble patch, the quantity $\xi H$ will count only the effect from string decays. 
Each benchmark scenario for $\xi_0 = 0.2$ and $\zeta = 2.12$ consists of 100 independent simulations, and individual elements are spread around the central values in $\xi$ and $q$ distributions. Over all samples, the normalized differential change of $\xi H$ in logarithmic time, $d\log(\xi H)/(d\log\frac{m_r}{H})$, and the spectral index $q$ are computed at some time slices for various choices of $\Delta\log$. An illustrative example taken at $\log \sim 7.5$ is presented in Fig.~\ref{fig:qVSxi:4096:dlog0.75} for the choice of $\Delta\log = 0.5$, and it reveals a strong correlation.
The correlation is slightly less pronounced with $\Delta\log = 0.25$, and more pronounced with $\Delta\log = 0.75$ in that scattered data points are slightly less and more densely aligned along the correlation axes for the cases with $\Delta\log = 0.25$ and 0.75, respectively.
While the correlation is not stable in early times, it becomes robust at late times, for instance, the orientation of the correlation axis stays same for $\log \gtrsim 7$. 
A higher decay rate of strings is correlated with a smaller $q$ according to Fig.~\ref{fig:qVSxi:4096:dlog0.75}. This looks consistent with the analytic study in~\cite{Drew:2019mzc,Battye:1993jv} that the faster string decay rates are associated with the emission of harder axions and vice versa. The detailed discussion is postponed to Appendix~\ref{app:sec:pos:correlation}.

\begin{figure}[tp]
\begin{center}
\includegraphics[width=0.48\textwidth]{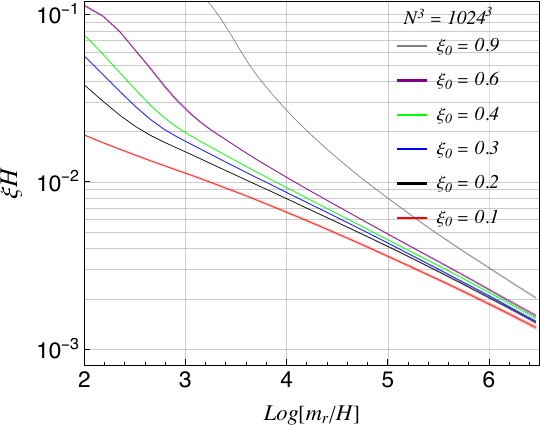}\quad
\includegraphics[width=0.48\textwidth]{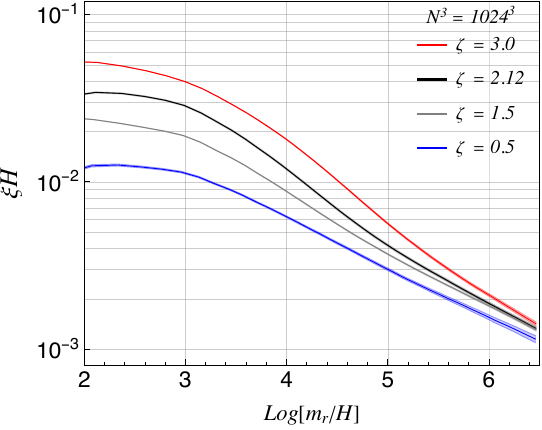}
\caption{\small The $\xi H$ evolution from simulations using the fat-string type pre-evolution (left) and thermal pre-evolution (right) on the lattice with $N^3 = 1024^3$ for various initial conditions.}
\label{fig:xiH:1024}
\end{center}
\end{figure}
Assuming that a similar correlation holds among different averaged $\xi$ curves from different initial conditions of $\xi_0$ or $\zeta$, the correlation in Fig.~\ref{fig:qVSxi:4096:dlog0.75} can imply that the spectral index will have similar values in the vicinity of the benchmark points. For instance, the distributions of $\xi H$ from simulations on the smaller lattice with $N^3 = 1024^3$ are illustrated in Fig.~\ref{fig:xiH:1024}. Although the maximal dynamic time range in Fig.~\ref{fig:xiH:1024} is smaller than $\log \sim 7$, it looks strong enough to support that the values of $d\log(\xi H)/(d\log\frac{m_r}{H})$ will become even closer at late times in the vicinity of the benchmark curves and thus the spectral index will not be significantly different from those from our benchmark simulations.

\section{Conclusion}

We have studied the dynamics of the global cosmic strings, originated from the PQ symmetry breaking, in the post-inflationary scenario where the definite prediction on the axion dark matter abundance can be made.
We performed independent numerical checks on recently discovered various logarithmic scalings in properties of the cosmic string network. Especially, we provided strong evidences for  two logarithmically growing scalings in the number of strings per Hubble patch and the spectral index of the power law for the axion spectrum against two differently prepared initial field configurations.
Having established the strong evidence of $q >1$ around the QCD scale, extrapolated from the simulation time, we pointed out a possible strong correlation between the axion spectrum and the string evolution with different initial conditions as a way of supporting the robust behavior of $q$ against nearby string evolution trajectories with different initial data. We showed that this strong correlation agrees with the theoretical prediction from a simple string vibration model. The IR dominated spectrum of axions radiated from cosmic strings in the scaling regime may indicate the insensitivity to UV physics of cosmic string modeling. It will be interesting to see if this property can lead us an extra handle for better understanding the nature of cosmic strings.

We surveyed various parametric behaviors of the axion abundance including the impact from the value of the spectral index, nonlinearities around the QCD scale, and average inter-string distances. This exercise should greatly help in extracting the lower bound on the axion mass accurately from the simulation result and comparing with different scenarios. Finally we estimated the lower bound on the axion mass based on our simulation.

\section*{Acknowledgments}
We thank Masahide Yamaguchi for useful discussions. We thank Marco Gorghetto for discussions about their previous work. We thank the department of physics at KAIST for supporting with the computing resources in the cluster center.
HK, JP, and MS were supported by National Research Foundation of Korea under Grant Number NRF-2021R1A2C1095430.
The simulation in this work was supported by KISTI National Supercomputing Center under Project Number KSC-2023-CRE-0078.

\appendix

\section{Equation of motion}
\label{app:sec:eom}
The equation of motion from the Lagrangian in Eq.~(\ref{eq:Lag:orig}) in the expanding universe with the scale factor of $R$ is given by
\begin{equation}
  \ddot{\phi} + 3 H \dot{\phi} - \frac{1}{R^2} \nabla ^2 \phi + \frac{m_r^2}{f_a^2} \phi \left ( |\phi|^2 - \frac{f_a^2}{2} \right ) = 0~,
\end{equation}
where dot is the differentiation with respect to the cosmic time $t$ and the gradient $\nabla$ is evaluated in the comoving coordinate. To simplify our code implementation and numerical operation, we rescale the field and spacetime coordinates to make them dimensionless in a way that $m_r^2/f_a^2$ appears:
\begin{equation}\label{app:eq:rescaling}
  \phi \rightarrow f_a\, \phi~,\quad t \rightarrow m_r^{-1} t~,\quad \vec{x} \rightarrow m_r^{-1} \vec{x}~,
\end{equation}
under which $H \rightarrow m_r H$.
After the rescaling, the equation of motion becomes
\begin{equation}
  \ddot{\phi} + 3 H \dot{\phi} - \frac{1}{R^2} \nabla ^2 \phi + \phi \left ( |\phi|^2 -\frac{1}{2}\right ) = 0~,
\end{equation}
where now dot is the differentiation with respect to the dimensionless cosmic time $t$ and the gradient $\nabla$ is evaluated in the dimensionless comoving coordinate. Similarly $H$ is dimensionless.

Upon the discretization, the field evolution in terms of $\phi$ is subject to the CFL condition, $\Delta t \lesssim R\Delta x$. This condition requires arbitrarily small time step size in the early stage of the evolution where the scale factor is small, and it is rather inefficient from the simulation point of view. Since we want more sampling points at later times, we re-express the equation of motion in terms of the rescaled field $\psi (\tau,\, \vec{x}) = R \phi (t,\, \vec{x})$ in the comoving spacetime coordinate,
\begin{equation}
  \psi'' - \nabla^2 \psi + \psi \left ( |\psi|^2 - \frac{R^2}{2} - \frac{R''}{R} \right ) = 0~,
\end{equation}
where the $H$-dependent term disappeared and it makes the simulation simpler. The CFL condition in this situation is relaxed to $\Delta \tau \lesssim \Delta x$. While the last term $R''/R$ vanishes in the radiation dominated era, it may not vanish in the universe with a generic scale factor.

In the approach using relaxation during thermal evolution, we add the thermal mass term to the original Lagrangian,
\begin{equation}
  \Delta \mathcal{L} = -\frac{m_r^2}{6 f_a^2} T^2 |\phi|^2~.
\end{equation}
After the rescaling in Eq.~(\ref{app:eq:rescaling}) along with $T \rightarrow f_a T$ and the replacement $\psi (\tau,\, \vec{x}) = R \phi (t,\, \vec{x})$, the equation of motion for $\phi$ in the radiation dominated era becomes
\begin{equation}
  \psi'' - \nabla^2 \psi + \psi \left ( |\psi|^2 - \frac{R^2}{2} + \frac{1}{6} R^2 T^2 \right ) = 0~,
\end{equation}
where $T^2 = 2 \zeta H$ (the definition of $\zeta$ will be given in Appendix~\ref{sec:app:option3}).

\section{Evolutions on the lattice}
\label{app:sec:detail:lattice}

The approach toward the scaling solution can be boosted by the relaxation which cleans up the noisy short-distance string structure.
The pre-evolution runs over the time interval $\tau = [\tau_{pi}, \, \tau_{pf}]$. The outcome of the pre-evolution is taken as the initial condition at $\tau =\tau_i$ for the actual physical string evolution. In our numerical simulation, three different time values, $\tau_{pi}$, $\tau_{pf}$, and $\tau_{i}$ are chosen to satisfy certain properties. In this section, we describe the detail of our simulation setup regarding two different types of the pre-evolutions.

In what follows, we present all the expressions in terms of the dimensionful spacetime to avoid any confusion.  To make it easier to convert the dimensionful spacetime to the dimensionless one in the $m_r$ unit convenient for the simulation, we adopt $R(t) = \sqrt{t/t_0}$ with $t_0 = 1/m_r$ for the radiation dominated era. In terms of the comoving time $\tau$, $R(\tau) = \frac{\tau}{2} m_r$ as $t=(\frac{\tau}{2})^2 m_r$. For the situation with the scaling of $R(t) \propto t$, we adopt $R(t) = t/t_0$ again with $t_0 = 1/m_r$. In terms of $\tau$, $R(\tau) = e^{m_r\tau}$ through $t = \frac{1}{m_r} e^{m_r \tau}$. One can pair $t$ and/or $x$ with $m_r$ to convert it to the dimensionless one in the $m_r$ unit.

\subsection{Common setup}
\label{sec:app:setup}
We first explain how we set $\Delta x$ which allows the longest simulation time in $m_r/H$. For given $N$, one can require the minimum number of the lattice points within the core at the final time, denoted by $n_c$, and the minimum number of the Hubble length within the simulation box $L$ at the final time, denoted by $n_H$:
\begin{equation}\label{app:eq:nc:nH}
   \frac{m_r^{-1}}{R \Delta x} \geq n_c~,\quad \frac{L}{H^{-1}} = \frac{N R \Delta x}{H^{-1}} \geq n_H~\quad \rightarrow \quad
   \frac{m_r}{H} \leq \frac{N}{n_c n_H}~,
\end{equation}
where the maximum time coverage in $m_r/H$ is reached when two inequalities in Eq.~(\ref{app:eq:nc:nH}) for $n_c$ and $n_H$ are saturated at the same time. 
The lattice spacing $\Delta x$ is set to guarantee the maximal $m_r/H$, and it is given by
\begin{equation}\label{app:eq:Deltax:static}
 \Delta x = \left . \frac{m_r^{-1}}{n_c R(t)} \right |_{t=(1/2H)_\text{max}} = \frac{m_r^{-1}}{n_c} \left ( \frac{2 n_c n_H}{N} \right )^{1/2}~.
\end{equation}
The step size of the conformal time $\Delta \tau$ is chosen to $\Delta \tau = r_\text{CFL}\, \Delta x$ where $r_\text{CFL} = 1/3 \sim 1/10$ depending on the simulation approach.

It is useful to compare this setup with that in the simulation assisted by the AMR. When turning on the AMR which recursively zooms in a local region up to the factor of $2^k$, one can repeat the similar steps as in Eq.~(\ref{app:eq:nc:nH}) with a replacement of either $n_c \rightarrow n_c/2^k$ or $N \rightarrow  2^k N$ (up to implication for $\Delta x$) to estimate the maximal time coverage.  In either choice, the maximum dynamic time range in $\log \frac{m_r}{H}$ is extended to (by $k\ln 2$) 
\begin{equation}
  \log \frac{m_r}{H} \leq \log \frac{N\cdot 2^k}{n_c n_H}~.
\end{equation}
Since $N \rightarrow  2^k N$ implies that the local resolution effectively increases by the factor of $2^k$, $\Delta x$ in Eq.~(\ref{app:eq:nc:nH}) corresponds to the lattice spacing of the finest grids in the comoving coordinate,
\begin{equation}\label{eq:dx:AMR:finest}
 \left . \Delta x \right |_{\text{Finest girds by }2^k}= \frac{m_r^{-1}}{n_c} \left ( \frac{2 n_c n_H}{2^k N} \right )^{1/2}~.
\end{equation}
In contrast, $n_c \rightarrow n_c/2^k$ in Eq.~(\ref{app:eq:nc:nH}) implies that the minimum number of grids required to be present within the core is relaxed by the factor of $2^k$ since the local region around the string core will be resolved later by up to $2^k$. In this case, $\Delta x$ in Eq.~(\ref{app:eq:nc:nH}) corresponds to the lattice spacing of the coarsest grids in the comoving coordinate,
\begin{equation}\label{eq:dx:AMR:coarse}
 \left . \Delta x \right |_{\text{Coarest girds}}= \frac{m_r^{-1}}{n_c} \left ( \frac{2 n_c n_H}{N} \right )^{1/2} \times 2^{k/2}~,
\end{equation}
which is $2^k \times \left . \Delta x \right |_{\text{Finest girds by }2^k}$ and it should be compared to the lattice spacing of the static simulation for the region away from strings. A peculiar feature of the scaling in Eq.~(\ref{eq:dx:AMR:coarse}) is that the lattice spacing away from strings in AMR is larger than that of the static case by the factor of $2^{k/2}$ for fixed $n_c$ and $N$. This can be compensated with a larger $n_c$.

\subsection{Fat string pre-evolution: $R(t) \propto \sqrt{t}$}
\label{sec:app:option1}

The first scheme introduced in~\cite{Gorghetto:2018myk} is that, during the fat-string pre-evolution, the core radius $m_{r,\text{fat}}^{-1}$ scales as $R \propto \sqrt{t}$ while the VEV of $\phi$ is fixed to be $f_a$, or equivalently, the string is made to be fat such that its size $r_\text{fat}$ in the comoving coordinate  stays the same during the evolution. The quartic coupling scales as $1/R^2$~\footnote{The decreasing quartic coupling in this scheme indicates that the field becomes less and less interacting with the evolution, and it might be the reason for why the simulation reach the attractor solution faster than the case for the physical string simulation.},
\begin{equation}
 \frac{m_{r,\text{fat}}^2}{2f^2_a} =\frac{1}{2 f_a^2 R^2 \left (m_{r,\text{fat}}^{-1}/R \right )^2}
 =\frac{1}{2 f_a^2 R^2 r_\text{fat}^2} \propto \frac{1}{R^2}~.
\end{equation}
We choose the string core size during the fat string pre-evolution to match the value at the initial time $\tau_{i}$ of the physical evolution, or
\begin{equation}\label{eq:app:fat1:coresize}
   r_\text{fat} (\tau_i)  =  \left . \frac{m_{r,\text{fat}}^{-1}}{R} \right |_{\tau=\tau_{pi}}= \frac{m_r^{-1}}{R(\tau_i)} = m_r^{-1}\times \frac{2}{m_r \tau_i}~.
\end{equation}
The number of lattice points within the string core at the initial time $\tau_i$ in our physical string simulation is given by
\begin{equation}\label{eq:app:fat1:core:Nr}
\begin{split}
 N_r =  \left . \frac{m_{r,\text{fat}}^{-1}}{\Delta x R} \right |_{\tau=\tau_{pi}}
= \frac{m_r^{-1}}{\Delta x}\times \frac{2}{m_r \tau_i}~.
\end{split}
\end{equation}
This allows us to start the physical evolution from the initial time $\tau_i$ for the given $\Delta x$ with $N_r$. 
We choose the initial time of the pre-evolution $\tau_{pi}$ such that the number of lattice points within the Hubble length at $\tau_{pi}$ is roughly of a similar order of the one within a wavelength of the string core, $2\pi m_r^{-1}$, at the time $\tau_{i}$ up to some constant $\alpha$,
\begin{equation}\label{eq:app:fat1:lattice}
\begin{split}
 \underbrace{\frac{\tau_{pi}}{\Delta x} = \left . \frac{H^{-1}}{\Delta x R} \right |_{\tau_{pi}}}_{{\substack{\text{Number of lattice within} \\ \text{the Hubble length} }}} 
= 
\alpha\times \underbrace{\left . \frac{m_r^{-1}}{\Delta x R} \right |_{\tau_{i}}}_{\substack{\text{Number of lattice} \\ \text{within the string core}}}~,
\end{split}
\end{equation}
where the second term is equal to $\alpha N_r$ which stays the same over the entire fat string evolution.
The role of the condition in Eq.~(\ref{eq:app:fat1:lattice}) is to achieve roughly a reasonable size of the correlation length of the cosmic string at $\tau_i$. The ambiguity in the identification in Eq.~(\ref{eq:app:fat1:lattice}) is taken into account by the constant $\alpha$ that can be empirically determined. The setting in Eq.~(\ref{eq:app:fat1:lattice}) ensures that $\frac{m_{r,\text{fat}} }{H} |_{\tau_{pi}} \sim \alpha$ which is taken to be order one.
The ending time of the fat-string evolution $\tau_{pf}$ is determined by demanding the requested value for the number of strings per Hubble patch $\xi_0$, namely, $\xi (\tau_{pf}) = \xi_0$\footnote{While the fat string evolution can be designed to be terminated at $\tau_{pf}$ when it reaches the same requested correlation length, the initial value $\xi_0$ will not be same in this case.}. The value of $\tau_{pf}$ varies depending on $\xi_0$, and the correlation length at $\tau_{pf}$ (which will be the property of the initial condition for the physical string evolution) is determined posteriori. A drawback is that the correlation length with respect to the core radius can significantly grow until $\tau_{pf}$ due to the scaling, $\frac{m_{r,\text{fat}}}{H}|_{\tau_{pf}} \propto \tau$. While the value of $\tau_i$ for the physical string evolution can be freely chosen irrespective of $\tau_{pf}$ ($\tau_i$ is not necessarily larger than $\tau_{pf}$), the implicit correlation length at $\tau_i$ is genuinely determined by the configuration of fields at $\tau_{pf}$ instead of being determined in terms of $H^{-1}(\tau_i)$. While the condition in Eq.~(\ref{eq:app:fat1:lattice}) looks a bit ad-hoc, it will be replaced by more consistent one in the second scheme.

\subsection{Fat string pre-evolution: $R(t) \propto t$}
\label{sec:app:option2}

The second scheme that we take as a default option for the pre-evolution was introduced in~\cite{Gorghetto:2020qws}.
In this scheme, during the fat-string evolution, the core radius $m_r^{-1}$ scales as $R_\text{fat}  \propto t \sim H_\text{fat}^{-1}$ while the VEV of $\phi$ is fixed to be $f_a$. Similarly to the option 1 in Section~\ref{sec:app:option1}, we match the core radii in the comoving coordinate in the transition from the pre-evolution to the physical one,
\begin{equation}\label{eq:app:fat2:coresize}
   r_\text{fat} (\tau_i)  =  \left . \frac{m_{r,\text{fat}}^{-1}}{R_\text{fat}} \right |_{\tau=\tau_{pi}}= \frac{m_r^{-1}}{R(\tau_i)} = m_r^{-1}\times \frac{2}{m_r \tau_i}~.
\end{equation}
The number of lattice points within the string core at the initial time $\tau_i$ will be the same as that during the pre-evolution.
Since $H_\text{fat}^{-1}/R_\text{fat}$ stays constant during the pre-evolution, we can maintain the exactly same correlation length in the comoving coordinate at $\tau_i$ as the one at $\tau_{pi}$ by demanding
\begin{equation}\label{eq:app:fat2:lattice}
\begin{split}
\left . \frac{H_\text{fat}^{-1}}{\Delta x R_\text{fat}} \right |_{\tau_{pi}}
=  
 \left . \frac{H^{-1}}{\Delta x R} \right |_{\tau_{i}}= \frac{\tau_i}{\Delta x}~,
\end{split}
\end{equation}
which sets the comoving Hubble length $H^{-1}/R$ during the pre-evolution to $\tau_i$. 
Combining two relations in Eqs.~(\ref{eq:app:fat2:coresize}) and (\ref{eq:app:fat2:lattice}), it is straightforward to see that the correlation length over the core size stays the same, that is,
\begin{equation}\label{eq:app:fat2:mH}
   \frac{m_{r,\text{fat}}(\tau)}{H_\text{fat} (\tau)}  = \frac{R_\text{fat}(\tau)}{m^{-1}_{r,\text{fat}}(\tau)} \frac{H^{-1}_\text{fat}(\tau)}{R_\text{fat} (\tau)}
   = \left . \frac{R}{m_r^{-1}} \right |_{\tau_i} \left . \frac{H^{-1}}{R} \right |_{\tau_i} =  \frac{m_r}{H(\tau_i)} ~,
\end{equation}
where the initial correlation length over the core size is given by $\frac{m_r}{H(\tau_i)} = \frac{1}{2} m_r^2 \tau_i^2$. 
The initial time for the physical string evolution $\tau_i$ is fixed by Eq. (\ref{eq:app:fat2:lattice})~\footnote{The freedom to choose any value of $\tau_i$ (or the freedom to choose any desired value of $\frac{m_r}{H(\tau_i)}$) effectively amounts to introducing $\alpha$ in $R(t) = \alpha (t/t_0)$, or $\tau_i = \frac{1}{\alpha m_r}$ from Eq. (\ref{eq:app:fat2:lattice}).} in this scheme which accordingly determines $N_r$ through Eq.~(\ref{eq:app:fat2:coresize}).
Similarly to the option 1 in Section~\ref{sec:app:option1}, the final time of the pre-evolution is determined by requesting a certain number of strings per Hubble patch $\xi_0$, namely, $\xi (\tau_{pf}) = \xi_0$. While the initial time $\tau_{pi}$ of the pre-evolution is apparently a free parameter, we set it to $\tau_{pi} = 0$.

\subsection{From pre-evolution to physical one}

Since the field equations during pre- and physical evolutions in options 2 and 3 are qualitatively different and not physically equivalent, there may not be a unique way of gluing two regions. 
In this work, we join two regions by two relations, 
\begin{equation}\label{app:eq:matching:pre:phys}
\begin{split}
 \left .  ( R^{-1}\psi ) \right |_{\tau =\tau_{pf}} &= \left .  ( R^{-1}\psi ) \right |_{\tau=\tau_{i}}~,
 \\[5pt]
 \left . \left ( -\frac{R'}{R^2}\psi + \frac{1}{R}\psi' \right ) \right |_{\tau=\tau_{pf}} &=
  \left . \left ( -\frac{R'}{R^2}\psi + \frac{1}{R}\psi' \right ) \right |_{\tau=\tau_{i}}~,
\end{split}
\end{equation}
where the second relation is the continuity of $(R^{-1} \psi)'$, or the velocity of $R^{-1}\psi$ with respect to the conformal time and it ensures that the energy density is smoothly connected between two regions~\footnote{While the velocity $\dot\phi$ is discontinuous in this scheme, we have numerically checked that the impact of the discontinuity on the scaling solution is negligible.}. We empirically find that maintaining the continuity of the energy density in the transition is strongly favored not to artificially introduce a discrete enhancement of the radial modes in the transition from the pre-evolution to the physical one.

\subsection{Thermal pre-evolution}
\label{sec:app:option3}

We continue presenting the relations in terms of the dimensionful parameters. 
The expressions in terms of the dimensionless quantities can be straightforwardly obtained as the previous discussion along with the rescaling $T$ by $f_a$.
The critical temperature $T_c$ is given by $T_c = \sqrt{3} f_a$. Using the relation for the Hubble parameter $H^2 = \frac{\pi^2}{90} g_* \frac{T^4}{M_p^2}$, the temperature can be expressed as
\begin{equation}
 T^4 = f_a^4 \times \frac{H^2}{m_r^2}\times 4\, \zeta^2~ \quad \text{with} \quad
 \zeta^2 = \frac{45 }{2\pi^2 g_*}\frac{M_p^2 m_r^2}{f_a^4}~,
\end{equation}
from which the correlation length over the core size at the critical temperature $T_c$ is given by
\begin{equation}\label{app:eq:option3:corrlength:tc}
  \left . \frac{m_r}{H} \right |_{T_c} = \frac{2}{3} \zeta~.
\end{equation}
While $\zeta \sim \mathcal{O}(1)$ ensures the order one size of $\frac{m_r}{H}$ at the phase transition where cosmic strings are formed, the natural size of it is given by
\begin{equation}\label{app:eq:zeta}
 \zeta =  2 \times 10^7 \left ( \frac{100}{g_*} \right )^{1/2} \left ( \frac{\lambda}{1} \right )^{1/2} 
\left ( \frac{M_p}{1. \times 10^{18} \text{GeV}} \right ) \left ( \frac{1. \times 10^{10} \text{GeV}}{f_a} \right )~.
\end{equation}
Eq.~(\ref{app:eq:zeta}) implies that the order one size of $\zeta$ corresponds to the symmetry breaking scale of $f_a \sim 10^{17}$ GeV, a huge number of relativistic degrees of freedom of $g_* \sim 10^{14}$, or an extremely small quartic coupling~\footnote{This looks like a similar unnatural situation in the fat-string type pre-evolution which relies on the quartic coupling scaling away as $R^{-2}$. }  $\lambda = \frac{m_r^2}{2 f_a^2} \sim 10^{-15}$ (or a combination of three factors). This might be considered to be a drawback of this option. The initial time of the simulation $\tau_i$ can be chosen to be any value earlier than the phase transition time $\tau_c$ which is given by, using Eq.~(\ref{app:eq:option3:corrlength:tc}),
\begin{equation}
   \tau_c^2 = m_r^{-2} \times \frac{4}{3} \zeta~.
\end{equation}
In our numerical simulation, we fix the initial time to $\tau_i = 0.1 \tau_c$ with the varying $\zeta$. The value of $\zeta$ can be fixed to have a specific Hubble length in the $m_r^{-1}$ unit at the critical temperature as was indicated in Eq.~(\ref{app:eq:option3:corrlength:tc}), for instance, $\zeta = \frac{3}{\sqrt{2}} \sim 2.12132$ for $ \left . \frac{m_r}{H} \right |_{\tau_c} = \sqrt{2}$, and it implies $\left . \frac{m_r}{H} \right |_{\tau_i} = (\frac{\tau_i}{\tau_c})^2\left . \frac{m_r}{H} \right |_{\tau_c}$ at the initial time $\tau_i$.

\subsection{Random initial conditions}
\label{app:sec:IC}

In fat string type relaxations explained in Sections~\ref{sec:app:option1} and \ref{sec:app:option2}, a random field configuration $\phi$ (and its velocity $\dot\phi$) is generated with the Fourier modes with the wavenumber truncated at $m_r$, or $|\vec{k}|_\text{max} = \frac{2\pi}{N\Delta x R (\tau_i)} |\vec{n}_k |_\text{max} = m_r$.
The field can be discretized as
$\phi ( \vec{x} = \Delta x\, \vec{n}_x)  = \sum_{\vec{n}_{k}} \frac{1}{(N\Delta x R)^3} \tilde \phi \left ( \frac{2 \pi}{N \Delta x R} \vec{n}_k \right )  \exp{\frac{2\pi i \vec{n}_k\cdot \vec{n}_x}{N}}$
where the Fourier mode $\tilde {\phi} (\vec{k})$ is randomly generated below the cutoff scale of the momentum as $\tilde{\phi}_0\, \frac{\tt{Rand}_1 + i \tt{Rand}_2}{\sqrt{2}}$ where $\tt{Rand}_{1,2}$ are random numbers following a normal distribution such that 
$\langle {\tt Rand}_i (\vec{n}_{k_1})\, {\tt Rand}_j (\vec{n}_{k_2}) \rangle = \delta_{ij}\delta_{\vec{n}_{k_1} \vec{n}_{k_2}} (N\Delta x R)^3$.
The overall size $\tilde{\phi}_0$ is chosen such that $\langle |\phi (\vec{n}_x) |^2\rangle = f_a^2 \times \frac{1}{2}$:
\begin{equation}
\langle \phi^* (\vec{n}_{x_1}) \phi (\vec{n}_{x_2}) \rangle 
 =  | \tilde{\phi}_0 |^2
  \sum_{\vec{n}_{k}} \frac{1}{(N\Delta x R)^{3}} \exp^{-\frac{2\pi i \vec{n}_k\cdot (\vec{n}_{x_1} - \vec{n}_{x_2} )}{N}}~,
\end{equation}
where the sum is truncated at $|\vec{n}_{k}|_\text{max}$.
The randomly generated Fourier mode $\tilde\phi$ is given by
\begin{equation}\label{app:eq:option12:phi:rand}
 \tilde \phi
 = \frac{f_a}{\sqrt{2}} \frac{(N\Delta x R)^{3/2}}{\sqrt{N_k}}
  \frac{{\tt{Rand}_1} (\vec{n}_k) + i {\tt{Rand}_2} (\vec{n}_k) }{\sqrt{2}}  \theta (|\vec{n}_k|_\text{max} -  |\vec{n}_k| )~,
\end{equation}
where $N_k$ is the total number of lattice points within the sphere in the momentum space, which is defined as $N_k \equiv \frac{1}{(\Delta k)^3} \int d^3\vec{k}\,  \theta (|\vec{n}_k|_\text{max} - |\vec{n}_k| )$ in terms of the notation in the continuum limit, and $\theta$ is the step function.
The random velocity distribution for $\phi'$ is similarly generated with the Fourier mode further weighted by  $R E$, compared to $\tilde{\phi}$ in Eq.~(\ref{app:eq:option12:phi:rand}), where $E (\vec{k} = 2\pi \vec{n}_{k}/(N\Delta x R)) = \sqrt{m_r^2 + \vec{k}^2}$.

In the thermal relaxation scheme in Section~\ref{sec:app:option3}, the initial conditions for the field and velocity are generated assuming the Gaussian random field configuration following the thermal distributions as in~\cite{Kawasaki:2018bzv}. Similarly to the fat string type relaxation, a random field configuration and its velocity are generated with an appropriate cut on the wavenumber.  
For the periodic boundary condition, the lowest wavenumber in the $m_r$ unit at the initial time is given by
$\left . \frac{k_\text{min}}{m_r} \right |_{t_i} = \left . \frac{2\pi}{m_r R N \Delta x} \right |_{t_i} = \frac{10 \sqrt{3}}{\sqrt{\zeta}} \frac{2\pi}{\left ( 2 n_H N/n_c\right )^{1/2}}$.
The cutoff $k_{max}$ on the physical momentum is imposed as a multiple of $m_r$. It corresponds to including only $|\vec{n}_k|_\text{max}$ number of lowest modes, or $k_\text{max} = \frac{2\pi}{R N \Delta x} |\vec{n}_k|_\text{max} = \kappa\, m_r$ at $\tau = \tau_i$ from which
$|\vec{n}_k|_\text{max} = \kappa \left . (k_\text{min}/m_r)^{-1} \right |_{\tau_i}$. For $\zeta = 2.12$ and $N=1024/2048/4096$ along with our default setting $n_H = 4^{1/3}$ and $n_c=1$, $\left . (k_\text{min}/m_r)^{-1} \right |_{\tau_i} \sim 0.763/1.079/1.526$. Therefore, the cut $k_\text{max}|_{\tau_i} \leq 10\, m_r$ is equivalent to including only $|\vec{n}_k|\lesssim$ 8/11/15 lowest modes, respectively. As $k_\text{max} \propto \tau^{-1}$, $k_\text{max}|_{\tau_i} \leq 10\, m_r$ implies $k_\text{max}|_{\tau_c} \leq m_r$ with the choice of $\tau_i = 0.1 \tau_c$ (note that $m_r$ denotes the constant string core scale).

\subsection{Tetrahedron-based string identification}
\label{app:sec:stringID}

We subdivide the entire lattice box into a set of tetrahedrons as in the left panel of Fig.~\ref{fig:cartoon:tetrahedron}, and decide whether a string penetrates a face or not by examining the pattern of the phases at three vertices. Once the criterion for the existence of a string is met, we simply determine the string core by the linear interpolation (although it is not guaranteed to be the truth position). In this approach, we estimate total string length in the simulation box by summing length of all the line segments connecting the string cores.
On the other hand, since the position of the string core over the plaquette (see right panel of Fig.~\ref{fig:cartoon:tetrahedron}) through which a string penetrates is likely randomly distributed from the statistics point of view, the total string length in the simulation box can also be estimated by just counting the number of plaquettes and multiplying it by a weighting factor~\cite{Fleury:2015aca}. 
Using the simulation on the grid with $N^3 = 1024^3$, we have compared our algorithm with the string identification in~\cite{Fleury:2015aca}, and the discrepancy $\Delta \xi /\xi$ over the entire simulation time was found to be smaller than a few \%.
Since we want to visualize the string formation and take snapshots at some selected times, we take our approach for the estimation of the total string length.
The number of long strings per Hubble patch in our simulation is calculated as
\begin{equation}
  \xi = \frac{(L_\text{Lattice unit} \Delta x R) t^2}{(N \Delta x R)^3} 
  = \frac{L_\text{Lattice unit} t^2}{N^3 \Delta x^2 R^2}~.
\end{equation}
where $L_\text{Lattice unit}$ is the total length of strings measured in a lattice unit.

In our prescription using the tetrahedron, the surface of the tetrahedron in the left panel of Fig.~\ref{fig:cartoon:tetrahedron} is the triangle with three vertices, call them $v_{i=1,2,3}$. Given the complex field values on three vertices, one can define $\theta_{ij} = \arg (\phi_j/\phi_i) = \arg (\phi_j \phi_i^*)$. A string core on the triangular surface is declared if $\theta_{123} \equiv \theta_{12} + \theta_{23} + \theta_{31}$ has the value of $\pm 2\pi$. Note that $\theta_{123}$ can have only $0,\, \pm 2\pi$. In practice, the computation of the arguments for all complex fields are time-consuming. To reduce the computational time, we replace the above criterion with more efficient equivalent step. $\theta_{123} = 2\pi$ ($- 2\pi$) is equivalent to the situation where all $\theta_{ij}$'s are non-negative (non-positive) while not all of them are zero. The sign of $\theta_{ij}$ is equivalently replaced with that of $\omega_{ij} = \Re\phi_i \Im\phi_j - \Im\phi_i\Re\phi_j$. 
Therefore, $\theta_{123} = 2\pi$ ($- 2\pi$) is equivalent to the case where all $\omega_{ij}$'s are non-negative (non-positive) while $\omega_{123}$ is positive (negative) ($\omega_{123} \equiv \omega_{12} + \omega_{23} + \omega_{31}$). Once a string core is identified, the location of the string core inside the triangle where the field value vanishes is assigned via the linear interpolation, or $(\omega_{23} v_1 + \omega_{31} v_2 + \omega_{12} v_3)/\omega_{123}$ where $v_i$ is the position of the $i$-th vertex on the triangle. It can be easily proven that the string core is strictly inside the triangle. This approach is well-defined as each tetrahedron includes either a pair of two faces on which string cores are found or none, and thus strings are guaranteed to be connected without any discontinuity. We find that our tetrahedron-based string identification algorithm is quite efficient in numerical simulations.

The connectedness of the string in our tetrahedron approach for the string identification can be rigorously proven as follows. 
One can introduce a complex field given as a linear function of coordinates, $\phi_\text{inter} (x,\, y,\, z) = \beta_0 + \beta_1 x + \beta_2 y + \beta_3 z$ with four complex coefficients which we fix them by matching the interpolating field $\phi_\text{inter}$ to field values at four vertices, labeled as $v_{i=0,1,2,3}$, of the tetrahedron, or $\phi_\text{inter} (v_i) = \phi_i$. That is, $\phi_\text{inter} (x,\, y,\, z)$ linearly interpolates the field values between the vertices of the tetrahedron in three-dimensional space. The linear interpolation of the field on the triangular face of the tetrahedron is equivalent to restricting $\phi_\text{inter}$ onto the triangle.
In this picture, the corresponding triangular face is declared to have a string core if and only if the triangular face and the locus of zero, $\phi_\text{inter} (x,\, y,\, z) =0$, intersect. $\phi_\text{inter} (x,\, y,\, z) =0$ represents a straight line as it is a collection of two constraints (for real and imaginary components of $\phi_\text{inter}$) in three-dimensional space.
A straight line can penetrate only a pair of triangular faces of the tetrahedron if it intersects unless the line passes through a vertex or an edge of the tetrahedron. Therefore, there must be none or exactly two triangular faces which are penetrated by a cosmic string. This prescription will work well for long strings compared to the size of the tetrahedron.

\subsection{Masking}
\label{app:sec:masking}

The axion energy density is approximated by $2\times \langle \frac{1}{2}\dot{a}^2 \rangle$ where $\langle \cdots \rangle$ denotes (with abuse of the notation) the spatial average over the physical space (denoted by $\vec{x}_p = R\, \vec{x}$),
\begin{equation}\label{eq:rho:axion}
  \rho_a = \frac{1}{L^3}\int d^3\vec{x}_p\, \dot{a}^2 (\vec{x}_p) = \frac{1}{L^3} \int \frac{d^3 \vec{k}}{(2\pi)^3}\, |\tilde{\dot{a}} (\vec{k})|^2~,
\end{equation}
from which the axion energy spectrum for the momentum $k$ is given by
\begin{equation}\label{eq:drhodk:axion}
  \frac{\partial\rho_a}{\partial k} = \frac{k^2}{2\pi^2 L^3} \int \frac{d\Omega_k (\hat{k})}{4\pi} \left | \tilde{\dot{a}} (\vec{k} = k\hat{k}) \right |^2~.
\end{equation}
Importantly, $\dot{a}(\vec{x}_p)$ in Eq.~(\ref{eq:rho:axion}) should be the axion field velocity with the string effect removed.
To this end, in presence of strings, the masking function needs to be applied to the unmasked axion velocity field to screen off the effect from string cores and extract the pure axion energy density,
\begin{equation}
   \tilde{\dot{a}}_\text{masked}(\vec{k}) = \int d^3 \vec{x}_p\, W(\vec{x}_p) \,\dot{a}_\text{unmasked} (\vec{x}_p) e^{-i\vec{k}\cdot \vec{x}_p}~,
\end{equation}
where $W(\vec{x}_p)$ is the masking function. 
We adopt the natural masking $W (\vec{x}_p) = (1+ r(\vec{x}_p)/f_a)$ for the smooth screening of the string core effect as a default. Another choice for the masking adopted in literature is the hard-masking which sets $W = 0$ for the region with distance from string $r < d_s m_r^{-1}$ (for some positive $d_s$) and $W=1$ outside the region. 
The masked Fourier transform $\tilde{\dot{a}}_\text{masked}(\vec{k})$ is related to the Fourier transform $\tilde{\dot{a}} (\vec{k})$ of the axion velocity field $\dot{a}(\vec{x}_p)$, now separated from the string effect, by the convolution,
\begin{equation}
  \tilde{\dot{a}}_\text{masked}(\vec{k})  = \int \frac{d^3\vec{k}}{(2\pi)^3}\, \tilde{W}(\vec{k} - \vec{k}')\, \tilde{\dot{a}}(\vec{k'})~,
\end{equation}
where $\tilde{W}(\vec{k})$ is the Fourier transform of the masking function $W(\vec{x}_p)$ itself. 
In our simulation, $\mu_{a, \text{masked}} (k) = \int \frac{d\Omega_k}{4\pi}  |\tilde{\dot{a}}_\text{masked} (\vec{k})|^2$ and similarly $\mu_W (k) = \int \frac{d\Omega_k}{4\pi}  |\tilde{W} (\vec{k})|^2$ are numerically computed. The value of $\mu_a (k) = \int \frac{d\Omega_k}{4\pi}  |\tilde{\dot{a}} (\vec{k})|^2$ for the axion velocity with the string core effect screened off can be obtained by inverting the following relation,
\begin{equation}\label{eq:mua}
  \mu_{a, \text{masked}} (k) = 
  \frac{1}{4\pi^2 k L^3} \int dk' k' \left ( \int^{k+k'}_{|k-k'|} dk'' k'' \mu_W (k'') \right ) \mu_{a} (k')~.
\end{equation}
This relation is derived under the assumption that each $\vec{k}$ modes of $\dot{a}$ are independent and has random amplitude from mean zero Gaussian ensemble whose variance only depends on the size of the momentum $k$. 
Here, $\mu_{a,\text{masked}}$ and $\mu_a(k)$ need to be the average over the ensemble to make the relation mathematically exact. Using the relation to get $\mu_a$ from the measured $\mu_{a,\text{masked}}$ is secretly relying on the law of large number applied to $\mu_{a,\text{masked}}(\vec{k})$ for $\vec{k}$'s with large number of different directions and the same magnitude. However, once space is discretized, and so is the momentum, there are only a few number of different momentum $\vec{k}$ for small magnitude $k$, and this may result in larger uncertainty at low momentum region while estimating the unmasked spectrum.
Using $\mu_a(k)$ extracted from Eq.~(\ref{eq:mua}), the axion energy spectrum for the momentum $k$ with the string core effect screened off is given by
\begin{equation}
  \frac{\partial\rho_a}{\partial k} = \frac{k^2}{2\pi^2 L^3} \mu_a (k)~.
\end{equation}
Similar masking is applied to the radial modes.

\section{Dependency on initial conditions and relaxations}
\label{app:sec:dependency}

\subsection{Scaling regime and fitting}
\label{app:sec:relaxation}

The underlying assumption behind the pre-evolution is that the string network eventually approaches the scaling solution. Any residual influence from the different pre-evolutions or/and different initial conditions should be faded away as the string network enters the scaling regime.  Any justification for it within the consistent numerical setup should be highly beneficial. To this end, we extend our discussion in Section~\ref{sec:SR} to include more comprehensive and detailed comparisons between results obtained by two different pre-evolution schemes for various initial data. We do this task mainly with simulations on the lattice of $N^3=1024^3$ to accumulate as many simulation runs as possible for various initial conditions within a reasonable time. As was discussed in Section~\ref{sec:SR} and illustrated in Fig.~\ref{fig:L1024:xi20:option2:vs:zeta1:option3}, this task helped us in choosing our benchmark points to be run on a larger lattice space for the study at much later dynamic time.

\begin{figure}[tp]
\begin{center}
\includegraphics[width=0.48\textwidth]{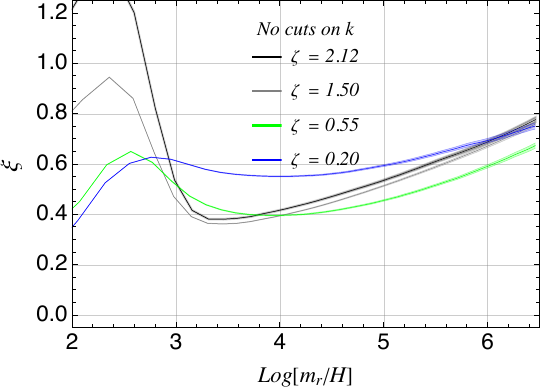}\quad
\includegraphics[width=0.48\textwidth]{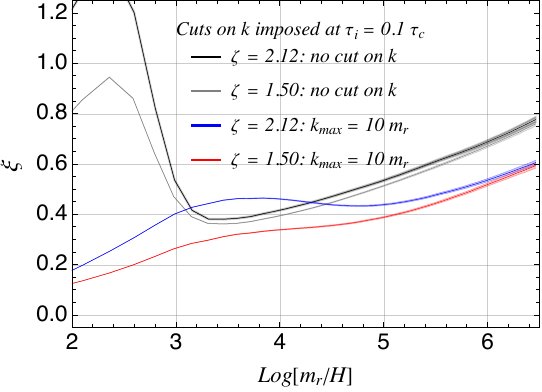}
\caption{\small The impact of higher modes above $k_\text{max} = 10 m_r$ on $\xi$ for physical string simulations with initial data generated with the thermal pre-evolution. Note that $k_\text{max} = 10 m_r$ at $\tau_i = 0.1 \tau_c$ is equivalent to $k_\text{max} = m_r$ at $\tau_c$.}
\label{fig:app:xi:1024:option2:vs:option3}
\end{center}
\end{figure}

The left panel of Fig.~\ref{fig:app:xi:1024:option2:vs:option3} is similar to the right panel of Fig.~\ref{fig:L1024:xi20:option2:vs:zeta1:option3} except that the cutoffs on the momentum were not imposed. That is, the initial configurations with $k$ modes between maximally allowed value (set by the lattice spacing and lattice size) and $10 m_r$ at $\tau_i = 0.1 \tau_c$ were included. While the $\xi$ curves in the left panel of Fig.~\ref{fig:app:xi:1024:option2:vs:option3} illustrate an approach toward the scaling regime, the $\xi$ values are overally larger than those with momentum cutoffs as is evident in the right panel of the same Fig.~\ref{fig:app:xi:1024:option2:vs:option3} where the dependence on the $k$ cutoffs is demonstrated for two selected $\zeta$ values.
%
%

\begin{figure}[tp]
\begin{center}
\includegraphics[width=0.48\textwidth]{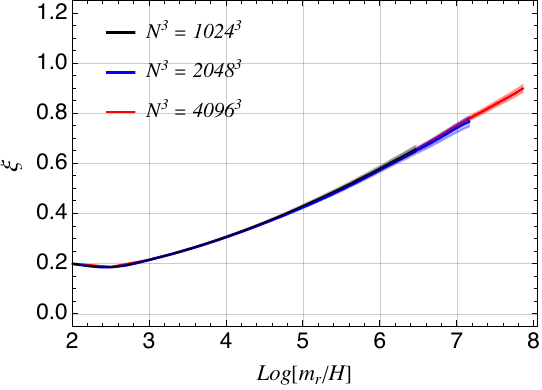} \quad
\includegraphics[width=0.48\textwidth]{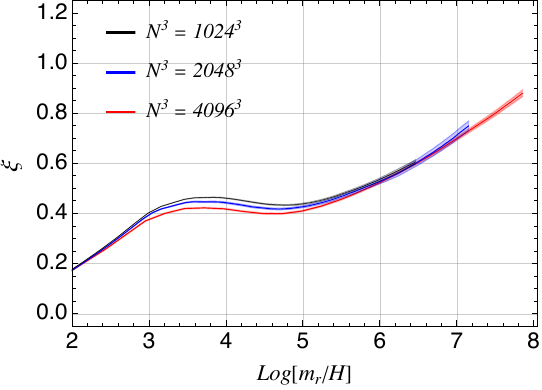}
\caption{\small The independence of $\xi$ curves on the lattice spacing $\Delta x$ (and thus lattice size $N$ as well) for simulations with the momentum cutoffs imposed.}
\label{fig:app:xi:Nindependence:option2:vs:option3}
\end{center}
\end{figure}

Since the maximally allowed $k$ modes varies with the lattice size, $\xi$ curves obtained from simulations with different lattice sizes do not overlap, and the simulation on a larger lattice populates more strings and takes longer time to be relaxed.
The cutoffs on the momentum removes this ambiguity as is illustrated in Fig.~\ref{fig:app:xi:Nindependence:option2:vs:option3} for two benchmark simulations: $\xi_0 = 0.2$ using fat-string pre-evolution (left panel) and $\zeta=2.12$ (with $\tau_i = 0.1\tau_c$) using thermal pre-evolution (right panel). Since the same momentum modes are initially randomly generated irrespective of the lattice size, the $\xi$ curves exactly overlap on top of each other~\footnote{The slight mismatch between $N^3=1024^3$ and $N^3=2048^3$ in the right panel of Fig.~\ref{fig:app:xi:Nindependence:option2:vs:option3} is due to the round-off error, for instance, the cutoff $k>10 m_r$ includes $k$ modes within the radius of roughly $\sim 8$ and $\sim 11$ for $N^3=1024^3$ and $N^3=2048^3$, respectively, (see Appendix~\ref{app:sec:IC}) and rounding-off causes an error up to 10\%. We have checked via independent simulations that increasing the cutoff, allowing more $k$ modes, makes the effect due to the round-off error negligible. A similar statement applies to the case with $N^3=4096^3$.}.
A good agreement between different lattice sizes in Fig.~\ref{fig:app:xi:Nindependence:option2:vs:option3} indicates that the impact of the lattice spacing is not relevant as long as the string core includes at least one grid at the final time, or $n_c =1$. At the final time of the simulation with $N^3=1024^3$ where $n_c^{1024} =1$, the number of grids within the string core for the simulation with $N^3 = 2048^3$ is $n_c^{2048}|_{t_f^{1024}\sim 6.3} = \sqrt{2}$.
%
%
%

To quantify the quality of the scaling solution, we fit the curves to the form in Eq.~(\ref{eq:xi:fit:m2top1}).
Given an ensemble of $N_\text{ens}$ independent runs for $\xi$ curves, we select the $N_\text{fit}$ number of sampling points in the logarithmic time $\log\frac{m_r}{H}$ for fitting the curve. The Gaussian likelihood function is constructed as 
\begin{equation}\label{app:eq:likeli:xi}
    L(c_0,c_1,d_1) =
    \frac{1}{(2\pi)^{N_\text{fit}/2}\sqrt{\det(S)}}
    e^{-\frac{1}{2}\sum_{i,j=1}^{N_\text{fit}}\left(\bar{\xi}_i - \xi_i^\text{sc} (c_0,\, c_1,\, d_1,\, d_2)  \right)\left(S^{-1}\right)_{ij}\left(\bar{\xi}_j - \xi_j^\text{sc} (c_0,\, c_1,\, d_1,\, d_2) \right)}~,
\end{equation}
where $\bar{\xi}_i$ is the mean of $\xi$ data at the time $\log_i = \log\frac{m_r}{H} |_i$ and the ansatz for the scaling solution is given by $\xi_i^\text{sc} (c_0,\, c_1,\, d_1,\, d_2) = c_0 + c_1\log_i + \frac{d_1}{\log_i} + \frac{d_2}{\log^2_i}$. 
The matrix $S$ in Eq.~(\ref{app:eq:likeli:xi}) is the $N_\text{fit} \times N_\text{fit}$ covariance matrix of the sampling points along the curve whose element is given by
$S_{ij} = \frac{1}{N_\text{ens}-1}\sum_{k=1}^{N_\text{ens}}\left(\xi_{i,k} - \bar{\xi}_i\right)\left(\xi_{j,k} - \bar{\xi}_j\right)$. 
The uncertainties $\sigma_{c_0}$, $\sigma_{c_1}$ and $\sigma_{d_1}$ for the $c_0$, $c_1$ and $d_1$ are estimated to be 
$\frac{1}{\sigma_X^2}=-\frac{\partial^2}{\partial X^2}\log L$ for $X=\{ c_0,\, c_1,\, d_1\}$ at the maximizing point.
If one ignores the correlation among $\xi_i$ at different times for a large enough separation $\Delta\log$ between two consecutive $\log_i$, or
\begin{equation}
    L(c_0,c_1,d_1)=\prod_{i=1}^{N_\text{fit}}\frac{1}{\sqrt{2\pi}\sigma_i}
    e^{-\frac{1}{2\sigma_i^2}\left( \bar{\xi}_i - \xi_i^\text{sc} (c_0, c_1, d_1, d_2) \right)^2 }~,
\end{equation}
where $\sigma_i^2=\frac{1}{N_\text{ens}-1}\sum_{j=1}^{N_\text{ens}}\left(\xi_{i,j} - \bar{\xi}_i\right)^2$. Our fitting result is presented in Table~\ref{tab:xi:fit:scaling}. We find that the fitting results are sensitive to the fitting ansatz and the fitting interval.
\begin{table}[tbh]
\centering
  \renewcommand{\arraystretch}{1.08}
      \addtolength{\tabcolsep}{0.35pt} 
\scalebox{0.80}{
\begin{tabular}{|c|c|c|c|c|c|c|c|}  
\hline
\multicolumn{4}{|c|}{Fat string pre-evolution} & \multicolumn{4}{c|}{Thermal pre-evolution}  \\[5pt]
\hline\hline
$\xi_0$ & Lattice size ($N^3$) & Cuts & $N_\text{runs}$ &  $\zeta$ & Lattice size ($N^3$) & Cuts at $\tau_i = 0.1 \tau_c$ & $N_\text{runs}$  \\[5pt]
\hline
0.1 & $1024^3$ & $k \leq m_r$&  160  & 0.5	&$1024^3$ & $k \leq 10m_r$	& 130 \\[5pt]
0.2 & $1024^3$ & $k \leq m_r$&  160  & 1.5	&$1024^3$& $k \leq 10m_r$	& 160\\[5pt]
0.3 & $1024^3$ & $k \leq m_r$&  160  & 1.5	&$1024^3$& no cuts			& 130\\[5pt]
0.4 & $1024^3$ & $k \leq m_r$&  160  & 2.12	&$1024^3$& $k \leq 10m_r$	& 160\\[5pt]
0.6 & $1024^3$ & $k \leq m_r$&  100  & 2.12	&$1024^3$& no cuts			& 130\\[5pt]
0.9 & $1024^3$ & $k \leq m_r$&  100  & 3.0	&$1024^3$& $k \leq 10m_r$	& 130\\[5pt]
\hline
0.2 & $2048^3$ & $k \leq m_r$&  100 & 2.12	&$2048^3$& $k \leq 10m_r$	& 100\\[5pt]
\hline
0.2 & $4096^3$ & $k \leq m_r$&  100 & 2.12	&$4096^3$& $k \leq 10m_r$	& 100\\[5pt]
\hline
%
%
\end{tabular}
}
\caption{\small Summary of our data set used for the analyses. Note that the cut $k \leq 10m_r$ at $\tau_i = 0.1 \tau_c$ is equivalent to $k \leq m_r$ at $\tau_c$}
\label{tab:xi:fit:scaling}
\end{table}
Based on our fitting result, we select two benchmark points for further simulations on the larger lattice. For the simulation with the fat-string type relaxation, $\xi_0 = 0.2$ is chosen. For the one with the thermal relaxation, $\zeta = 2.12$ and $\tau_i = 0.1 \tau_c = 0.1414$ is chosen.

\subsection{Cosmological evolution of string network}
%

\subsubsection{Energy budget}
\label{app:sec:Ebudget}
%
%

\begin{figure}[tp]
\begin{center}
\includegraphics[width=0.47\textwidth]{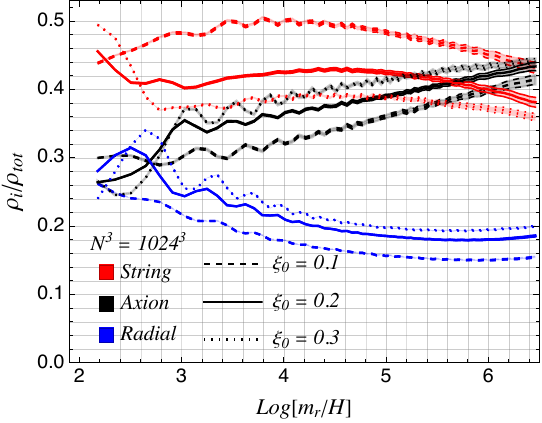}\quad
\includegraphics[width=0.47\textwidth]{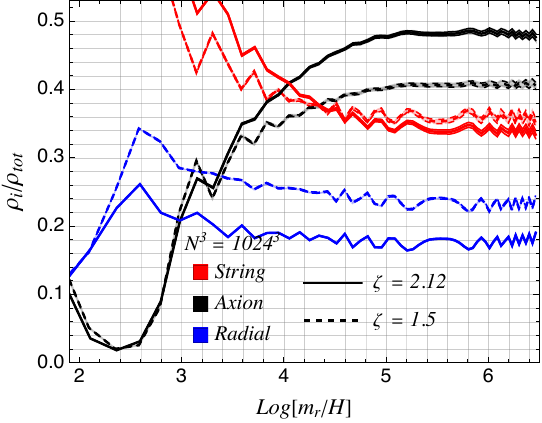}
\caption{\small Energy budget for axion, radial, and string modes from simulations with different lattice sizes: $N^3 = 1024^3$ (solid), $N^3 = 2048^3$ (dashed), and $N^3 = 4096^3$ (dotted), which equivalently represents the dependence on $m_r \Delta$.}
\label{fig:app:Ebudget:varyingN}
\end{center}
\end{figure}
In addition to illustrations in Section~\ref{sec:energybudget}, Fig.~\ref{fig:app:Ebudget:varyingN} shows the dependence of the energy budgets on the initial conditions for two benchmark simulations using our large ensemble data from simulations on the lattice of $N^3 = 1024^3$. As is evident in Fig.~\ref{fig:app:Ebudget:varyingN}, the fraction of the string energy density, $\rho_s/\rho_\text{tot}$ ($\rho_s = \rho_\text{tot} - \rho_a - \rho_r$), tends to have a larger value for a bigger size of $\xi_0$ or $\zeta$ in early times. However, the hierarchy switches over the time, namely $\rho_s/\rho_\text{tot}$ eventually tends to have a smaller value for a bigger $\xi_0$ or $\zeta$ at late times. The transition time for switching the hierarchy, or $\log\frac{m_r}{H} \sim 2.5$ (fat-string pre-evolution) and $\log\frac{m_r}{H} \sim 4.5$ (thermal pre-evolution), looks consistent with the evolution of $\xi$ in Fig.~\ref{fig:L1024:xi20:option2:vs:zeta1:option3}. While the decreased fractional string energy density, when increasing $\xi_0$, goes democratically into axions and radial modes for the case with the fat-string pre-evolution (see left panel of Fig.~\ref{fig:app:Ebudget:varyingN}), the situation, when increasing $\zeta$, in the case with the thermal pre-evolution (see right panel of Fig.~\ref{fig:app:Ebudget:varyingN}) looks more difficult to interpret due to a large migration of radial modes to axions (see also more noisy spectrum in $F$ for a smaller $\zeta$ in the right panel of Fig.~\ref{fig:app:F:1024:ini}). More detailed study on the distinctive behaviors of energy budgets originated from different relaxations will be interesting.

\subsubsection{Instantaneous emission}

\begin{figure}[tph]
\begin{center}
\includegraphics[width=0.48\textwidth]{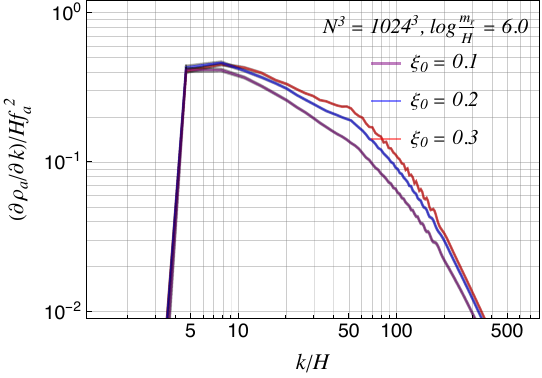}\quad
\includegraphics[width=0.48\textwidth]{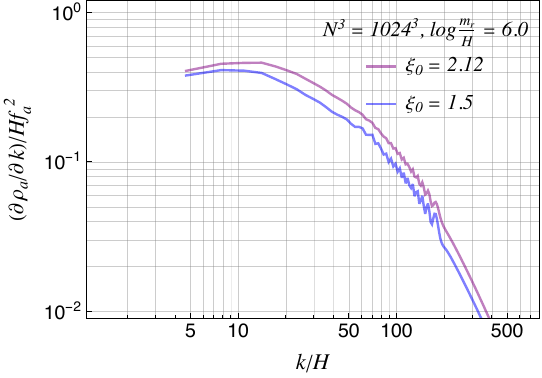}
\caption{\small The differential axion energy spectrum rescaled by $Hf_a^2$ at the time slice $\log \frac{m_r}{H}= 6$ for various initial conditions in two different simulation setups: fat-string type pre-evolution (left) and thermal pre-evolution (right).}
\label{fig:app:drhodk:1024:ini}
\end{center}
\end{figure}

\begin{figure}[tph]
\begin{center}
\includegraphics[width=0.47\textwidth]{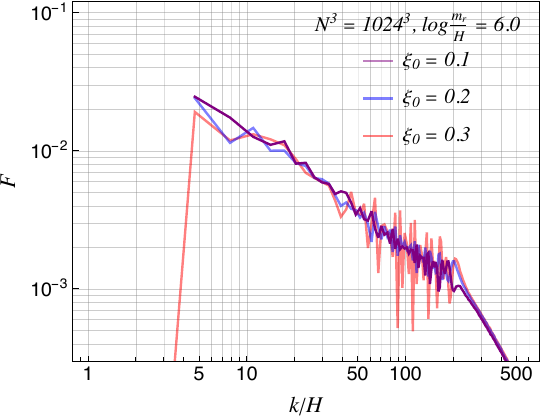}\quad
\includegraphics[width=0.47\textwidth]{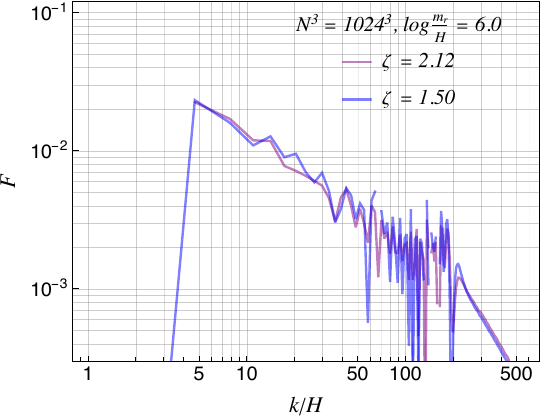}
\caption{\small The instantaneous emission $F$ at the time slice $\log \frac{m_r}{H}= 6$ for various initial conditions in two different simulation setups: fat-string type pre-evolution (left) and thermal pre-evolution (right).}
\label{fig:app:F:1024:ini}
\end{center}
\end{figure}

We continue investigating the dependency of our simulation result on the initial configurations and relaxation schemes with the large ensemble  data from simulations on the lattice of $N^3=1024^3$.
The differential axion energy spectrum $\partial\rho_a/\partial k$ and the instantaneous emission $F$ for the curves with various initial values of $\xi_0$ and $\zeta$ are illustrated in Figs.~\ref{fig:app:drhodk:1024:ini} and~\ref{fig:app:F:1024:ini}, respectively. Overall, the instantaneous emission obtained by the simulation with the thermal relaxation shows more noisy spectrum.
While the larger (smaller) value of $\xi$ populates more (less) pronounced high momentum modes which cause more (less) oscillatory instantaneous emission spectrum in the left panel of Fig.~\ref{fig:app:F:1024:ini}, the larger value of $\zeta$ appears to populate more broad range of momentum modes (see the right panel of Figs.~\ref{fig:app:drhodk:1024:ini}) and the strength of the oscillation amplitudes in the corresponding instantaneous emissions for both $\zeta = 2.12$ and 1.5 look similar (although the case with $\zeta = 1.5$ is more noisy) as is evident in the right panel of Fig.~\ref{fig:app:F:1024:ini}.
The axion emission spectra in Fig.~\ref{fig:app:F:1024:ini} look more oscillatory than the case in~\cite{Gorghetto:2018myk} where the average over samples with finite-ranged initial times was considered. It may be reasonable to assume a stochastic distribution of the Hubble length during the phase transition, instead of a one specific value, and consider its influence on the instantaneous emission spectrum.
It can be implemented by sampling physical string simulations over finite-ranged initial times and taking an average over them to derive the emission spectrum as was done in~\cite{Gorghetto:2018myk}. We also performed the simulation taking into account the stochastic distribution of the initial Hubble length, and we confirmed that distributions become significantly smoothened. 
While this effect is visibly pronounced in early times where the higher modes were not relaxed enough, it gets diminished in later times beyond the time coverage of $N^3=1024^3$, for instance, it becomes less relevant in the simulation with the lattice size of $N^3 = 4096^3$.

\begin{figure}[tph]
\begin{center}
\includegraphics[width=0.47\textwidth]{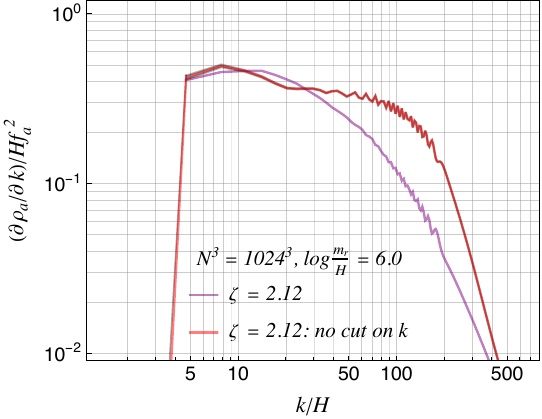}\quad
\includegraphics[width=0.47\textwidth]{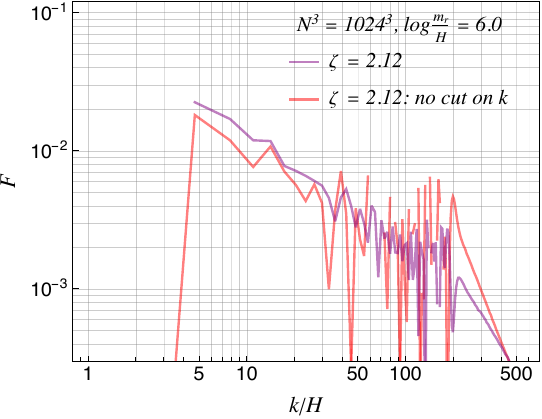}
\caption{\small The differential axion energy spectrum (left) and instantaneous emission $F$ (right) at the time slice $\log \frac{m_r}{H}= 6$ for  two situations with and without the cuts on UV modes above $m_r$ in the simulation using the thermal pre-evolution.}
\label{fig:app:F:1024:Kcut}
\end{center}
\end{figure}

Finally we demonstrate in Fig.~\ref{fig:app:F:1024:Kcut} the impact on the differential axion energy spectrum and the instantaneous emission from UV modes shorter than string core length. The momentum cutoff set by the lattice spacing is higher than the string core scale $m_r$ and generating those UV modes above the $m_r$ scale significantly distorts the spectrum as is seen in Fig.~\ref{fig:app:F:1024:Kcut}. Although the initial field configurations above $m_r$ are not generated at the beginning of the simulation, those higher momentum modes above $m_r$ show up later in the simulation through the equations of motion and their effects are much smaller compared to those from directly generated UV modes from initial random fields configurations.

\section{More on axion spectrum}

\subsection{Fitting instantaneous emission function}
\label{app:sec:fit:F}

\begin{figure}[tp]
\begin{center}
\includegraphics[width=0.47\textwidth]{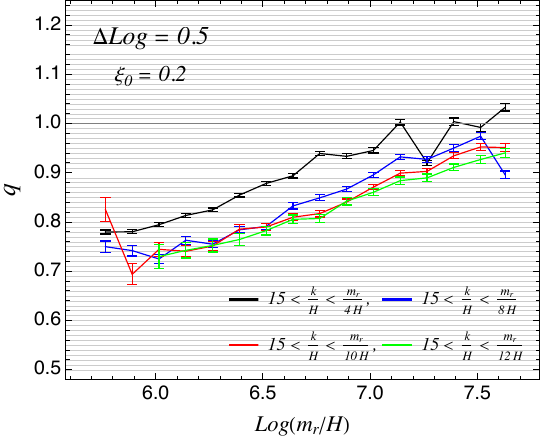}\quad
\includegraphics[width=0.47\textwidth]{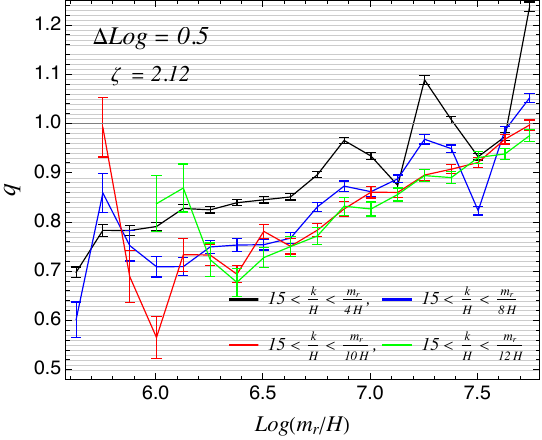}
\\[10pt]
\includegraphics[width=0.47\textwidth]{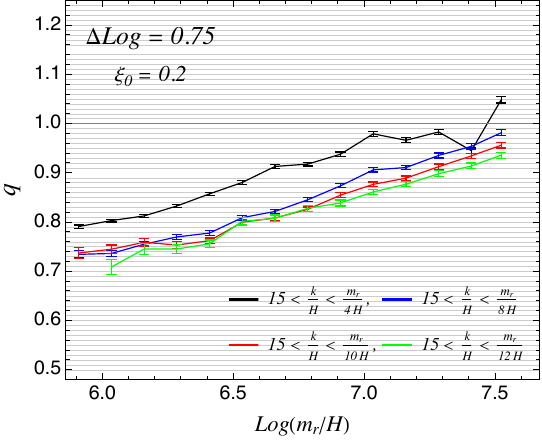}\quad
\includegraphics[width=0.47\textwidth]{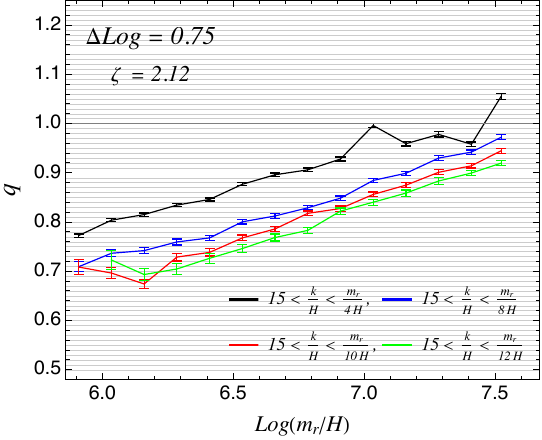}
\caption{\small The spectral index $q$ of the instantaneous emission $F$ fitted in the interval with the varying $x_\text{UV}$ while $x_\text{IR} = 15$. The error bars are statistical.}
\label{fig:q:4096:dlogvaried:UVvaried}
\end{center}
\end{figure}

\begin{figure}[tp]
\begin{center}
\includegraphics[width=0.47\textwidth]{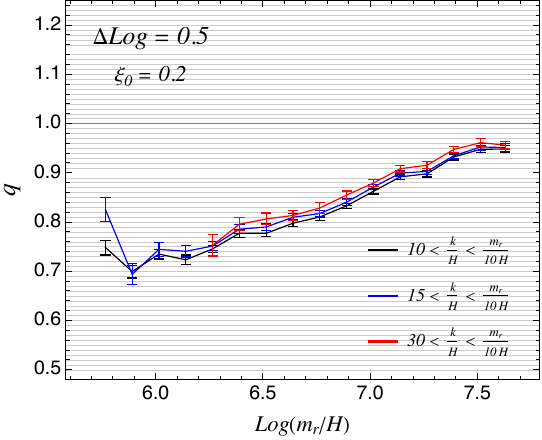}\quad
\includegraphics[width=0.47\textwidth]{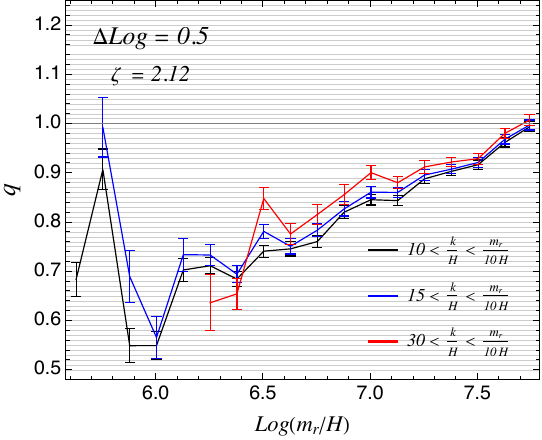}
\\[10pt]
\includegraphics[width=0.47\textwidth]{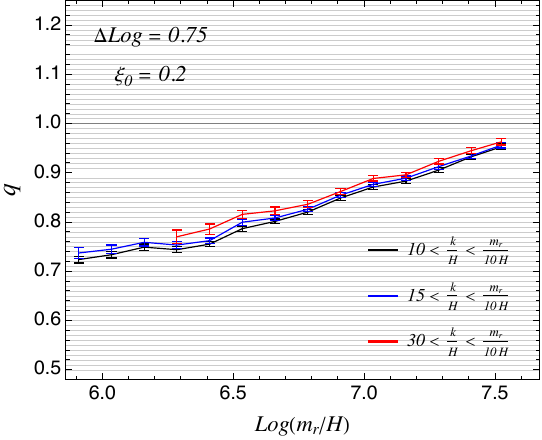}\quad
\includegraphics[width=0.47\textwidth]{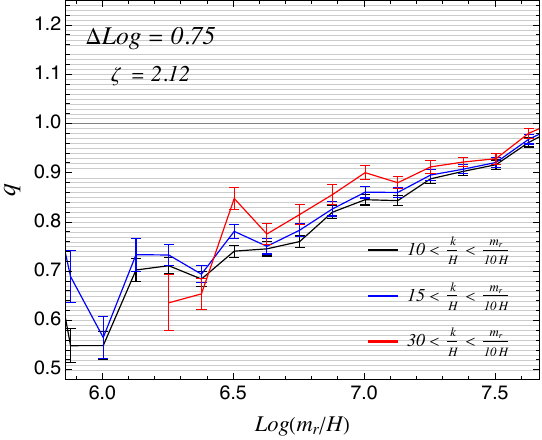}
\caption{\small The spectral index $q$ of the instantaneous emission $F$ fitted in the interval with the varying $x_\text{IR}$ while $x_\text{UV} = 10$. The error bars are statistical.}
\label{fig:q:4096:dlogvaried:IRvaried}
\end{center}
\end{figure}

We explain how the instantaneous emission was fitted using the ansatz for the power law fall-off profile.
The spectral index $q$ for each individual simulation is fitted with the ansatz for $F(x,y)$ assuming the Gaussian distribution with the mean $a  x^{-q}$ and the standard deviation $\sigma(x,B,p,C)=Bx^{-p}+C$ (similarly to~\cite{Buschmann:2021sdq}) where $x=\frac{k}{H}$ is the momentum in the Hubble unit. The fit is performed  within the finite range $[x_\text{IR},\, \frac{y}{x_\text{UV}}]$ at a given time $y\equiv \frac{m_r}{H}$, and the fit parameters $a$, $q$, $B$, $p$ and $C$ are estimated by maximizing the likelihood function, 
\begin{equation}\label{app:eq:L:fitF}
    L(a,q,B,p,C)=\prod_{j=1}^{N_\text{fit}}\frac{1}{\sqrt{2\pi}\sigma(x_j,B,p,C)}e^{-\frac{1}{2\sigma(x_j,B,p,C)^2}\left(F_j-a x_{j}^{-q}\right)^2}
\end{equation}
where $\{ x_j,F_j \}$ are the $N_\text{fit}$ number of data points within the fitting interval $x_\text{IR}\leq x\leq y/x_\text{UV}$. Our ansatz in Eq.~(\ref{app:eq:L:fitF}) assumes no correlation among the measurements of instantaneous power spectrum at different momenta to be consistent with string core masking procedure in which the mode independence is assumed.
The results from individual fits are averaged to yield the spectral index for the corresponding ensemble. They are illustrated in Figs.~\ref{fig:q:4096:dlog0.25:UVvaried},~\ref{fig:q:4096:dlog0.25:IRvaried},~\ref{fig:q:4096:dlogvaried:UVvaried} and~\ref{fig:q:4096:dlogvaried:IRvaried}. 
The fit results of the spectral index in Figs.~\ref{fig:q:4096:dlogvaried:UVvaried} and~\ref{fig:q:4096:dlogvaried:IRvaried} differs from those in Figs.~\ref{fig:q:4096:dlog0.25:UVvaried} and~\ref{fig:q:4096:dlog0.25:IRvaried} in Section~\ref{sec:spectralIndex} only by the $\Delta \log$. Increasing the time separation from the default value, $\Delta \log = 0.25$,  to $\Delta \log = 0.5$ and 0.75 makes the distributions smoother.
As was illustrated in Figs.~\ref{fig:q:4096:dlog0.25:UVvaried} and~\ref{fig:q:4096:dlog0.25:IRvaried}, $x_\text{UV} \sim 10$ looks good enough to get the stable result, and the distribution is less sensitive to $x_\text{IR}$.

While those in Figs.~\ref{fig:q:4096:dlog0.25:UVvaried},~\ref{fig:q:4096:dlog0.25:IRvaried},~\ref{fig:q:4096:dlogvaried:UVvaried} and~\ref{fig:q:4096:dlogvaried:IRvaried} are our fiducial fit results, the fits of $q$ with no $Bx^{-p}$ term in $\sigma$ are carried out as well by setting $B=0$ by hand. It leads to $30\%$ deviation in slope $q_1$ compared to our fiducial result. Nevertheless, the conclusion from linear growth model fit expecting $q\sim O(10)$ at $\log{\frac{m_r}{H}}\sim 70$ is not altered.

\begin{figure}[tp]
\begin{center}
\includegraphics[width=1.0\textwidth]{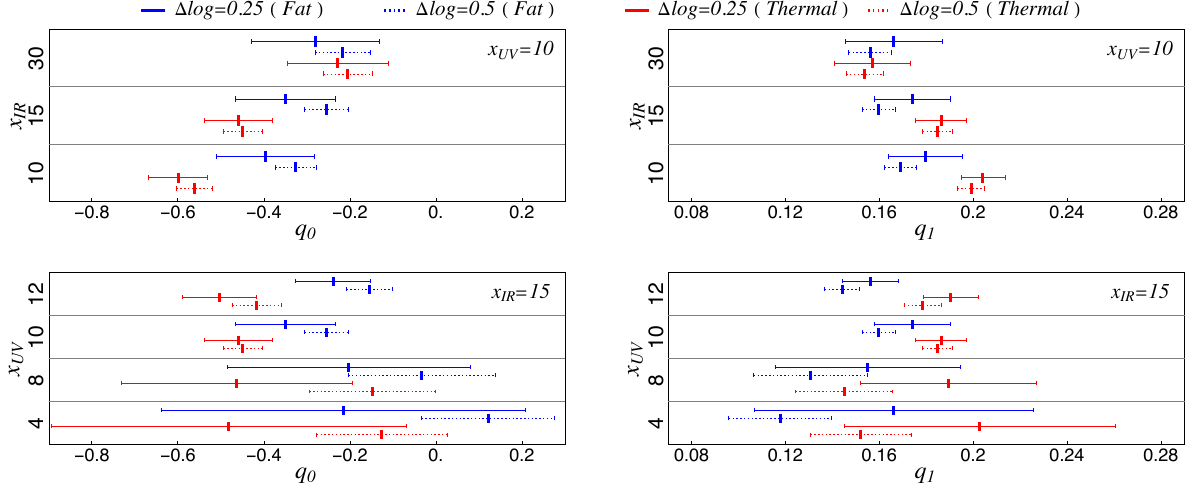}
\caption{\small Our fit results on the spectral index with log-hypothesis ($q = q_0 + q_1 \log\frac{m_r}{H}$), assuming the standard deviation of the form $\sigma = B x^{-p}+C$, for two types of relaxations (fat-string and thermal pre-evolutions) while varying $x_\text{IR}$ or $x_\text{UV}$. See Appendix~\ref{app:sec:fit:F} for the detail.}
\label{fig:q:summary:B}
\end{center}
\end{figure}

\begin{figure}[tp]
\begin{center}
\includegraphics[width=1.0\textwidth]{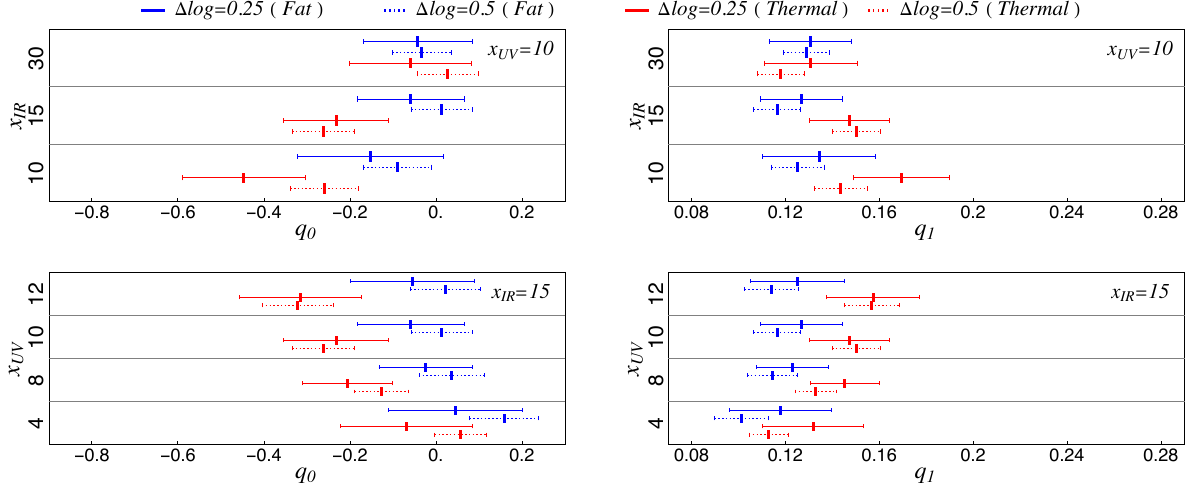}
\caption{\small Our fit results on the spectral index with log-hypothesis ($q = q_0 + q_1 \log\frac{m_r}{H}$), assuming the constant standard deviation for two types of relaxations (fat-string and thermal pre-evolutions) while varying $x_\text{IR}$ or $x_\text{UV}$. See Appendix~\ref{app:sec:fit:F} for the detail.}
\label{fig:q:summary:noB}
\end{center}
\end{figure}

The final fit results with two different forms of the standard deviations as well as varying UV, IR cutoffs and the time interval $\Delta \log$ are summarized in Figs.~\ref{fig:q:summary:B} and~\ref{fig:q:summary:noB}.

\subsection{Analytic understanding of positive correlation}
\label{app:sec:pos:correlation}

\begin{figure}[tp]
\begin{center}
\includegraphics[width=0.47\textwidth]{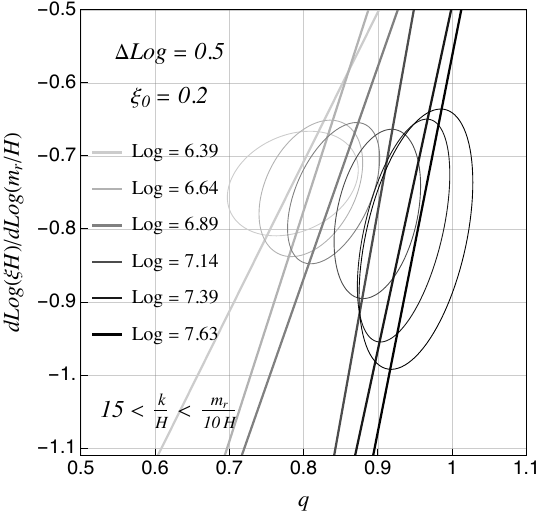}\quad
\includegraphics[width=0.47\textwidth]{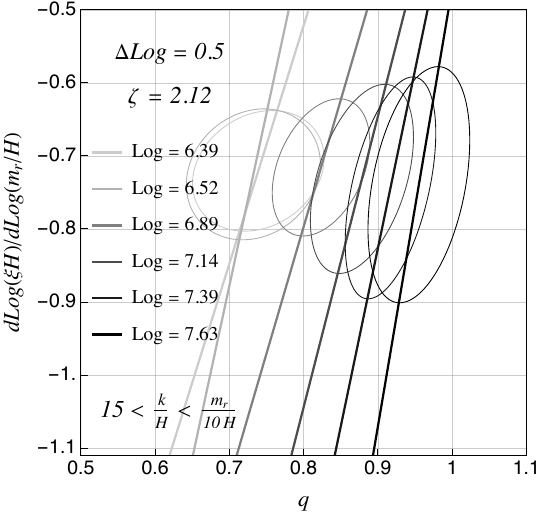}
\caption{\small The correlation between the differential rate of the comoving number of strings per Hubble with respect to the logarithmic time change, $d\log(\xi H)/(d\log\frac{m_r}{H})$, and the spectral index $q$ at various instant times from two benchmark simulations. The straight lines are major axes of the ellipses and each ellipse denotes 1-sigma contour of the Gaussian distribution over 100 individual runs.}
\label{fig:q:4096:dlog05:distr}
\end{center}
\end{figure}

Fig.~\ref{fig:q:4096:dlog05:distr} shows the correlation between the spectral index $q$ and $d\log(\xi H)/(d\log\frac{m_r}{H})$ (the differential rate of the comoving number of strings per Hubble with respect to time) for multiple instant time slices to illustrate that the orientation of the correlation axes get saturated at late times while the spectral index $q$ increases with the logarithmic time. The ellipse represents the 1$\sigma$ contour around the central values, assuming the Gaussian distribution of the 100 ensemble samples at each instant time. 
In both simulations using the fat-string and thermal pre-evolutions, the correlation axes, which are taken from the major axes of the ellipses, similarly stay constant at later times for $\log >7$. This correlation can hint us on how the spectral index might have shifted when a different initial condition following a different $\xi$ evolution curve was chosen.

The positive correlation between $q$ and $d\log(\xi H)/(d\log\frac{m_r}{H})$ can be qualitatively understood by the analytic study on the axion radiation power from sinusoidal string oscillations with small-amplitudes~\cite{Drew:2019mzc,Battye:1993jv}. 
The time averaged radiation power per unit length by long strings aligned in $z$-direction is given by the second harmonic at leading-order,
\begin{equation}
    \frac{\partial P}{\partial z}\simeq\frac{\beta f_a^2}{8\lambda_\text{str}}\varepsilon^4~,
\end{equation}
in terms of the wavelength of the string $\lambda_\text{str}$, the relative amplitude $\varepsilon\equiv 2\pi A/\lambda_\text{str}$ which is the ratio of the amplitude $A$ to its wavelength $\lambda_\text{str}$, and a constant $\beta$ encoding the string geometry, for instance, $\beta = \pi^3/4$ for the sinusoidal oscillation. 
The relative amplitude $\varepsilon$ is expected to not vary much with respect to $\lambda_\text{str}$ for being a normalized amplitude.
Assuming no energy transfer to radial modes, the radiation power can be related to the decay rate of the string energy per unit length, $\partial E_\text{str}/\partial z$, via the energy conservation,
\begin{equation}
    \frac{\partial}{\partial t}\frac{\partial E_\text{str}}{\partial z}=-\frac{\partial P}{\partial z}~.
\end{equation}
With the relation of the string energy to the string length assuming the constant string tension $\mu_\text{eff}$,  $\frac{\partial E_\text{str}}{\partial z}=\mu_\text{eff}\frac{\partial L_\text{str}}{\partial z}$, the back-reaction leads to shortening of string length $L_\text{str}$,
\begin{equation}\label{app:eq:Lstr:decayrate}
    \frac{1}{L_\text{str}}\frac{\partial L_\text{str}}{\partial t}=-\frac{1}{\partial E_\text{str}/\partial z}\frac{\partial P}{\partial z}\simeq -\frac{\beta f_a^2\varepsilon^4}{8\mu_\text{eff}\lambda_\text{str}}~.
\end{equation}
As the energy is mostly radiated into axions of wavelength $\lambda_\text{str}/2$~\cite{Drew:2019mzc}, the relation in Eq.~(\ref{app:eq:Lstr:decayrate}) indicates more rapid decay rate of the string length for higher-frequency (or shorter wavelength) fluctuations, which are associated with smaller spectral indices, and vice versa.

\section{More on axion abundance}
\label{app:sec:axion:na:detail}

The parametric behavior of the axion abundance crucially depends on the size of the spectral index $q$ (whether $q>1$ or $q=1$), the average axion field values in the $f_a$ unit (whether $\langle a^2 \rangle^{1/2}/f_a \ll 1$ or $\langle a^2 \rangle^{1/2}/f_a \gtrsim 1$), and the lower bound of the momentum integral (whether $k_\text{IR} = H$ or $H \sqrt{\xi}$).  The nature of the evolution after the scaling regime ends until the axions become nonrelativistic crucially relies on the value of $\langle a^2 \rangle^{1/2}/f_a$. The average inter-string distance is of the order of $1/(H\sqrt{\xi})$ as opposed to $H^{-1}$. The discrepancy in literature is basically originated from different set of those factors. Therefore, sorting out all possible parametric behaviors should be highly beneficial for a clearer comparison. In this section, we follow the similar computation to~\cite{Gorghetto:2020qws} not only filling more details, but also extending to other cases.

The axion spectrum is given by
\begin{equation}\label{app:eq:drhodk}
  \frac{\partial \rho_a}{\partial k} 
 = \int^t dt' \frac{\Gamma(t')}{H(t')} \left ( \frac{R(t')}{R(t)} \right )^3 F \left [ \frac{k'}{H(t')},\, \frac{m_r}{H(t')} \right ]~,
\end{equation}
where $k' = (R(t)/R(t'))k$ and the normalized instantaneous emission in the interval $x = [x_\text{IR},\, y/x_\text{UV}]$ is given by, depending on $q$, 
\begin{equation}\label{app:eq:F:original}
\begin{split}
F \left [x,\, y \right ] = &
\left\{ 
\begin{array} {lll}
\displaystyle\frac{q-1}{x_\text{IR}^{1-q} -\left (\frac{y}{x_\text{UV}} \right )^{1-q} }\, \frac{1}{x^q} & \text{for} & q > 1
\\[25pt]
\hspace{0.14in}  \displaystyle\frac{1}{\log\frac{y}{x_\text{UV} x_\text{IR}}}\, \frac{1}{x^q} & \text{for} & q = 1~.
\end{array}
\right.
\end{split}
\end{equation}
The transfer rate from strings to axions $\Gamma$ is approximated at late times by $\Gamma \sim \xi \mu_\text{eff}/t^3 \sim 8 \pi H^3 f_a^2 \xi \log \frac{m_r}{H}$~\cite{Gorghetto:2018myk} where $\langle \gamma \rangle \mu_0 = \pi f_a^2$ in Eq.~(\ref{eq:mueff:th}) was used in the second relation.

\subsection{Estimation for $q>1$}
\label{app:sec:q:large}

Following similar steps to~\cite{Gorghetto:2020qws}, we show that modifying IR momentum cutoff can induce non-negligible change in the overall size of the axion abundance due to the slightly changed parametric behavior. 
Using the approximation $\xi \sim c_1 \log\frac{m_r}{H}$ at late times, the axion spectrum in Eq.~(\ref{app:eq:drhodk}) can be rewritten as
\begin{equation}\label{app:eq:drhodk:yvariable}
  \frac{\partial \rho_a}{\partial k} (k,\, t)
 =  \frac{4\pi f_a^2 c_1 m_r}{y^{3/2}} \int^y_{y_0} dy' \frac{(\log y')^2}{{y'}^{1/2}} F \left [ \left ( \frac{y'}{y} \right )^{1/2} x,\, y' \right ]~,
\end{equation}
where $x = \frac{k}{H}$, $y = \frac{m_r}{H(t)}$, and the integration over $t'$ in the interval $[t_0,\, t]$ was converted in terms of $y'=\frac{m_r}{H(t')}$ with the interval $[y_0,\, y]$. We take into account the modified IR cutoff by setting $x_\text{IR} = x_0 \sqrt{\xi}$. $x_\text{UV} = 1$ is chosen similarly to~\cite{Gorghetto:2020qws} and the result is insensitive to this choice. The instantaneous emission in Eq.~(\ref{app:eq:F:original}) for $q > 1$ can be approximated as
\begin{equation}
\begin{split}
F \left [x,\, y \right ] \sim &
\left\{ 
\begin{array} {lll}
\displaystyle\frac{q-1}{x_\text{IR}^{1-q}  }\, \frac{1}{x^q} & \text{for} &  x_0 \sqrt{\xi} = x_\text{IR} \leq x \leq y
\\[15pt]
\hspace{0.14in}  0 & & \text{otherwise}~.
\end{array}
\right.
\end{split}
\end{equation}
In terms of the arguments in Eq.~(\ref{app:eq:drhodk:yvariable}), the finite support of the function $F$ is defined by the interval $(x_0^2\, c_1 \log y')^{1/2} \leq (y'/y)^{1/2} x \leq y'$ and, subsequently, it defines the integration range $y_1 \leq y' \leq y_2$ in which $F$ in Eq.~(\ref{app:eq:drhodk:yvariable}) is nonvanishing. There are two choices of $(y_1,\, y_2)$ depending on the value of $y^{-1/2}x = \frac{k}{\sqrt{m_r H}}$. 

When $y^{-1/2}x$ is large (large momentum region), the integration is done over $x^2/y = y_1 \leq y' \leq y_2 = y$ where $x^2/y$ is the intersection between two curves $y'$ and $(y'/y)^{1/2} x$. 
The integration in Eq.~(\ref{app:eq:drhodk:yvariable}) can be done straightforwardly:
\begin{equation}\label{app:eq:drhodk:yvariable:int}
\begin{split} 
  \frac{\partial \rho_a}{\partial k} (k,\, t)
 &=  8\pi f_a^2 c_1 m_r (x_0^2 c_1 )^{\frac{q-1}{2}} y^{\frac{q-3}{2}}\frac{1}{x^q} \left ( \frac{q-1}{2} \right )^{-\frac{q+3}{2}}
 \\[5pt]
  &\quad \times \left [ \Gamma \left ( \frac{q+5}{2},\, \frac{q-1}{2} \log \frac{x^2}{y} \right ) - \Gamma \left ( \frac{q+5}{2},\, \frac{q-1}{2} \log y \right ) \right ] ~,
\end{split}
\end{equation}
where $\Gamma(s,x) = \int_x^\infty dt\, t^{s-1} e^{-t}$ is the incomplete Gamma function. The variable $x^2/y = \frac{k^2}{m_r H}$ for the fixed momentum $k$. When $\frac{m_r}{H}$ is taken to be large, the second argument of $\Gamma$ function in Eq.~(\ref{app:eq:drhodk:yvariable:int}) becomes large as well. Since $\Gamma(s,\, x) \rightarrow x^{s-1} e^{-x}$ as $x \rightarrow \infty$, the result in Eq.~(\ref{app:eq:drhodk:yvariable:int}) can be approximated as, in the large $m_r/H$ limit,
\begin{equation}\label{app:eq:drhodk:yvariable:int:approx}
\begin{split} 
  \frac{\partial \rho_a}{\partial k} (k,\, t)
 \approx \frac{8 H \mu_\text{eff} \sqrt{\xi}}{x_0} \left ( \frac{Hx_0 \sqrt{\xi}}{k} \right )^q f \left ( \frac{m_r}{H},\, \frac{k}{m_r} \right )~,
\end{split}
\end{equation}
where $\pi f_a^2 \log y \sim \mu_\text{eff}$ and $c_1\log y \sim \xi$ and the function $f(y,\, u)$ is defined as
\begin{equation}
  f(y,\, u) = \left ( \frac{\log (yu^2)}{u^2 \log y} \right )^{\frac{q+3}{2}} u^4 - 1~.
\end{equation}
Since $f(y,\, 1) = 0$ and $\partial f/\partial u < 0$ for $\exp \frac{q+3}{2(q-1)} < y^{1/2} u = \frac{k}{\sqrt{m_r H}}$, the function $f(y,\, u)$ is positive decreasing function in $u$ for $u < 1$. Therefore, we see that $\partial \rho_a/\partial k$ rapidly decays faster than $\sim k^{-q}$ for the momentum $k \gtrsim e^{ \frac{q+3}{2(q-1)} } \sqrt{m_r H}$, and its contribution to the axion abundance will be accordingly suppressed.
 
When $y^{-1/2}x$ is small (low momentum region), the integration is done over $- \frac{x_0^2 c_1 y}{x^2} W_{-1}(-\frac{x^2}{x_0^2 c_1 y}) = y_1 \leq y' \leq y_2 = y$ where $y_1$ in terms of Lambert $W$ function is the intersection between two curves $(y'/y)^{1/2} x$ and $(x_0^2 c_1 \log y')^{1/2}$.
The integration in Eq.~(\ref{app:eq:drhodk:yvariable}) can also be done straightforwardly and it is given by
\begin{equation}\label{app:eq:drhodk:yvariable:int:lowk}
\begin{split} 
  \frac{\partial \rho_a}{\partial k} (k,\, t)
 & =  8\pi f_a^2 c_1 m_r (x_0^2 c_1 )^{\frac{q-1}{2}} y^{\frac{q-3}{2}}\frac{1}{x^q} \left ( \frac{q-1}{2} \right )^{-\frac{q+3}{2}}
 \\[5pt]
  &\times \left [ \Gamma \left ( \frac{q+5}{2},\, \frac{q-1}{2} \log \left ( - \frac{x_0^2 c_1 y}{x^2} W_{-1} \left (-\frac{x^2}{x_0^2 c_1 y} \right ) \right ) \right ) - \Gamma \left ( \frac{q+5}{2},\, \frac{q-1}{2} \log y \right ) \right ] ~,
  \\[5pt]
 & =  8\pi f_a^2 c_1 m_r (x_0^2 c_1 )^{\frac{q-1}{2}} y^{\frac{q-3}{2}}\frac{1}{x^q} \left ( \frac{q-1}{2} \right )^{-\frac{q+3}{2}}
 \\[5pt]
  &\times \left [ \Gamma \left ( \frac{q+5}{2},\, -\frac{q-1}{2}  W_{-1} \left (-\frac{x^2}{x_0^2 c_1 y} \right ) \right ) - \Gamma \left ( \frac{q+5}{2},\, \frac{q-1}{2} \log y \right ) \right ] ~,
\end{split}
\end{equation}
where $\log(-w W_k (-w^{-1})) = - W_k (-w^{-1})$ was used in the second relation. 
It can be easily shown that $-W_{-1}(-w^{-1})$ becomes large at large $y$ for the low momentum.
Using $\Gamma(s,\, x) \rightarrow x^{s-1} e^{-x}$ as $x \rightarrow \infty$ as before, the expression in Eq.~(\ref{app:eq:drhodk:yvariable:int:lowk}) can similarly be approximated as, in the large $\log\frac{m_r}{H}$ limit,
\begin{equation}\label{app:eq:drhodk:yvariable:int:approx:lowk}
\begin{split} 
  \frac{\partial \rho_a}{\partial k} (k,\, t)
 \approx \frac{8 H^2 \mu_\text{eff} \xi}{k}
 \left [ \left ( \frac{-W_{-1} \left ( -\frac{H}{m_r}\log \frac{m_r}{H} \left ( \frac{k}{x_0 H \sqrt{\xi}} \right )^2  \right ) }{\log\frac{m_r}{H}}\right )^2  - \left ( \frac{k}{x_0 H \sqrt{\xi}} \right )^{1-q}\right ]~,
\end{split}
\end{equation}
where the identity $\exp [-W_k(-w^{-1})] = - w W_k (- w^{-1})$ was used and $\pi f_a^2 \log y \sim \mu_\text{eff}$ and $c_1\log y \sim \xi$ were used for a neat expression. The above expression can be further simplified by approximating the Lambert function,
\begin{equation}\label{app:eq:drhodk:yvariable:int:approx2:lowk}
\begin{split} 
  \frac{\partial \rho_a}{\partial k} (k,\, t)
 \approx \frac{8 H^2 \mu_\text{eff} \xi}{k}
 \left [ \left ( 1 - 2 \frac{\log \frac{k}{x_0 H\sqrt{\xi}}}{\log\frac{m_r}{H}} \right )^2  - \left ( \frac{k}{x_0 H \sqrt{\xi}} \right )^{1-q}\right ]~.
\end{split}
\end{equation}
From Eq.~(\ref{app:eq:drhodk:yvariable:int:approx2:lowk}), we see that axion spectrum $\partial \rho_a/\partial k$ scales as $\sim k^{-1}$ in the low momentum region $x_0 H\sqrt{\xi} \lesssim k \lesssim z_k \sqrt{m_r H}$
\footnote{$z_k = (- \frac{x_0^2 c_1}{2} W_{-1}(-\frac{2}{x_0^2 c_1}))^{1/4}$ can be read off from the equality $- \frac{x_0^2 c_1 y}{x^2} W_{-1}(-\frac{x^2}{x_0^2 c_1 y}) = \frac{x^2}{y}$, provided that $\frac{x^2}{x_0^2 c_1 y} < e^{-1}$, for our relevant choice of parameters, and it is found to be order one constant. The large and low momentum regions are separated below and above $\sqrt{x_0 m_r H}$ when the IR cutoff of $H$ is used. }. Comparing with~\cite{Gorghetto:2020qws}, the net effect of changing the IR cutoff from $x_0 H$ to $x_0 H\sqrt{\xi}$ is equivalent to shifting $k^0 = x_0 H$ (in their notation) to $k^0 = x_0 H \sqrt{\xi}$ without changing the overall factor.

Another place where the modified IR cutoff can affect is the average axion field value, $\langle a^2 \rangle^{1/2}/f_a$ where
$\langle a^2(t) \rangle = \int dk k^{-2} (\partial\rho_a (k,\, t)/\partial k)$. Using our result in Eq.~(\ref{app:eq:drhodk:yvariable:int:approx2:lowk}), we can evaluate $\langle a^2(t) \rangle$ (or by a simple power counting to get the leading term) with the modified cutoff $k_\text{IR} = x_0 H \sqrt{\xi}$, and it gives rise to
$\langle a^2(t) \rangle \approx \frac{8\mu_\text{eff}}{x_0^2} (1/2 - (1+q)^{-1} - \log^{-1}\frac{m_r}{H} + \log^{-2}\frac{m_r}{H})$.
That is, $\langle a^2(t) \rangle \sim 4 \mu_\text{eff}$ at late times for $q > 1$, as opposed to $\langle a^2(t) \rangle \sim 4 \mu_\text{eff} \xi$ when the IR cutoff is $\sim H$. The ratio $\langle a^2 \rangle^{1/2}/f_a$ is reduced by the factor of $\sqrt{\xi}$. While $\langle a^2(t_\star) \rangle \sim 4 \mu_{\text{eff}\star} \gg 1$ is still expected around the time $t_\star$ of the QCD crossover based on our simulation, it may be informative to consider the opposite case with $\langle a^2(t) \rangle^{1/2}/f_a \ll 1$ and see how the modified IR cutoff by $\sqrt{\xi}$ can change the parametric of the axion abundance. In this hypothetical situation, the nonlinearities due to the axion potential will be suppressed, and the axion number density at later time $t > t_\star$ will be simply given by $n_a^{\text{str}}(t) = (H(t)/H_\star)^{3/2}\, n_a^{\text{str}}(t_\star)$ where $n_a^{\text{str}}(t_\star) = \int dk k^{-1} (\partial \rho_a/\partial k) \approx 8H_\star \mu_{\text{eff}\star} \sqrt{\xi_\star}$, as opposed to 
$8H_\star \mu_{\text{eff}\star} \xi_\star$ when $k_\text{IR} \sim H$ was assumed. That is, the shift by $\sqrt{\xi}$ in the cutoff suppresses the abundance by the factor of $\sqrt{\xi}$.

Now we get back to our situation where a large $\langle a^2(t_\star) \rangle^{1/2}/f_a$ is expected and the nonlinearities arising from the axion potential can not be neglected. The axions after time $t_\star$ still evolves as free relativistic fields, and the axion energy density at time $t > t_\star$ is accordingly redshifted and it is given by, in the range of $x_0 \sqrt{\xi_\star H_\star H} < k < z_k \sqrt{m_r H}$,
\begin{equation}\label{app:eq:drhodk:yvariable:int:approx2:lowk:redshift}
\begin{split} 
  \frac{\partial \rho_a}{\partial k} (k,\, t)
 \approx \frac{8 H^2 \mu_{\text{eff}\star} \xi_\star}{k}
 \left [ \left ( 1 - 2 \frac{\log \frac{k}{x_0 \sqrt{\xi_\star H_\star H}}}{\log\frac{m_r}{H_\star}} \right )^2  
 - \left ( \frac{k}{x_0 \sqrt{\xi_\star H_\star H}} \right )^{1-q}\right ]~.
\end{split}
\end{equation}
The evolution continues until the transition time, denoted by $t_\ell$, where the axion energy density stored in the gradient terms becomes comparable with the axion potential and during which the axion energy density is assumed to be promptly converted into non-relativistic ones, namely
\begin{equation}\label{app:eq:rhoIR:eq:axionpot}
   \rho_\text{IR}(t_\ell) = \int_{x_0 \sqrt{\xi_\star H_\star H}}^{c_m m_a (t_\ell)} 
    dk \frac{\partial \rho_a}{\partial k} (k,\, t_\ell) 
    = c_V m_a^2 (t_\ell ) f_a^2~,
\end{equation}
where $c_m$ and $c_V$ are order one parameters that have to be determined by numerical simulation. Axions with $k > c_m m_a$ will decay faster than those contributing to the dominant abundance. 
The relation in Eq.~(\ref{app:eq:rhoIR:eq:axionpot}) leads to the condition (differs in the definition of $\kappa$ compared to~\cite{Gorghetto:2020qws}),
\begin{equation}\label{app:eq:rhoIR:eq:axionpot:con1}
 8 H^2_\ell \mu_{\text{eff}\star} \xi_\star \left [ \log \kappa 
 \left ( 1 - 2 \frac{\log \kappa}{\log\frac{m_r}{H_\star}} + \frac{4}{3} \frac{\log^2 \kappa}{\log^2 \frac{m_r}{H_\star}} 
 - \frac{1-\kappa^{1-q}}{q-1}
  \right )  \right ]
  = c_V m_a^2 (t_\ell ) f_a^2~,
\end{equation}
where $\kappa = \frac{c_m m_a (t_\ell )}{x_0 \sqrt{\xi_\star H_\star H}}$ and $H_\ell = H(t_\ell)$. 
Parametrizing the axion mass as $m_a(t) = H_\star (H_\star / H)^{\alpha/4}$ with $m_a (t_\star) = H_\star$ and introducing~\cite{Gorghetto:2020qws},
\begin{equation}\label{app:eq:delaypar}
  z  \equiv \left ( \frac{m_a(t_\ell)}{H_\star} \right )^{1+\frac{6}{\alpha}}~,
\end{equation}
we obtain, keeping only the first term as the dominant one in Eq.~(\ref{app:eq:rhoIR:eq:axionpot:con1}) at late times $\log \gg 1$,
\begin{equation}\label{app:eq:cond:on:z:qbig}
  \frac{8 \mu_{\text{eff}\star} \xi_\star}{c_V f_a^2} \log \left ( \frac{c_m}{x_0 \sqrt{\xi_\star}} z^{\frac{\alpha + 2}{\alpha + 6}} \right )
  =
  z^{\frac{2(\alpha + 4)}{\alpha + 6}}~.
\end{equation} 
Comparing with the result in~\cite{Gorghetto:2020qws}, modifying the IR momentum cutoff amounts to shifting $x_0$ to $x_0\sqrt{\xi_\star}$.  
The axion abundance at $t = t_\ell$ is estimated as 
$n_a^{\text{str}} (t_\ell) = c_n \frac{\rho_\text{IR} (t_\ell)}{m_a(t_\ell)} = c_n c_V m_a (t_\ell ) f_a^2$
whereas the contribution from the misalignment with order one angle is 
$n_a^\text{mis}(t_\ell) = c'_n m_a (t_\star) f_a^2 (H_\ell/H_\star)^{3/2}$. 
Expressing $m_a (t_\ell)$ in terms of the solution $z$ using Eq.~(\ref{app:eq:delaypar}) and taking ratio of the two different types of contributions, we obtain
\begin{equation}\label{app:eq:natonmis:master}
 \frac{n_a^{\text{str}} (t_\ell)}{n_a^\text{mis}(t_\ell) } = \frac{c_n c_V}{c'_n} z~,
\end{equation}
where the solution $z$ of Eq.~(\ref{app:eq:cond:on:z:qbig}) is given by
\begin{equation}\label{app:eq:z:sol:exact}
 z = \left [ - \frac{\alpha +2}{2(\alpha + 4)} \frac{8 \mu_{\text{eff}\star} \xi_\star}{c_V f_a^2} 
 W_{-1} \left ( -\frac{2(\alpha +4)}{\alpha + 2} \frac{c_V f_a^2}{8 \mu_{\text{eff}\star} \xi_\star}  \left ( \frac{c_m}{x_0\sqrt{\xi_\star}} \right )^{-\frac{2(\alpha +4)}{\alpha+2}} \right )
  \right ]^{\frac{\alpha + 6}{2(\alpha + 4)}}~,
\end{equation}
which is the same as that in~\cite{Gorghetto:2020qws} except for $x_0 \rightarrow x_0\sqrt{\xi}$. Using the relation, 
$-W_{-1} (- w^{-1} ) = \log (w \log (w \log (\cdots)))$, the above ratio of the two different contributions is finally given by (in a similar form to~\cite{Gorghetto:2020qws})
\begin{equation}
 \frac{n_a^{\text{str}, q>1} (t_\ell)}{n_a^\text{mis}(t_\ell) } = \frac{c_n c_V}{c'_n}
 \left [ 
  \frac{4 \mu_{\text{eff}\star} \xi_\star}{c_V f_a^2} \frac{\alpha+2}{\alpha + 4} 
  \log \left (  \frac{4 \mu_{\text{eff}\star} \xi_\star}{c_V f_a^2} \frac{\alpha+2}{\alpha + 4} \left (\frac{c_m}{x_0\sqrt{\xi_\star}} \right )^{\frac{2(\alpha+4)}{\alpha +2} }  \log (\cdots) \right )
 \right ]^{\frac{\alpha+6}{2(\alpha +4)}}~.
\end{equation}
Comparing to~\cite{Gorghetto:2020qws}, the final result differs by the replacement of $x_0 \rightarrow x_0\sqrt{\xi_\star}$ without changing the overall factors. Upon replacements of $\mu_{\text{eff}\star}  \sim \pi f_a^2 \log_\star$, $\xi_\star  \sim c_1 \log_\star$ inside $\log$ at late times, or $\log_\star \gg 1$, the shifted momentum cutoff induces a modification of the overall coefficient of the axion abundance (and sub-leading terms) as
\begin{equation}
 \frac{n_a^{\text{str}, q>1} (t_\ell)}{n_a^\text{mis}(t_\ell) } \approx \frac{c_n c_V}{c'_n}
 \left [ 
  \frac{4 \mu_{\text{eff}\star} \xi_\star}{c_V f_a^2} \frac{\alpha}{\alpha + 4} 
  \left ( \log \log\frac{m_r}{H_\star} +  \mathcal{O} (\log\log\log\frac{m_r}{H_\star})\right ) 
 \right ]^{\frac{1}{2}\left (1+\frac{2}{\alpha+4} \right )}~,
\end{equation}
where the factor $\frac{\alpha}{\alpha+4}$ will change to $\frac{2(\alpha + 2)}{\alpha+4}$ when $k_\text{IR} = x_0 H$ is used as in~\cite{Gorghetto:2020qws}.  For instance, it can cause roughly 20\% reduction of the overall rate for the typical choice $\alpha = 8$.

\subsection{Estimation for $q = 1$}
\label{app:sec:na:qone}
We extend the previous discussion to the case with $q=1$  for the sake of completeness (and for a clear comparison with literature). We primarily present the estimate with IR cutoff $x_\text{IR}=x_0\sqrt{\xi}$ while commenting on the case with $x_\text{IR}=x_0$ when it is relevant. $x_\text{UV}=1$ is chosen as before. 
The instantaneous emission for $q=1$ is given by
\begin{equation}
\begin{split}
F \left [x,\, y \right ] \sim &
\left\{ 
\begin{array} {lll}
\displaystyle\frac{1}{\log{\frac{y}{x_\text{IR}}}}\, \frac{1}{x} & \text{for} &  x_0 \sqrt{\xi} = x_\text{IR} \leq x \leq y
\\[15pt]
\hspace{0.14in}  0 & & \text{otherwise}~.
\end{array}
\right.
\end{split}
\end{equation}
Since the finite support of $F$ is the same as in Section~\ref{app:sec:q:large}, the integration in Eq.~(\ref{app:eq:drhodk:yvariable}) can be done similarly for large and small momenta $k$, or $y^{-1/2}x$ over similar integration ranges to Section~\ref{app:sec:q:large}. 

When $y^{-1/2}x$ is large, the integration gives rise to
\begin{equation}\label{app:eq:drhodk:q1:yvariable:int:highk}
\begin{split} 
  \frac{\partial \rho_a}{\partial k} (k,\, t)
 &=  \frac{2\pi f_a^2 c_1 m_r}{yx}\left(\log^2{y}-\log^2{\frac{x^2}{y}}\right)
 \approx\frac{8H^2\mu_\text{eff}\xi}{k}\frac{\log\frac{k}{H}}{\log\frac{m_r}{H}}\left(1-\frac{\log\frac{k}{H}}{\log\frac{m_r}{H}}\right) ~,
\end{split}
\end{equation}
where $\pi f_a^2\log{y}\sim\mu_\text{eff}$ and $c_1\log{y}\sim\xi$ were used. 
While $\partial\rho_a/\partial k$ scales as $\sim k^{-1}$ as expected, it is multiplied by a logarithmic suppression $\frac{\log{(k/H)}}{\log{(m_r/H)}}$. Similarly, when $y^{-1/2}x$ is small, the integration is given by
\begin{equation}\label{app:eq:drhodk:q1:yvariable:int:lowk}
\begin{split} 
  \frac{\partial \rho_a}{\partial k} (k,\, t)
 &=\frac{2\pi f_a^2 c_1 m_r}{yx} \left [ \log^2{y}-\log^2\left(-\frac{{x_0}^2c_1y}{x^2}W_{-1}\left(-\frac{x^2}{{x_0}^2c_1y}\right) \right) \right ]
 \\[5pt]
  &=\frac{2\pi f_a^2 c_1 m_r}{yx} \left [ \log^2{y}-\left(-W_{-1}\left(-\frac{x^2}{{x_0}^2c_1y}\right) \right)^2 \right ] ~,
\end{split}
\end{equation}
where the relation $\log(-w W_k (-w^{-1})) = - W_k (-w^{-1})$ is used in the second relation. 
Using the approximations $\pi f_a^2\log{y}\sim\mu_\text{eff}$ and $c_1\log{y}\sim\xi$ and approximating the Lambert function, the expression in Eq.~(\ref{app:eq:drhodk:q1:yvariable:int:lowk}) is further simplified as
\begin{equation}\label{app:eq:drhodk:q1:yvariable:int:approx:lowk}
\begin{split} 
  \frac{\partial \rho_a}{\partial k} (k,\, t)
 \approx \frac{8 H^2 \mu_\text{eff} \xi}{k}
 \frac{\log{\frac{k}{x_0H\sqrt{\xi}}}}{\log{\frac{m_r}{H}}}\left(1- \frac{\log{\frac{k}{x_0H\sqrt{\xi}}}}{\log{\frac{m_r}{H}}}\right)~.
\end{split}
\end{equation}
The axion spectrum $\partial\rho_a/\partial k$ at small momentum also scales as $\sim k^{-1}$, as is expected, with a logarithmic suppression. Unlike the situation with $q>1$, the current case with $q=1$ shows the power law behavior $\sim k^{-1}$ over the entire momentum range as well as the logarithmic suppression. These properties remain the same even if $x_\text{IR}=x_0$ is taken. It does not change $\partial\rho_a/\partial k$ except for replacing $x_0H\sqrt{\xi}$ with $x_0H$ and, subsequently, the splitting of the momentum range into $x_0H<k\leq \sqrt{x_0m_rH}$ (for low) and $\sqrt{x_0m_rH}<k$ (for high).

Switching from $q>1$ to $q=1$ case also affects the average axion field value. 
Using the result in Eq.~(\ref{app:eq:drhodk:q1:yvariable:int:approx:lowk}) with cutoff $k_\text{IR}=x_0H\sqrt{\xi}$, the average axion field value is given by
\begin{equation}
 \langle a^2(t) \rangle \approx \frac{2\mu_\text{eff}}{x_0^2\log{\frac{m_r}{H}}} \quad \text{for}\quad x_\text{IR} = x_0 \sqrt{\xi}~,
\end{equation}
that is, $\langle a^2(t) \rangle \sim 2\pi f_a^2$ at late times with no $\log{\frac{m_r}{H}}$ enhancement unlike the $q>1$ case. However, this behavior is IR cutoff sensitive, and taking $x_\text{IR}=x_0$ instead gives 
\begin{equation}
 \langle a^2(t) \rangle \approx \frac{2\mu_\text{eff}\xi}{x_0^2\log{\frac{m_r}{H}}} \quad \text{for}\quad x_\text{IR} = x_0~,
\end{equation}
which has a $\log$ enhancement. 
It implies that the size of $\langle a^2(t_\star) \rangle^{1/2}/f_a$ is not obvious a priori when $q=1$, and an
explicit numerical check may be necessary as in~\cite{Buschmann:2021sdq}. Here we discuss the parametric behavior of the axion number density for both cases, one with $\langle a^2(t_\star) \rangle^{1/2}/f_a\ll1$ and the other one with $\langle a^2(t_\star) \rangle^{1/2}/f_a\gg1$. In the former case, the scaling in $\log{\frac{m_r}{H_\star}}$ is sensitive to the IR cutoff. 
As was discussed in Section~\ref{app:sec:q:large}, the axion number density at a later time $t>t_\star$ is given by $n_a^{\text{str}}(t) = (H(t)/H_\star)^{3/2}\, n_a^{\text{str}}(t_\star)$ where $n_a^{\text{str}}(t_\star) = \int dk k^{-1} (\partial \rho_a/\partial k) \approx \frac{8\pi f_a^2 H_\star}{x_0}\sqrt{\xi_\star}$ for $x_\text{IR}=x_0\sqrt{\xi}$. 
The abundance is enhanced by the factor of $\sqrt{\xi_\star}$, when $x_\text{IR}=x_0$, or $n_a^{\text{str}}(t_\star) \approx \frac{8\pi f_a^2 H_\star\xi_\star}{x_0}$. However, this may not cause a large deviation in the abundance $n_a^{\text{str}}(t_\star)$ since $\xi_\star$ is limited to $\xi_\star\lesssim\frac{x_0^2}{2\pi}$ to satisfy $\langle a^2(t_\star) \rangle^{1/2}/f_a\lesssim1$.

Now we discuss the latter case, where $\langle a^2(t_\star) \rangle^{1/2}/f_a$ is large. Similarly to the case for $q>1$, the axion radiation around the transition time $t_\ell$ is estimated by solving the relation in  Eq.~(\ref{app:eq:rhoIR:eq:axionpot}) with the axion energy density at time $t>t_\star$, 
\begin{equation}\label{app:eq:drhodk:q1:yvariable:int:approx:lowk:redshift}
\begin{split} 
  \frac{\partial \rho_a}{\partial k} (k,\, t)
 \approx \frac{8 H^2 \mu_{\text{eff}\star} \xi_\star}{k}
  \frac{\log{\frac{k}{x_0\sqrt{\xi_\star H_\star H}}}}{\log{\frac{m_r}{H_\star}}}\left(1- \frac{\log{\frac{k}{x_0\sqrt{\xi_\star H_\star H}}}}{\log{\frac{m_r}{H_\star}}}\right)~,
\end{split}
\end{equation}
in the range of $x_0\sqrt{\xi_\star H_\star H} < k < z_k \sqrt{m_r H}$. Solving the relation in Eq.~(\ref{app:eq:rhoIR:eq:axionpot})  and expressing in terms of $\kappa = \frac{c_m m_a (t_\ell )}{x_0 \sqrt{\xi_\star H_\star H}}$ and $H_\ell = H(t_\ell)$ gives rise to
\begin{equation}\label{app:eq:rhoIR:q1:eq:axionpot:con1}
8\pi f_a^2 H_l^2\xi_{\star}(\log\kappa)^2\left(\frac{1}{2}-\frac{1}{3}\frac{\log{\kappa}}{\log{\frac{m_r}{H_\star}}}\right)=c_Vm_a^2(t_\ell)f_a^2~.
\end{equation}
With the same parametrization of the axion mass $m_a(t)$, introducing the $z$ parameter as in Section~\ref{app:sec:q:large}, we obtain, keeping only the first term in the parenthesis on the left hand side of Eq.~(\ref{app:eq:rhoIR:q1:eq:axionpot:con1}),
\begin{equation}\label{app:eq:cond:on:z}
  \left(\frac{4\pi\xi_\star}{c_V}\right)^{\frac{1}{2}}\log\left(\frac{c_m}{x_0\sqrt{\xi_\star}}z^{\frac{\alpha+2}{\alpha+6}}\right)=z^{\frac{\alpha+4}{\alpha+6}}~,
\end{equation} 
where choosing $x_\text{IR}=x_0$ amounts to replacing $x_0\sqrt{\xi_\star}$ with $x_0$. As in Section~\ref{app:sec:q:large}, the relative axion abundance with respect to the misalignment can be estimated using the solution $z$ to Eq.~(\ref{app:eq:cond:on:z}) which is given by
\begin{equation}\label{app:eq:zsol:qunit}
 z = \left [ - \frac{\alpha +2}{\alpha + 4} \left(\frac{4\pi \xi_\star}{c_V}\right)^{\frac{1}{2}} 
 W_{-1} \left ( -\frac{\alpha +4}{\alpha + 2}\left(\frac{c_V}{4\pi\xi_\star}\right)^{\frac{1}{2}} \left ( \frac{c_m}{x_0\sqrt{\xi}} \right )^{-\frac{\alpha +4}{\alpha+2}} \right )
   \right ]^{\frac{\alpha + 6}{\alpha + 4}}~.
\end{equation}
Unlike the case for $q>1$, the above solution $z$ in Eq.~(\ref{app:eq:zsol:qunit}) does not hold for an arbitrarily large $\xi_\star$ for given fixed value of $c_m$. The value of $-W_{-1}(-w^{-1})$ exists only for $w\geq e$, and the solution in Eq.~(\ref{app:eq:zsol:qunit}) makes sense only for $\xi_\star$ smaller than $\left(\frac{1}{e}\frac{\alpha+2}{\alpha+4}\sqrt{\frac{4\pi}{c_V}}\right)^{\alpha+2}\left(\frac{c_m}{x_0}\right)^{\alpha+4}$, denoted by $\xi_{\star\text{max}}$. 
The solution to Eq.~(\ref{app:eq:zsol:qunit}) does not exist for larger values of $\xi_\star > \xi_{\star\text{max}}$ since the low momentum range $x_0\sqrt{\xi_\star H_\star H}<k\leq c_mm_a(t)$ (that dominantly contributes to $\rho_{IR}$) becomes too narrow so that $\rho_{IR}(t)$ can not be large enough to satisfy Eq.~(\ref{app:eq:zsol:qunit}). 

A large $\xi_\star$ may become compatible with the solution to Eq.~(\ref{app:eq:zsol:qunit}) if $c_m$ is allowed to depend on $\xi_\star$. 
While the actual values of $c_m$ should be determined from numerical simulations, applying the property $-W_{-1}(-w^{-1})\geq 1$ to the solution $z$ in Eq.~(\ref{app:eq:zsol:qunit}), we obtain
\begin{equation}
    z \geq \left(\frac{\alpha+2}{\alpha+4}\left(\frac{4\pi\xi_\star}{c_V}\right)^{\frac{1}{2}}\right)^{\frac{\alpha+6}{\alpha+4}}~,
\end{equation}
where switching inequality to equality amounts to restoring the multiplicative $-W_{-1}(-w^{-1})$ function which depends at most logarithmically on $w$. Finally, the axion abundance from strings in the scaling regime is estimated as
\begin{equation}
 \frac{n_a^{\text{str}, q=1} (t_\ell)}{n_a^\text{mis}(t_\ell) } \approx \frac{c_n c_V}{c'_n}
 \left [ \frac{\alpha+2}{\alpha + 4}
  \left(\frac{4 \pi \xi_\star}{c_V}\right)^{\frac{1}{2}}
 \right ]^{1+\frac{2}{\alpha+4}}~,
\end{equation}
with a possible multiplicative factor scaling at most as $\log\log\frac{m_r}{H_\star}$.

\newpage

{\small
\bibliography{lit}{}
\bibliographystyle{JHEP}}

\end{document}